\documentclass{aa}  
\usepackage{graphicx}
\usepackage{txfonts}
\usepackage{xcolor}
\usepackage{blindtext}

\usepackage{booktabs,rotating}
\usepackage{lscape}
\usepackage{multirow}

\usepackage[flushleft]{threeparttable}

\newcommand{\red}{\textcolor{red}}

\usepackage{subcaption}
\usepackage{placeins} 

\begin{document}

   \title{The detection of cluster magnetic fields via radio source depolarisation\thanks{Full Tables \ref{tab:polsources}, \ref{tab:unpolsources} and \ref{tab:clusters} are only available at the CDS via anonymous ftp to cdsarc.u-strasbg.fr (130.79.128.5) or via http://cdsarc.u-strasbg.fr/viz-bin/cat/J/A+A/665/A71}}

   \author{E. Osinga
          \inst{1}
          \and 
          R. J. van Weeren\inst{1}
          \and
          F. Andrade-Santos\inst{2,9}
          \and 
          L. Rudnick\inst{3}
          \and
          A. Bonafede\inst{4,5}
          \and
          T. Clarke\inst{6}
          \and
          K. Duncan\inst{7}
          \and
          S. Giacintucci\inst{6}
          \and 
          Tony Mroczkowski\inst{8}
          \and
          H. J. A. R\"ottgering\inst{1}
          }

   \institute{Leiden Observatory, Leiden University, PO Box 9513, NL-2300 RA Leiden, The Netherlands  \email{osinga@strw.leidenuniv.nl} 
    \and
    Center for Astrophysics | Harvard \& Smithsonian, 60 Garden Street, Cambridge, MA 02138, USA 
    \and
    Minnesota Institute for Astrophysics, University of Minnesota, 116 Church St SE, Minneapolis, MN 55455, USA 
    \and
    DIFA - Universit\`a di Bologna, via Gobetti 93/2, I-40129 Bologna, Italy 
    \and
    NAF - IRA, Via Gobetti 101, I-40129 Bologna, Italy; IRA - INAF, via P. Gobetti 101, I-40129 Bologna, Italy 
    \and 
    Naval Research Laboratory, 4555 Overlook Avenue SW, Code 7213, Washington, DC 20375, USA 
    \and
    Institute for Astronomy, Royal Observatory, Blackford Hill, Edinburgh, EH9 3HJ, UK 
    \and
    European Southern Observatory (ESO), Karl-Schwarzschild-Strasse 2, Garching 85748, Germany 
    \and
    Department of Liberal Arts and Sciences, Berklee College of Music, 7 Haviland Street, Boston, MA 02215, USA 
    }

   \date{Received 11-03-2022; accepted 04-07-2022}

 
 
  \abstract{
    It has been well established that galaxy clusters have magnetic fields. The exact properties and origin of these magnetic fields are still uncertain even though these fields play a key role in many astrophysical processes. Various attempts have been made to derive the magnetic field strength and structure of nearby galaxy clusters using Faraday rotation of extended cluster radio sources. This approach needs to make various assumptions that could be circumvented when using background radio sources. However, because the number of polarised radio sources behind clusters is low, at the moment such a study can only be done statistically. In this paper, we investigate the depolarisation of radio sources inside and behind clusters in a sample of 124 massive clusters at $z<0.35$ observed with the Karl G. Jansky Very Large Array.  We detect a clear depolarisation trend with the cluster impact parameter, with sources at smaller projected distances to the cluster centre showing more depolarisation. By combining the radio observations with ancillary X-ray data from Chandra, we compare the observed depolarisation with expectations from cluster magnetic field models using individual cluster density profiles. The best-fitting models have a central magnetic field strength of $5-10\,\mu$G with power-law indices between $n=1$ and $n=4$. We find no strong difference in the depolarisation trend between sources embedded in clusters and background sources located at similar projected radii, although the central region of clusters is still poorly probed by background sources. We also examine the depolarisation trend as a function of cluster properties such as the dynamical state, mass, and redshift. We see a hint that dynamically disturbed clusters show more depolarisation than relaxed clusters in the $r>0.2R_\mathrm{500}$ region. In the core region, we did not observe enough sources to detect a significant difference between cool-core and non-cool-core clusters. Our findings show that the statistical depolarisation of radio sources is a good probe of cluster magnetic field parameters. Cluster members can be used for this purpose as well as background sources because the local interaction between the radio galaxies and the intracluster medium does not strongly affect the observed depolarisation trend. 
    }


   \maketitle
%

\section{Introduction}\label{sec:introduction}
Through observations of diffuse synchrotron emission such as radio halos \citep[e.g.][for a recent review]{2019SSRv..215...16V} and Faraday rotation measures (RMs) of polarised radio sources \citep[e.g.][for a recent review]{2018PASJ...70R...2A}, it has been proven that galaxy clusters have magnetic fields. These fields play a key role in many astrophysical processes such as heat conduction, gas mixing, and cosmic ray propagation, but the exact properties and origin of these magnetic fields are still uncertain \citep[see][for reviews on magnetic fields in galaxy clusters]{2002ARA&A..40..319C,2018SSRv..214..122D}. 
Estimates of the magnetic field strength from observations of diffuse synchrotron emission (i.e. radio halos) place the magnetic field strengths of galaxy clusters around the $\mu$G level \citep[][]{2008SSRv..134...93F}. 
Recently, observations of high-redshift radio halos have revealed that clusters at $z>0.6$ might have similar magnetic field strengths to local galaxy clusters \citep{2021NatAs...5..268D}, implying that magnetic field amplification should happen fast during cluster formation.  However, estimates of the magnetic field strength from diffuse synchrotron emission require various assumptions as to the energy spectrum and distribution of relativistic particles \citep[e.g. equipartition or minimum energy;][]{2005AN....326..414B}. 

The most promising method to derive magnetic field properties in clusters is through Faraday rotation of polarised radio emission \citep[see][for a review]{2004IJMPD..13.1549G}. 
Various studies have constrained the magnetic field strength and structure of nearby galaxy clusters using the RM of extended radio sources \citep[e.g.][]{2004A&A...424..429M,2006A&A...460..425G,2008A&A...483..699G,2008MNRAS.391..521L,2010A&A...513A..30B,2010A&A...514A..50G,2012A&A...540A..38V,2017A&A...603A.122G}. These studies have found central magnetic field strengths of the order of 1-10 $\mu$G, and a magnetic field power spectrum index between $n=2$ and $n=4$.

The depolarising effect of Faraday rotation can also be used to constrain magnetic field properties \citep[e.g.][]{2006MNRAS.368.1500T,Bonafede2011,2019A&A...622A..16O,2020A&A...638A..48S,2020ApJ...903...36S,2021ApJ...911....3D,2022A&A...659A.146D,2022A&A...657A...2R}. Since we observe radio sources with a finite spatial resolution, the differential Faraday rotation between different lines of sight within a single beam reduces the observed degree of polarisation. This beam depolarisation effect depends on the correlation scales of the magnetic field and the magnetic field strength. 
In this way, the average properties of magnetic fields in clusters can be investigated, and differences can be studied between various cluster properties, such as the presence or absence of a cool core \citep[][]{Bonafede2011}. The advantage of using fractional polarisation over the RM of radio sources is that unpolarised sources can also be taken into account, as upper limits on the polarisation fraction can be estimated.

A drawback in most studies of Faraday rotation and the resulting depolarisation is that the polarised radio sources are often cluster members. This introduces a small uncertainty because the location of the radio sources inside the cluster cannot be determined accurately, but a larger uncertainty is introduced by the gas in the intracluster medium (ICM), whose properties are usually not known in detail. Often, it is assumed that the interaction between the ICM gas around the radio source and the radio plasma is negligible. However, it is debated to what extent this assumption is true, with some studies showing evidence for local Faraday rotation being induced in radio lobes \citep[e.g.][]{2003ApJ...588..143R} and other studies finding no evidence for this \citep[e.g.][]{2003ApJ...597..870E}. The ICM could be locally compressed around cluster radio sources, causing higher densities and thus also higher depolarisation, potentially biasing results. Additionally, bent-tailed radio galaxies which are often seen in clusters might not have the same intrinsic polarisation as classic double-lobed radio galaxies \citep[][]{1998A&A...331..475F}. 

In this paper, we aim to alleviate these problems through a study of the polarisation properties of sources inside and  behind clusters. Because the number of polarised radio sources (behind clusters) is typically low \citep[e.g.][]{2014ApJ...785...45R}, such a study will be most often statistical. Although polarisation properties of sources behind clusters have been investigated for some single clusters \citep[e.g.][]{2010A&A...513A..30B}, this has not yet been studied thoroughly in a sample of clusters. 
This paper focuses on the beam depolarisation effect and considers only the implications of the fractional polarisation measurements of the radio sources. In a follow-up paper, we extensively study the Faraday RM of the polarised sources. 

Samples of galaxy clusters can be selected relatively unbiased through the thermal Sunyaev-Zel'dovich effect, which imprints a redshift-independent distortion on the spectrum of the cosmic microwave background \citep[][]{1970CoASP...2...66S,1972CoASP...4..173S}. The sample we use as a starting point for this work is the Planck Early Sunyaev Zel'dovich sample \citep{2011A&A...536A...8P}. This provides mass-selected samples of galaxy clusters up to high redshifts. We obtained observations with the Karl G. Jansky Very Large Array (VLA), which are detailed in Section \ref{sec:data}, to study the linear polarisation properties of radio sources located inside and behind ESZ clusters. The source finding, determination of the polarisation properties, host galaxy identification and redshift estimation process is explained in Section \ref{sec:methods}. Theoretical depolarisation expectations are derived through modelling of the magnetic fields as Gaussian random fields in Section \ref{sec:modelling} and results are shown in Sections \ref{sec:results} and \ref{sec:results2}. Finally, we conclude with a discussion and summary in Sections \ref{sec:discussion} and \ref{sec:conclusion}. Possible biases are discussed in Appendix \ref{sec:AppendixA}.
Throughout this paper, we assume a flat $\Lambda$CDM model with $H_0=70$ kms$^{-1}$Mpc$^{-1}$, $\Omega_m=0.3$ and $\Omega_\Lambda=0.7$. We refer to the intensity of linearly polarised light simply as the polarised intensity.


\section{Data}\label{sec:data}

\subsection{Chandra-Planck ESZ sample}
The Planck Early Sunyaev-Zel'dovich (ESZ) results presented 189 cluster candidates all-sky \citep{2011A&A...536A...8P}, of which 163 clusters are at a redshift of $z<0.35$. The Chandra-Planck Legacy Program for Massive Clusters of Galaxies\footnote{\url{http://hea-www.cfa.harvard.edu/CHANDRA_PLANCK_CLUSTERS/}} observed all 163 clusters with sufficient exposure time to collect at least 10,000 source counts per cluster. This makes it (one of) the largest relatively unbiased samples of galaxy clusters with high-quality X-ray data available. The Chandra observations of 147 clusters from the ESZ sample are presented in \citet{2017ApJ...843...76A,2021ApJ...914...58A}, where the sample has been reduced by 16 because six clusters are too close to point sources, nine clusters are classed as multiple objects and one system was too large to allow for a reliable background estimate in the Chandra field of view. High-quality X-ray data is particularly important for polarisation studies to be able to break the degeneracy between electron density and magnetic field. In this paper, we used the thermal electron density profiles which were calculated from the fitting procedure detailed in \citet{2017ApJ...843...76A}.

\subsection{Observations and data reduction}
We have obtained VLA L-band (1-2 GHz) observations of 126 Planck clusters at $z<0.35$ and DEC>-40$^\circ$ (VLA project code 15A-270). The redshift cut is made because the angular size of higher redshift clusters on the sky becomes too small to find a significant number of polarised background sources. Out of these 126 clusters, 102 are from the ESZ catalogue and 24 clusters are new detections in the PSZ1 \citep[][]{2015A&A...581A..14P} and PSZ2 \citep[][]{2016A&A...594A..27P} catalogues that have been added to the sample. The observations were taken in the B(nA) array configuration, with the BnA configuration employed for targets located at DEC $<$ -15$^\circ$ or DEC $>$ +75$^\circ$ to match the resolution of targets observed in favourable declination ranges. Targets are observed for $\sim 40$ minutes each. The full L-band comprises 16 spectral windows, each consisting of 64 channels before frequency averaging.

The calibration of the radio data was done using the Common Astronomy Software Application ($\mathtt{CASA}$; \citealt{2007ASPC..376..127M}) and proceeded in the following fashion. For each observing run, the initial data calibration was done per spectral window, such that bad spectral windows could be identified and flagged. We Hanning smoothed the spectral axis to reduce the effect of Gibbs ringing due to strong radio frequency interference (RFI) in the L-band. Shadowed antennas were flagged and the initial flagging of RFI was done with the $\mathtt{CASA}$ \textit{TFCrop} algorithm. The effect of the elevation on the antenna gain and efficiency was calculated and antenna position corrections were applied. The flux scale was set to the \citet{PerleyButler2017} scale. We calculated initial-bandpass calibration solutions using a large solution interval and initially calibrated the complex gains with the central channels of the spectral window. The antenna delay terms were then calculated and applied, after which the final-bandpass solutions could be calculated. A polarised calibrator (either 3C138 or 3C286) was used to solve for a global cross-hand delay and an unpolarised calibrator (3C147) was used to calibrate on-axis polarisation leakage. Subsequently, the polarised calibrator was then used to calibrate the polarisation angle. Off-axis polarisation leakage due to a time, frequency, and polarisation-dependent primary beam becomes important as the distance from the pointing centre increases but is known to be less important in Stokes Q and U than in Stokes V \citep[][]{2008A&A...486..647U}. Typically for VLA L-band observations, the leakage from Stokes I into Q and U is around 1\% at the primary beam full-width half maximum \citep[][]{2017AJ....154...56J}. While this effect can mimic depolarisation due to the frequency dependence of the primary beam, we do not consider it to be a major issue for this study because all clusters are observed near the pointing centre. We discuss off-axis leakage in more detail in Section \ref{sec:caveats}. Finally, the antenna-based complex gain solutions were calculated using the calibrator sources, and another round of automatic flagging was performed using the $\mathtt{CASA}$ \textit{TFcrop} and \textit{Rflag} algorithms. All spectral windows were then combined and the resulting data were averaged to 8 MHz channels and 6-second timesteps. Leftover RFI was then flagged with the \textit{AOflagger} \citep{2012A&A...539A..95O} and a custom strategy to flag RFI in the cross-hand correlation (rl,lr) plane was employed. Spectral window 8 was fully lost to RFI in every observing run, resulting in a total of 90 frequency channels after initial calibration. 

To remove residual amplitude and phase errors in the direction of the target fields and increase the quality of the final images, we further performed six rounds of self-calibration, automatically calculating the solution interval based on the mean flux density in each field. This was done to ensure enough signal-to-noise during the calibration steps, with larger solution intervals used for fields with fainter sources. The imaging and cleaning were done using $\mathtt{WSclean}$ version 2.7.3 with the options \textit{-join-polarizations} and  \textit{-squared-channel-joining} for Stokes Q and U imaging \citep{2014MNRAS.444..606O}.
The six rounds of self-calibration involve three phase-only calibration rounds and three amplitude and phase rounds, decreasing the solution interval each round. For the majority of targets, this automatic self-calibration pipeline proved sufficient to obtain high-quality images of the target fields. A small number of target fields needed manual tweaking of parameters or flagging of RFI. For those clusters, one or two additional rounds of self-calibration were performed after the pipeline. 

\begin{figure}[t]
    \centering
    \includegraphics[width=\columnwidth]{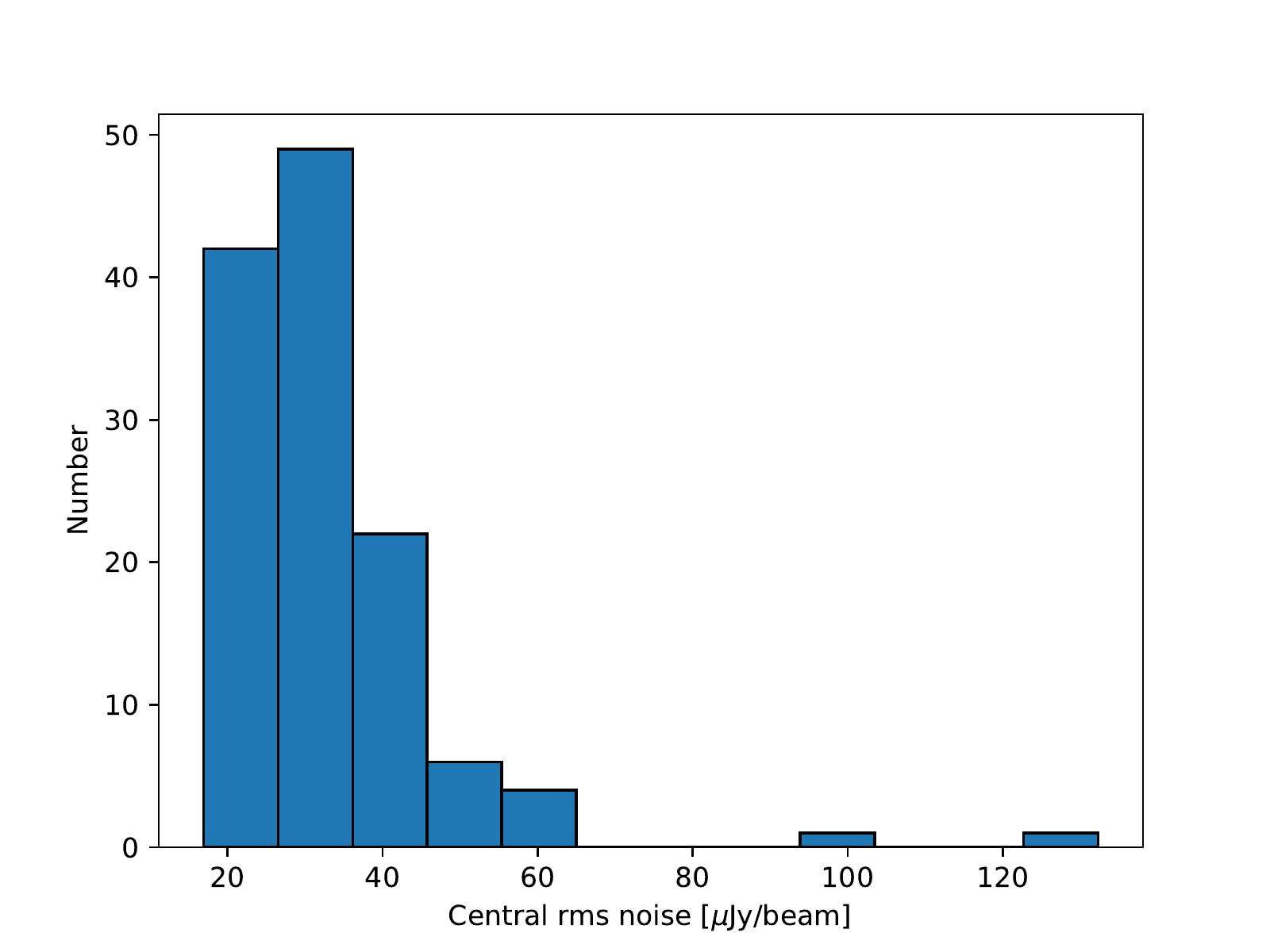}
    \caption{Central RMS noise in Stokes I in the 124 observed target fields.}
    \label{fig:rms}
\end{figure}

Each 8 MHz channel was corrected for the VLA primary beam attenuation, and all channels were smoothed to a circular Gaussian restoring beam at the resolution of the lowest frequency channel, to ensure that all channels have the same angular resolution. This resulted typically in a synthesised beam size of 6-7 arcseconds. The distribution of central root-mean-square (RMS) noise in the full-band Stokes I images is given in Figure \ref{fig:rms}. Most targets have an RMS noise of around 20-30 $\mu$Jy/beam in the centre of the field. Two clusters, G033.46-48.43 and G226.17-21.91, have been removed from the sample. Calibration artefacts from a bright radio source were completely dominating the G033.46-48.43 field and during observations of G226.17-21.91 most of the data was lost to interference by a thunderstorm. 

We found significant flux density variations between the spectral windows in all observations, also noted by \citet[][]{2021ApJ...911....3D}, probably related to bandpass calibration or deconvolution uncertainties. To mitigate this problem as much as possible, we aligned the flux scale per observing run by fitting a simple power-law model to all bright Stokes I sources with at least a signal-to-noise ratio of 100,
\begin{equation}\label{eq:powlaw}
    I_\nu = I_0 \nu^{\alpha},
\end{equation}
where $\alpha$ represents the spectral index and $I_\nu$ is the Stokes I intensity. Correction factors for each spectral window were determined per observing run by averaging the correction factors of individual sources. These correction factors were usually of the order of 5-10\%. The corrections were applied to the Stokes I, Q and U fluxes. 

\begin{figure*}[tbh]
    \centering
    \includegraphics[width=0.85\textwidth]{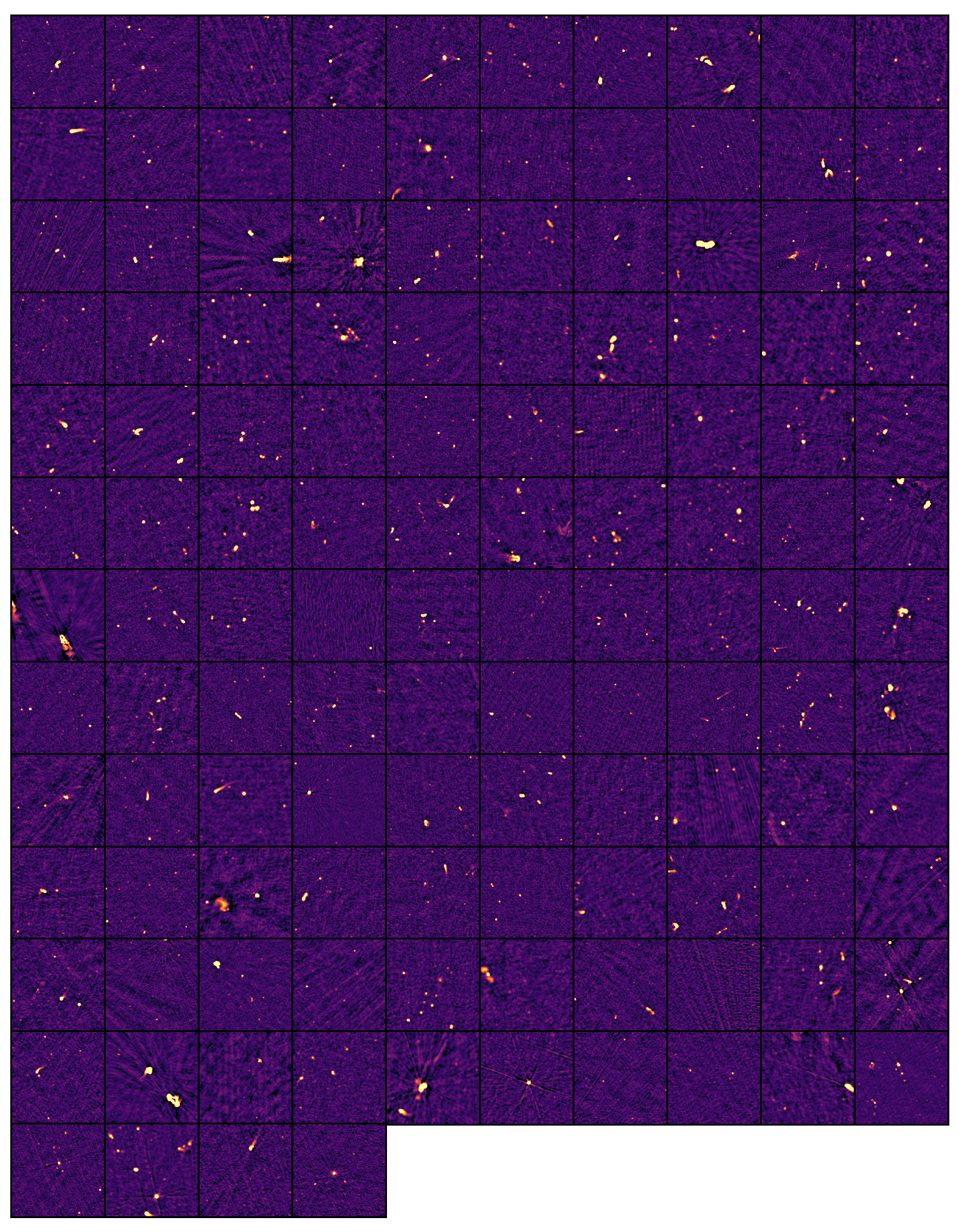}
    \caption{VLA 1-2 GHz primary beam corrected total intensity images of the 124 Planck clusters. The images are smoothed to the resolution of the lowest frequency channel (typically 6$^{\prime\prime}$) and the size is equal to the field-of-view at 2 GHz ($0.35\times0.35$ deg$^2$). The colour scale is logarithmic with the scale range determined individually per cluster for visualisation purposes. The order of clusters follows the order in the Table \ref{tab:clusters}, in row-major order.}
    \label{fig:allclusters}
\end{figure*}

The final 124 calibrated radio images are shown as a mosaic in Figure \ref{fig:allclusters}. The five fields with RMS noise higher than 60 $\mu$Jy/beam in Figure \ref{fig:rms} are caused by calibration artefacts from bright sources at the edge of the fields, and in one case in the centre of the field. Direction-dependent calibration \citep[e.g.][]{2014A&A...566A.127T} could improve the quality of the images affected by bright off-axis sources, but these few fields should not significantly affect the results presented here. We decided to keep all 124 fields for our analysis because even in the five fields with bright artefacts 27 polarised radio sources were still relatively unaffected by those artefacts and could be used in the analysis. 

\section{Methods}\label{sec:methods}
This section details the source finding of both polarised and unpolarised radio sources and the determination of their polarisation properties. Finally, we explain the optical counterpart identification and estimation of the redshift of the sources, such that we can classify the host galaxies as background sources, cluster members or foreground sources.

\subsection{Polarised source finding}\label{sec:sourcefinding}
Linear polarisation can be expressed as a complex quantity by a combination of Stokes Q and Stokes U or written as a complex vector
\begin{equation}\label{eq:polint}
    P(\lambda^2) = Q+iU = p_0 I \exp(2i\chi),
\end{equation}
where $\lambda$ indicates the observed wavelength, $p_0$ the polarisation fraction, $I$ refers to the Stokes I intensity and 
\begin{equation}\label{eq:chi}
\chi(\lambda^2)=\frac{1}{2}\arctan\left(\frac{U}{Q}\right)   
\end{equation}
is the polarisation angle. Faraday rotation introduces a wavelength-dependent rotation of the polarisation angle $\chi$. In the general case, the Faraday depth of a source is defined as \citep{1966MNRAS.133...67B,BrentjensBruyn}
\begin{equation}\label{eq:fdepth}
    \phi(\textbf{r}) = 0.81 \int n_e \textbf{B} \cdot d\textbf{r} \;\; \left[\mathrm{rad\: m}^{-2}\right],
\end{equation}
where $n_e$ is the electron density in parts per cm$^{-3}$, $\textbf{B}$ is the magnetic field in $\mu$Gauss and d$\textbf{r}$ the infinitesimal path length increment along the line of sight in parsecs. We adhere to the definition that $\phi(\textbf{r})>0$ implies that the magnetic field is pointing towards the observer.

In the simplest case, where only one source is present along the line of sight without internal Faraday rotation, the Faraday depth $\phi$ is equal to the RM of a source, and the observed rotation can be expressed as
\begin{equation}\label{eq:RMwavelength}
\chi(\lambda^2) = \chi_0 + \phi \lambda^2.
\end{equation}

Faraday rotation may cause polarised sources to be undetected in the wide-band Stokes Q and U images or in the linearly polarised intensity ($\sqrt{Q^2+U^2}$) images. This is because the Stokes Q and U intensities can be both positive and negative, resulting in averaging out the frequency integrated signal if the RM is significant. To solve this problem, we used the Faraday RM-synthesis \citep[][]{BrentjensBruyn} technique. RM-synthesis aims to approximate the Faraday dispersion function $F(\phi)$ by Fourier inversion of the following equation
\begin{equation}\label{eq:FDF}
P(\lambda^2) = \int_{-\infty}^{+\infty} F(\phi) e^{2i\phi\lambda^2} d\phi,
\end{equation}
where $P(\lambda^2)$ is the complex polarised surface brightness (Eq. \ref{eq:polint}) as a function of the observing wavelength (squared) and $\phi$ is the Faraday depth of the source (Eq. \ref{eq:fdepth}). Calculating the Faraday dispersion function $F(\phi)$ essentially corresponds to de-rotating polarisation vectors to their position at an arbitrary wavelength $\lambda_0^2$. However, we note that RM-synthesis only approximates the Faraday dispersion function because we cannot sample all wavelengths. The limitations of our frequency setup can be expressed with the three following quantities \citep[][]{BrentjensBruyn}. The maximum Faraday depth to which we have more than 50\% sensitivity is given by the channel width: $\delta\lambda^2$
\begin{equation}\label{eq:bandwidth_depol}
    ||\phi_{max}|| \approx \frac{\sqrt{3}}{\delta\lambda^2} \approx 1200 \;\; \left[\mathrm{rad\: m}^{-2}\right].
\end{equation}
The resolution in $\phi$ space is determined by our wavelength coverage, with the full-width half-maximum given by
\begin{equation}
\delta\phi \approx \frac{2\sqrt{3}}{\Delta\lambda^2} \approx 52 \;\; \left[\mathrm{rad\: m}^{-2}\right].
\end{equation}
The maximum scale we can resolve in $\phi$ space (analogous to resolving-out extended radio sources in synthesis imaging) is given by the shortest observable wavelength
\begin{equation}
    \mathrm{maximum\,scale} \approx \frac{\pi}{\lambda^2_{min}} \approx 140 \;\; \left[\mathrm{rad\: m}^{-2}\right].
\end{equation}
Because the resolution in $\phi$ space is smaller than the maximum scale we can resolve, we are technically able to detect slightly extended sources in Faraday space (i.e. Faraday thick sources). Typical values of RM found in clusters are usually of the order of $10^2$ rad m$^{-2}$, going up to $10^3$ rad m$^{-2}$ in dense cool-core clusters \citep[e.g. Abell 780 and Cygnus A;][]{1990ApJ...360...41T,2020ApJ...903...36S}. Thus with the current frequency setup, we are sensitive to the typical amount of Faraday rotation in clusters. 

\begin{figure*}[tbh]
    \centering
    \includegraphics[width=0.85\textwidth]{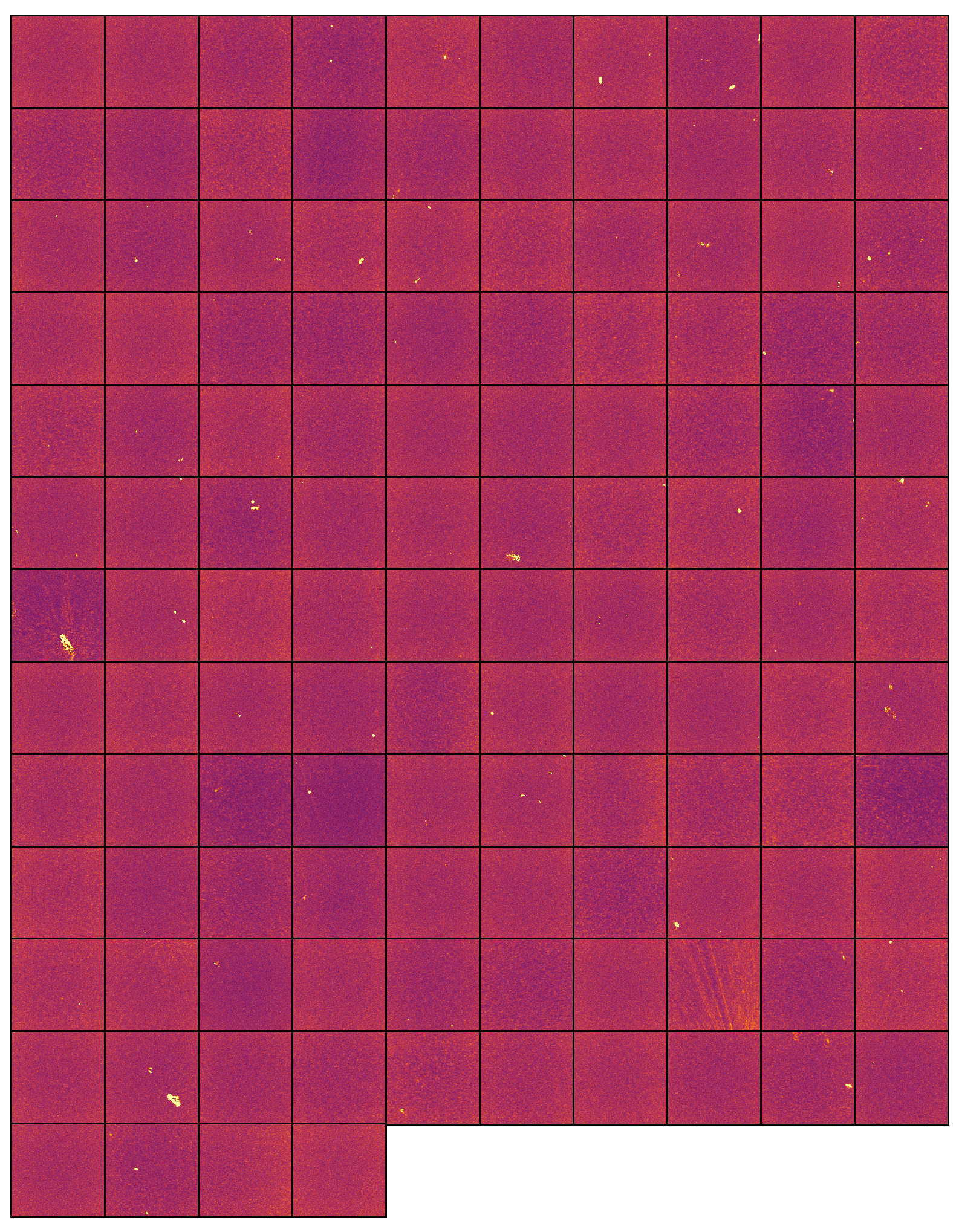}
    \caption{Same as Figure \ref{fig:allclusters}, but for peak polarisation intensity. The colour scale used here is $\mathtt{arcsinh}$. We note that the calibration artefacts visible in cluster position 107 (zero-based row-major index [10,7]) were not used in the analysis, but the field was kept as two polarised sources at the edge of the primary beam were relatively unaffected by the artefacts.}
    \label{fig:allclusters_polint}
\end{figure*}

We performed RM-synthesis using the \texttt{pyrmsynth}\footnote{\url{http://www.github.com/mrbell/pyrmsynth}} module, weighting by the inverse RMS noise of the channels and ignoring bad channels. The result is an `RM-cube' with two spatial axes and a Faraday depth $\phi$ axis, that contains the polarised intensity at each pixel location as a function of the Faraday depth, sampled from $\phi=-2000$ to $\phi=2000$ rad m$^{-2}$ in steps of 10 rad m$^{-2}$. The peak polarised intensity map is then made by taking the maximum value along the $\phi$ axis. The peak polarised intensity map for all clusters is shown as a mosaic in Figure \ref{fig:allclusters_polint}. 

To find polarised source candidates automatically, we used the source finder program PyBDSF \citep{2015ascl.soft02007M} on both the Stokes I images and the peak polarised intensity maps after RM-synthesis. We set the parameters $\mathtt{thresh\_pix}=5.0$ and $\mathtt{thresh\_isl}=3.0$, 
meaning that a five-sigma threshold was used for the source detection and a three-sigma threshold was used during the fitting of the total intensity source properties. The background noise was calculated over the image in a box with a size of 3 arcminutes in steps of 1 arcminute to account for the varying background noise to the primary beam.

We found that PyBDSF performed better when de-correcting the peak polarised intensity maps for the primary beam, such that an approximately flat-noise image was used for the source finding. More involved methods for polarised source finding were considered \citep[e.g. moment analysis; ][]{2018MNRAS.474.3280F}, but our simple method was found to be sufficient given the still relatively small data size which allowed for visual inspection of the polarised source candidates. 
All polarised source candidates were cross-matched with sources found in the Stokes I images and source candidates that lie inside the extent of the Stokes I source were retained as real polarised sources. We defined the extent of the sources in the Stokes I map as twice the full width at half maximum (FWHM) of the fitted Gaussian, which is empirically found to be a good estimate of source sizes \citep[e.g.][]{2019A&A...622A..12H}. This source finding process proved successful in most of the observations, but all fields were also visually inspected and some manual intervention was needed for rare cases, such as clear polarised source candidates that were positioned just outside of the extent of the source in the Stokes I map. In total, PyBDSF found 6\,807 source candidates in Stokes I and 819 source candidates in polarisation over the 124 target fields. 

\subsection{Fractional polarisation measurement}\label{sec:fracpolmeasurement}
To determine the polarisation properties such as the intrinsic polarisation fraction $p_0$ and the Faraday depth $\phi$ of polarised source candidates, we can model the polarised emission as
\begin{equation}\label{eq:polint_nodepol}
    P(\lambda^2) = p_0 I \exp\left[2i(\chi_0 + \phi \lambda^2)\right].
\end{equation}
However, if a source emits at different Faraday depths along the same line of sight it suffers from depolarisation due to the differential Faraday rotation causing the emission from the far side of the source to be rotated more than emission from the nearby side of the source. This internal depolarisation can be modelled as \citep[see][for details]{1998MNRAS.299..189S} 
\begin{equation}\label{eq:indepol}
    P = p_0 I \left[ \frac{1- \exp(-2\Sigma^2_{\mathrm{RM}}\lambda^4)}{2\Sigma^2_{\mathrm{RM}}\lambda^4}\right]\exp[2i(\chi_0+\phi \lambda^2)],
\end{equation}
where $\Sigma^2_{\mathrm{RM}}$ represents the amount of depolarisation. 
A similar effect happens because we observe the sources with a finite spatial resolution. If the magnetic field in an external Faraday screen (e.g. the ICM) changes on scales smaller than the restoring beam sources are partly depolarised by beam depolarisation. This is an external depolarisation effect and can be modelled as \citep[see][for details]{1998MNRAS.299..189S}
\begin{equation}\label{eq:extdepol}
    P = p_0 I\exp (-2\sigma^2_{\mathrm{RM}}\lambda^4)\exp[2i(\chi_0+\phi \lambda^2)],
\end{equation}
where $\sigma^2_{\mathrm{RM}}$ models the amount of depolarisation. Finally, if the polarisation angle rotates significantly in a single frequency channel, bandwidth depolarisation occurs. This limits the maximum observable RM, as is given in Equation \ref{eq:bandwidth_depol}.

Distinguishing between internal and external depolarisation effects can be done by measuring the spectral index of the polarised emission at lower frequencies with high resolution because external depolarisation effects are stronger at low frequencies \citep[][]{2011MNRAS.418.2336A}. In reality, there are probably both internal and external Faraday effects at play and a combination of the models could be used to fit the data. However, for this study distinguishing exactly between polarisation mechanisms is not important, as we are only interested in the polarisation fraction trend. The internal depolarisation of radio sources should not affect the general trend and can be found from the depolarisation ratio of sources at cluster outskirts (see Sec. \ref{sec:results}.) Therefore we decided to fit only the external depolarisation model given by Equation \ref{eq:extdepol}. 
We fitted this model to the Stokes Q and U channels simultaneously and the total intensity (Stokes I) spectrum was modelled as a simple power-law (Eq. \ref{eq:powlaw}). 
Fitting the Stokes I, Q and U channels directly has the advantage that we can assume Gaussian likelihoods because these channels have Gaussian noise properties, unlike the polarised intensity maps, whose distribution is Ricean. We fitted the integrated Q and U flux densities of each polarised source candidate, where the integration was performed over the extent of the polarised source as defined in Section \ref{sec:sourcefinding}. This means that separate polarised components of the same physical source (e.g. two polarised lobes of a single radio galaxy) were treated as separate sources during fitting. The uncertainty in the integrated flux density per channel was calculated as
\begin{equation}
\sigma_i = \sqrt{ (\sigma_{\mathrm{rms}}\times \sqrt{N})^2 + (\delta_\mathrm{cal} \times f_i)^2  },
\end{equation}
where N is the number of beams covered by the source and $\sigma_{\mathrm{rms}}$ the background RMS noise in the corresponding channel. The second term accounts for the flux density variations explained in Section \ref{sec:data} by assuming a $\delta_\mathrm{cal}=5\%$ error on the measured flux densities per channel, denoted by $f_i$.

\begin{figure*}[tbh]
\centering
\begin{subfigure}{0.3\paperwidth}
  \centering
  \includegraphics[width=0.8\linewidth]{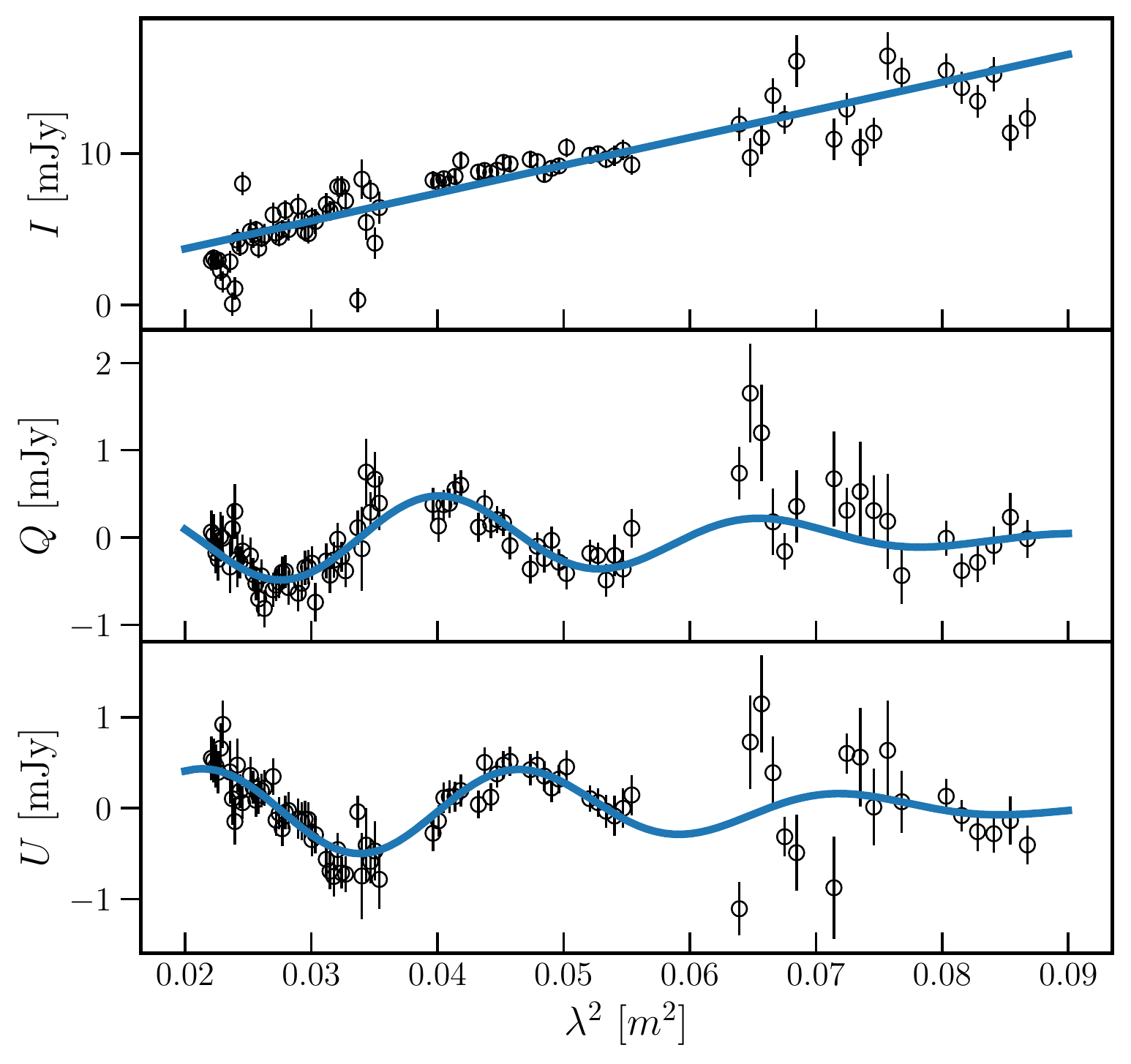}
  \caption{ }
  \label{fig:examplefit1}
\end{subfigure}%
\begin{subfigure}{.3\paperwidth}
  \centering
  \includegraphics[width=0.8\linewidth]{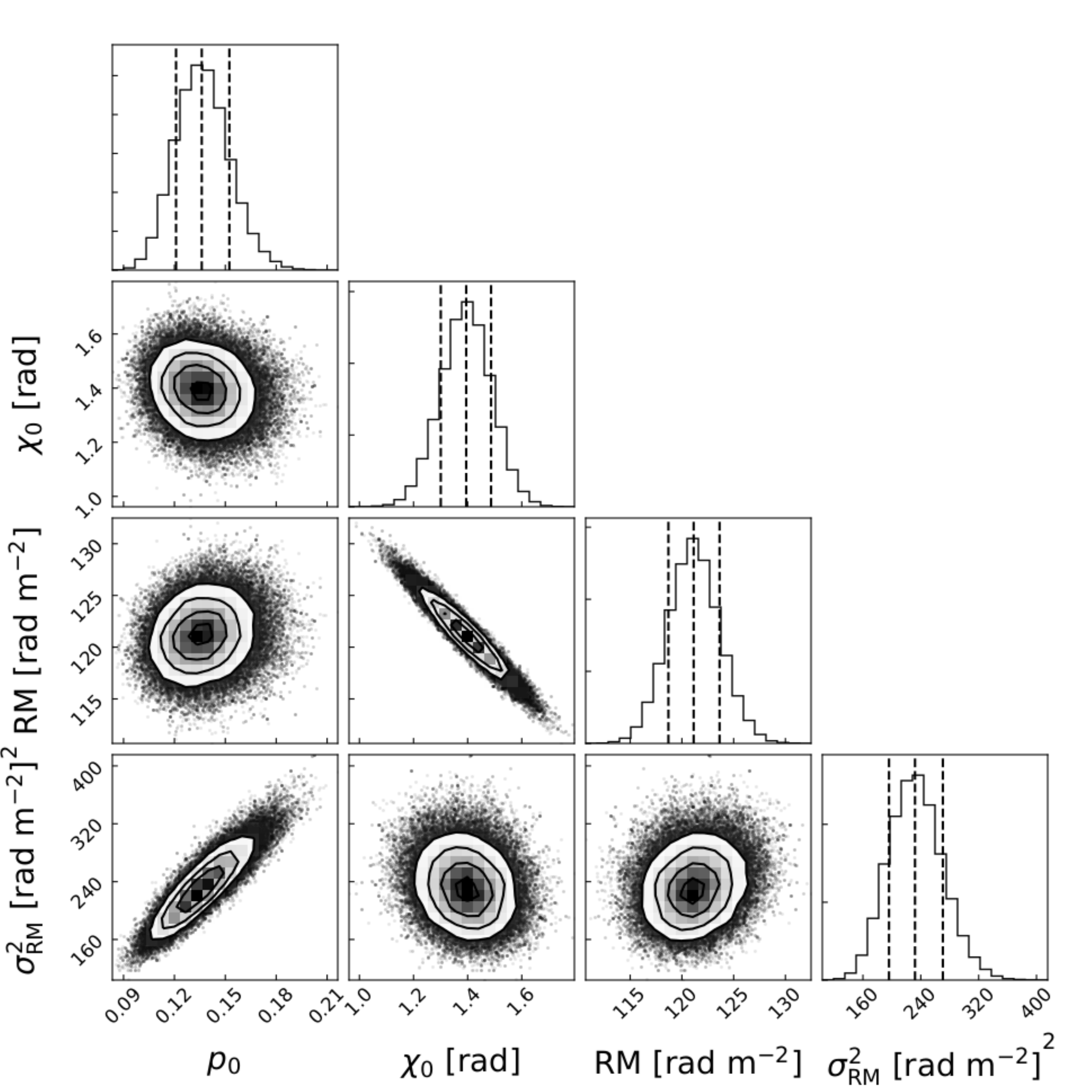}
  \caption{ }
  \label{fig:examplefit2}
\end{subfigure}%
\begin{subfigure}{.3\paperwidth}
  \centering
  \includegraphics[width=0.8\linewidth]{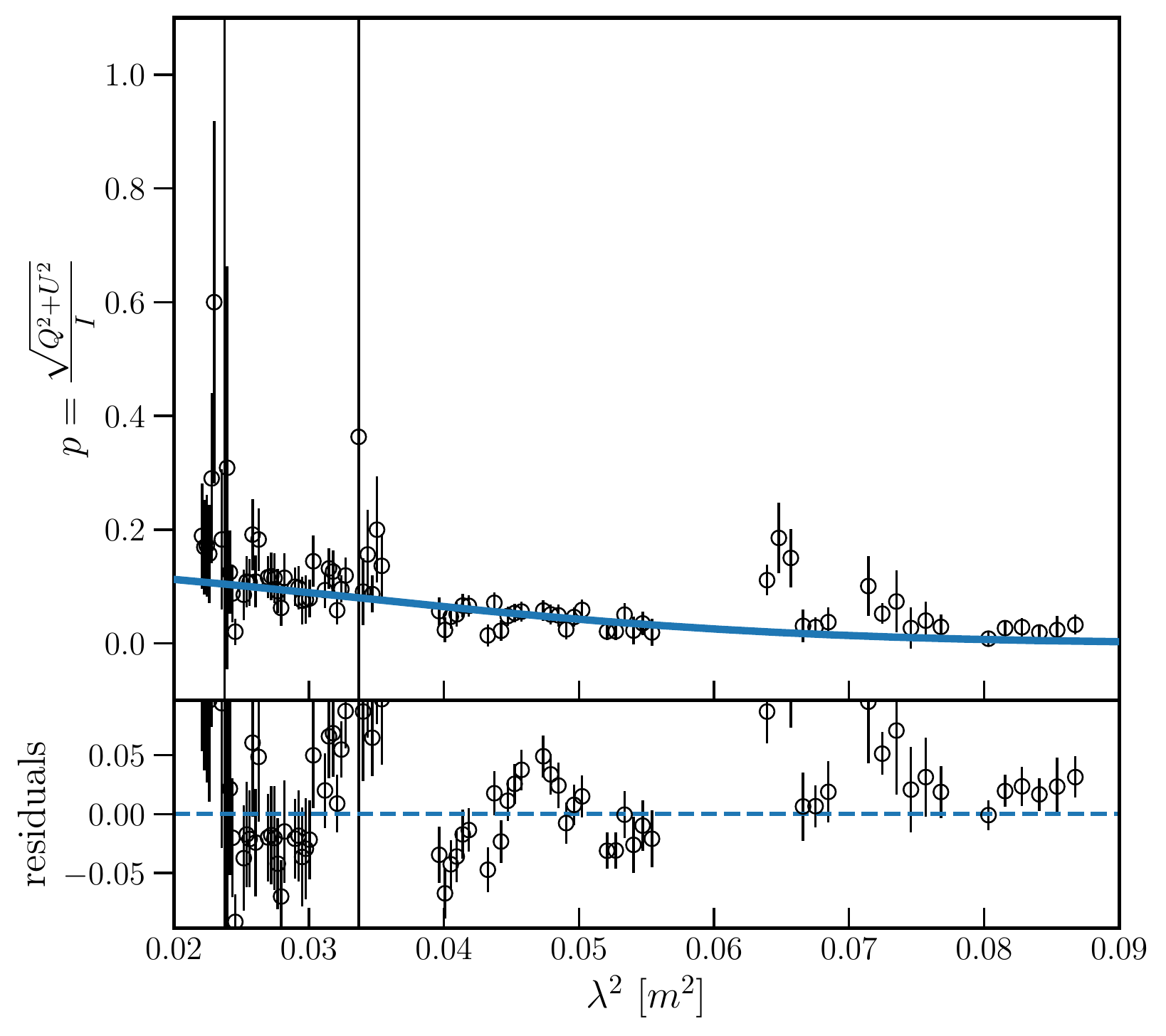}
  \caption{ }
  \label{fig:examplefit3}
\end{subfigure}
\caption{Results of the MCMC fitting of the external depolarisation model on the Stokes I, Q and U channels to a polarised source with good signal-to-noise ratio. Figure (a) shows the measured flux densities per channel in black and the best-fit model in blue. Figure (b) shows the posterior distribution of the model parameters visualised as one and two-dimensional projections in a corner plot. Figure (c) shows the inferred polarisation fraction from the data and the best-fit model together with the residuals.}
\label{fig:examplefit}
\end{figure*}

The fitting was done using a Monte Carlo Markov Chain (MCMC) fitting code developed by \citet[][]{2021ApJ...911....3D} to sample the posterior probability. The following uniform priors were assumed:

\begin{equation}
\begin{cases}
I_0 \sim \mathcal{U}(0,\infty) \\

\alpha \sim \mathcal{U}(-\infty,\infty) \\

p_0 \sim \mathcal{U}(0,1) \\

\chi_0 \sim \mathcal{U} (0,\pi) \\ 

\phi \sim \mathcal{U}(-2000,2000) \\ 

\sigma_{\rm RM}^2 \sim \mathcal{U}(0,\infty), \\
\end{cases}
\end{equation}
The initial values for the parameters were found through a least-square fit, using the $\phi$ as obtained from the RM-synthesis method as the initial guess for the Faraday depth. The posterior was sampled with 200 walkers for 1\,000 steps and a burn-in period of 200 steps was removed from each chain. The one-sigma uncertainties on the best-fit parameters are given by the 16th and 84th percentile of the chain. An example of the results on a polarised source with a good signal-to-noise ratio is shown in Figure \ref{fig:examplefit}.

\begin{figure}[t]
    \centering
    \includegraphics[width=1.0\columnwidth]{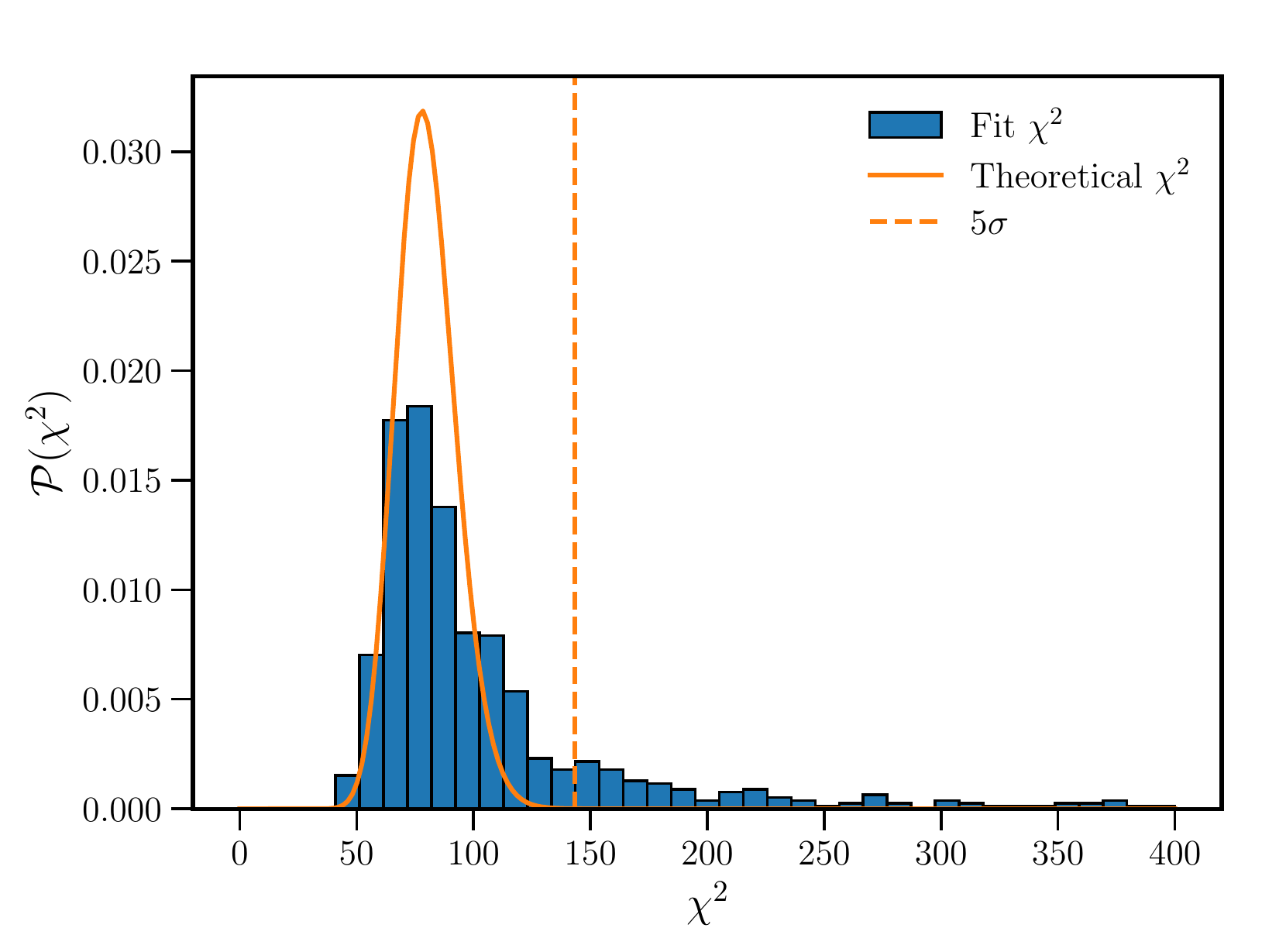}
    \caption{Distribution of the sum of the squared residuals ($\chi^2$) of the model fits to all polarised components. The plot is truncated at $\chi^2=400$ for visibility.}
    \label{fig:chi2}
\end{figure}

To judge whether the model $m$, given by Equation \ref{eq:extdepol}, is a good fit to the data points $y_i$ (i.e. the Stokes Q and U flux densities), we inspected the normalised residuals,
\begin{equation}
 R_i=\frac{y_i - m(\lambda^2,\theta)}{\sigma_i},
\end{equation}
where $\sigma_i$ is the uncertainty on the polarised flux. It is not possible to determine analytically the number of degrees of freedom $k$ in the external depolarisation model because it is a non-linear model, which is why we are not able to determine the reduced chi-squared value. In Figure \ref{fig:chi2} we plot the distribution of the sum of the squared residuals (i.e. the $\chi^2$ value) of the best fitting external depolarisation model to each polarised component. Most polarised sources have around 84 to 89 data points (i.e. channels) after masking the bad channels. This would give 80 to 85 degrees of freedom if the model was linear with 4 parameters.
For comparison, we show also the theoretical $\chi^2$ distribution with 80 degrees of freedom. The main peak of the sum of the squared residuals shows good agreement with the theoretical $\chi^2$ distribution, indicating that most sources have acceptable fits. There is however a long tail of large $\chi^2$ values, mainly caused by bright Stokes I sources, where residual calibration artefacts are more noticeable. To automatically discard bad fits, we decided to cut all fits with a $\chi^2$ value that is $5\sigma$ away from the theoretical distribution, indicated by the dashed line in the Figure. This cut removed 148 polarised components. A few of these components are possibly Faraday complex sources, for which the simple model with a single RM component is not sufficient. We note that there are no apparent correlations between the $\chi^2$ parameter and the derived best-fitting parameters $p_0,\chi0,\phi,\sigma_\mathrm{RM}^2$, or the projected radius to the cluster centre, so we are not biasing our analysis by removing these sources.
Additionally, for sources with low signal-to-noise polarised emission, a good fit (according to the $\chi^2$ parameter) can be found by artificially large values of $\sigma_\mathrm{RM}$. For these sources, the best-fit $\sigma_\mathrm{RM}$ is basically unconstrained, with large error bars. Therefore we decided to also cut sources where the fractional uncertainty on the best-fit $\sigma_\mathrm{RM}$ is larger than two. This cuts 45 additional sources, so in total 193 polarised components have bad fits. These components are indicated in the polarised source Table \ref{tab:polsources} by the column `Flagged', where we have also flagged 11 components that are part of a radio relic.

To calculate the upper limit on the fractional polarisation of sources detected only in Stokes I, we followed a method similar to \citet{Bonafede2011}. We randomly sampled empty regions with a size of 10$^{\prime\prime}$, a bit larger than the synthesised beam. For these `noise sources' we computed the polarised surface brightness and compared the distribution of the polarised surface brightness of the `noise sources' to the distribution of real sources, taking into account the varying background noise level due to the primary beam of the VLA. We put the noise-dependent threshold of the surface brightness where noise sources constitute 10\% of the real sources. The resulting threshold ($P_t$) as a function of the RMS noise is 0.04, 0.05, 0.06 and 0.08 mJy beam$^{-1}$ for sources with background RMS values 0-29, 29-36, 36-46 and 46+ $\mu$Jy/beam respectively, where the background noise bins are chosen such that in each bin there are an equal amount of simulated noise sources. 
For comparison, the threshold $P_t$ calculated independently of the background noise level gives a value of $P_t=0.06$ mJy beam$^{-1}$. The one-sigma upper limit on the fractional polarisation value is then calculated as
\begin{equation}
F_p \leq \sqrt{ \frac{P_t^2}{I^2-\sigma_I^2} },
\end{equation}
where $I$ is the surface brightness of the unpolarised source and $\sigma_I$ is the background RMS noise. This method gives conservative upper limits for extended unpolarised sources because the Stokes I surface brightness is computed over the entire extent of the source (rather than e.g. per lobe). 

\subsection{Optical counterparts}\label{sec:optical}
To determine the redshift of the radio sources, each radio source needs to be associated with an optical counterpart. PyBDSF is known to occasionally split up components of a single physical radio source \citep[e.g.][]{2019A&A...622A...2W}. This particularly happens for large and extended sources, and often when the source has multiple disconnected patches of emission. To group PyBDSF Stokes I source candidates into single physical sources and to identify the optical counterpart, radio-optical overlays were created and every source was visually inspected. We used the $g,r,z$ filters from the Legacy Survey \citep[][]{2019AJ....157..168D} where available and used the Pan-STARRS survey \citep[][]{2016arXiv161205560C} for the 32 fields outside of the Legacy survey sky coverage. 

As a first guess of the optical host galaxy, the nearest optical neighbour to the radio source was marked. This proved a good guess in 5\,806 out of 6\,807 total intensity source candidates. For the remaining 1\,001 sources, the best candidate optical counterpart position was manually marked from visual inspection of radio-optical overlays.

The source association was done in the same visual inspection step as the host galaxy identification. Out of the 6\,807 total intensity source candidates, 411 candidates were components of another source, leaving 6\,396 physical sources detected in total intensity. This indicates that PyBDSF in most cases correctly identified the total extent of the Stokes I source. We did not perform the source association step for the polarised components. Because different parts of a radio source can have different RM determinations and thus polarised intensities, we decided to treat separate polarised components as separate polarised sources, as for example was also done in \citet[][]{2016A&A...596A..22B}.

\subsection{Redshift estimation}\label{sec:redshifts}
With the best-estimated location of the host galaxies determined, we employed different methods to estimate the source redshift. First, we checked whether a source has a spectroscopic redshift measurement available by cross-matching the host galaxy positions to the \textit{NASA/IPAC Extragalactic Database} (NED\footnote{\url{https://ned.ipac.caltech.edu/}}) and the Sloan Digital Sky Survey \citep[SDSS DR16;][]{2020ApJS..249....3A} with a matching radius of 0.5 and 3 arcseconds respectively. If a spectroscopic redshift was found, but no uncertainty was quoted the redshift uncertainty is set to 0 in the catalogue. 

If no spectroscopic redshift was found, a photometric redshift estimation was done from the Legacy Imaging Surveys Data Release 8 \citep[][]{2019AJ....157..168D},
which is the most sensitive optical survey covering the majority of our clusters. This approach is detailed fully in \citet[][]{2022MNRAS.512.3662D} and provides high quality redshifts for galaxies at $z<1$. For sources outside of the Legacy Survey, we calculated the photometric redshift using the Pan-STARRS \textit{grizy} bands. We followed the method and used the code provided by \citet[][]{2020A&A...642A.102T}, which estimates redshifts through local linear regression in a five-dimensional colour and magnitude space. The five-dimensional space consists of (r, g-r, r-i, i-z, z-y) where the letters indicate the extinction corrected Kron magnitudes \citep[][]{1980ApJS...43..305K} of galaxies in the PanSTARRS \textit{grizy} bands. The correction for interstellar extinction used the maps from \citet[][]{1998ApJ...500..525S} and is described in detail in Section 2.3 of \citet[][]{2020A&A...642A.102T}. To compute the photometric redshifts from the Pan-STARRS band, we found the 100 nearest neighbours in the five-dimensional space for each source, from a training set composed of 2\,313\,724 galaxies with spectroscopic redshifts, constructed by \citet[][]{2020A&A...642A.102T}. The redshift was computed for all sources where at least four out of five colours were available. We note that missing features most often occur in very faint galaxies, which makes it likely that the sources are at a redshift $z > 0.35$, and are thus background sources. The quality of the Pan-STARRS photometric redshifts was checked by comparing to spectroscopic redshifts for sources where spectroscopic redshifts were available. Using standard literature metrics for the robust scatter $\sigma_\mathrm{NMAD}$ and outlier fraction OLF \citep[cf.][]{2013ApJ...775...93D,2022MNRAS.512.3662D} we find that the photometric redshifts have good quality, with $\sigma_\mathrm{NMAD}=0.025$ and OLF$=0.075$. 

The combination of all methods resulted in a redshift estimate for 77\% (632/819) of the polarised sources and 67\% (4\,544/6\,807) of the unpolarised sources. The distribution of redshifts estimates is given in Table \ref{tab:photz_results}.
The final catalogues of polarised and unpolarised radio sources are provided with this paper, shown in Tables \ref{tab:polsources} and \ref{tab:unpolsources} respectively. These tables contain the polarised and unpolarised source properties, the best estimate for the redshift of the sources and the method used to get this estimate.

\begin{table}[]
\caption{Redshift estimates of all polarised ($N_\mathrm{pol}$) and unpolarised sources ($N_\mathrm{I}$).}
\label{tab:photz_results}
\begin{tabular}{llll}
\hline
$z_\mathrm{best}$ source\tablefootmark{a} & Source & $N_\mathrm{pol}$ & $N_\mathrm{I}$ \\ \hline
0 & NED/Literature (spectroscopic) & 248 & 1059 \\
1 & SDSS (spectroscopic) & 21 & 208 \\
2 & Legacy (photometric) & 260 & 2097 \\
3 & PANSTARRS (photometric) & 101 & 1131 \\
4 & SDSS (photometric) & 2 & 49 \\
- & No redshift available & 187 & 2263 \\ \hline
Total &  & 819 & 6807 \\ \hline
\end{tabular}
\tablefoot{\tablefoottext{a}{The `$z_\mathrm{best}$ source' key is used in the catalogue presented in Table \ref{tab:polsources} to indicate the origin of the redshift estimate.}}
\end{table}

\section{Magnetic field modelling}\label{sec:modelling}
To compare observations with theoretical expectations, we modelled the magnetic field as a three-dimensional Gaussian random field, characterised by a single power-law spectrum. We followed the approach proposed by \citet[][]{1991MNRAS.253..147T}, used in various works in the literature \citep[e.g.][]{2004A&A...424..429M,2008A&A...483..699G,Bonafede2011,2013MNRAS.433.3208B,2012A&A...540A..38V,2017A&A...603A.122G,2021MNRAS.502.2518S}. This approach starts with generating the vector potential of the magnetic field, $A$, in Fourier space, denoted by $\tilde{A}$. The amplitude and phase of the components of the vector potential were generated such that the phases are randomly distributed between $0$ and $2\pi$ and the amplitudes follow a power-law given by 
\begin{equation}
    |A_k|^2 \propto k^{-\xi},
\end{equation}
where $k$ denotes the magnitude of the three-dimensional wave-vector $\vec{k}$. 
The wave numbers $k$ are related to the spatial scales $\Lambda$ as $k = 0.5 \cdot \frac{2\pi}{\Lambda}$, where $\Lambda$ refers to the reversal scale of the magnetic field, following the definition used by \citet[][]{2004A&A...424..429M}
Because the vector potential and magnetic field are real quantities, we made sure that the matrix $\tilde{A}$ is Hermitian (i.e. equal to its conjugate transpose). The components of the Fourier transform of the magnetic field are then given by the following cross product
\begin{equation}
    \tilde{B}(k) = ik \times \tilde{A}(k).
\end{equation}
This results in the magnetic field $B$, which is simply calculated by (fast) Fourier transform, being divergence-free, isotropic and component-wise Gaussian random, with a power-law spectrum
\begin{equation}
|B_k|^2 \propto k^{-n},
\end{equation}
where $n=\xi-2$. A power-law spectral index of $n=3$ implies that the magnetic field energy density is scale-invariant, for $n<3$ the energy density is larger on smaller scales and for $n>3$ the energy density is mostly in the larger scales 
\citep[][]{2004A&A...424..429M}. The range of spatial scales $\Lambda$ that can be explored is given by the size of the computational grid. The simulated maximum scale on which the magnetic field reverses is equal to $\Lambda_{\mathrm{max}}=\pi/k$ while the minimum scale $\Lambda_\mathrm{min}$ that can be probed is determined by the cell size. 

The normalisation of the magnetic field was set after the Fourier transform such that the magnetic field strength approximately follows an assumed magnetic field profile. Like previous literature, we assumed that the magnetic field profile is proportional to the gas density profile, which is expected to happen during cluster formation from simulations \citep[][]{2008SSRv..134..311D},
\begin{equation}\label{eq:Bfieldprofile}
B(r) = B_0 \left(\frac{n_e(r)}{n_e(0)}\right)^\eta,
\end{equation}
where $B_0$ is the average magnetic field strength at the cluster centre, $n_e$ is the thermal electron gas density profile and $\eta$ denotes the proportionality between the magnetic field strength and electron density. For $\eta=0.5$, the magnetic field energy density is linearly proportional to the thermal gas density. The thermal electron density profile is available for every cluster in the Chandra-Planck sample, from the X-ray observations presented in \citet{2017ApJ...843...76A}, where the fitted profile was assumed to follow a modified double $\beta$ model (see \cite{2006ApJ...640..691V} for more details):

\begin{eqnarray}\label{eq:doubleBmodel}
n_{\rm e}n_{\rm p} &=& n_0^2
\frac{ (r/r_{\rm c})^{-\alpha}}{(1+r^2/r_{\rm c}^2)^{3\beta-\alpha/2}}
\frac{1}{(1+r^\gamma/r_{\rm s}^\gamma)^{\epsilon/\gamma}} \nonumber \\
&+& \frac{n^2_{02}}{(1+r^2/r_{\rm c2}^2)^{3\beta_2}},\label{eq:nenp}
\end{eqnarray} 

Given the modelled magnetic field and observed electron density profile, we calculated the expected Faraday rotation in the clusters by numerical integration of Equation \ref{eq:fdepth}. Then, assuming an intrinsic polarisation $p_0$, and a single polarisation angle $\chi_0$ for a polarised background screen, we computed the polarisation angle of the radio emission at 1.5 GHz at the cluster redshift ($\chi_\mathrm{obs}$) with Equation \ref{eq:RMwavelength}. The predicted Stokes Q and U intensities were then obtained by inversion of $P=\sqrt{Q^2+U^2}$ and Equation \ref{eq:chi}: 
\begin{equation}
Q =  \pm \sqrt{\frac{p_0^2}{1+\tan^2(2\chi_\mathrm{obs})}} \\ 
U =  \pm \sqrt{p_0^2 - Q^2}.
\end{equation}
Using the convention that Stokes Q is positive for $-\pi/2 \leq \chi_\mathrm{obs} \leq \pi/2$ and Stokes U is positive for $0 \leq \chi_\mathrm{obs} < \pi$. 
Finally, the images were convolved with a beam corresponding to a $6^{\prime\prime}$ FWHM at the cluster redshift. From the convolved Stokes Q and U images, we calculated the expected depolarisation fraction at 1.5 GHz in the cluster rest-frame. 

\section{Results - Observations}\label{sec:results}

\begin{figure}[thb]
    \centering
    \includegraphics[width=1.0\columnwidth]{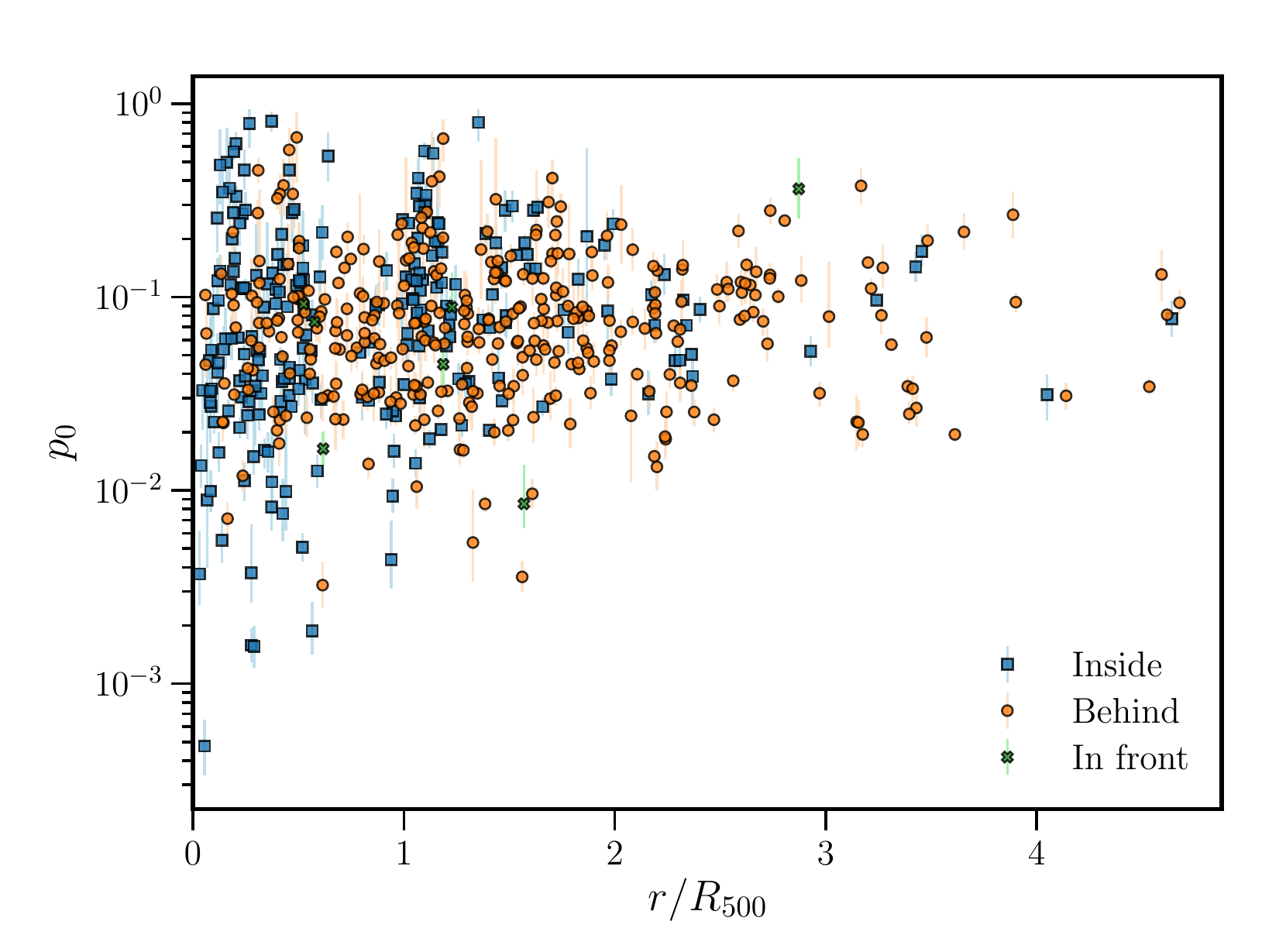}
    \caption{Best-fit intrinsic polarisation fraction against the projected distance to the cluster centre, normalised by the cluster $R_{500}$. Polarised sources are coloured based on their position along the line-of-sight with respect to the nearest cluster, defined in Section \ref{sec:sourcez}.}
    \label{fig:p0}
\end{figure}

The intrinsic polarisation fraction of radio sources is often assumed to be the same for sources irrespective of their projected distance from the cluster centre. To test whether this is a good assumption, we plot in Figure \ref{fig:p0} the intrinsic polarisation fraction $p_0$ (see Eq. \ref{eq:extdepol}) as a function of projected radius to the cluster centre. As the figure shows, there is a relatively large scatter in the intrinsic polarisation fraction of radio sources. This indicates that the assumption does not hold for this dataset and that the intrinsic polarisation fraction should be taken into account when estimating the amount of depolarisation. 

To minimise the effect of the scatter introduced by source-dependent intrinsic polarisation, we calculated for every source the depolarisation ratio $DP$. We defined this as the ratio of the polarisation fraction at 1.5 GHz in the cluster rest-frame $p_\mathrm{1.5GHz}$ to the intrinsic polarisation fraction $p_0$, using the best-fit model (Eq. \ref{eq:extdepol}). In this way, we do not assume the same intrinsic polarisation fraction for all radio sources and take into account the cosmological redshift. 

We combined the information from the upper limits (unpolarised sources) with the depolarisation ratio of polarised sources using the Kaplan-Meier (KM) estimator (\citealt{Feigelson1985}; see also \citealt{Bonafede2011}). The KM estimator is a non-parametric statistic used to estimate the complement of the cumulative distribution function, called the survival function. With $x_1<x_2<..<x_r$ denoting distinct, ordered, observed values, the survival function is given by:
\begin{equation}
S_{KM}(x) = P(X \geq x) = 1 - F(x),
\end{equation}
where F(x) denotes the cumulative distribution function of the random variable $x$, in our case the random variable is the depolarisation ratio measured in the centre of the band (i.e. $DP$). The KM estimator of the survival function is given by
\begin{equation}
\hat{S}_{KM}(x) = 
\begin{cases}
\begin{split}
= &\Pi_{i,x_i<x} \left(1-\frac{d_i}{n_i}\right)^{\delta_i} & \mathrm{\, for \,x> x_1} \\
= & 1 &  \mathrm{\, for \,x \leq x_1},
\end{split}
\end{cases}
\end{equation}
with $x_i$ the observed or censored depolarisation fraction of source $i$, $d_i$ the number of sources with fractional polarisation equal to $x_i$, $n_i$ the number of sources with (upper limits on) fractional polarisation $\geq x_i$ and $\delta_i = (1,0)$ if $x_i$ is polarised or unpolarised, respectively. The KM estimator is here expressed in the case of a right-censored sample, and most algorithms indeed only support right-censored data. Thus, we transformed our left-censored data to right-censored data by subtracting the data from a constant, following \citet[][]{Feigelson1985}.

For unpolarised sources, we calculated upper limits on $p_\mathrm{1.5GHz}$ as explained in Section \ref{sec:fracpolmeasurement}, so an assumption on the intrinsic polarisation fraction must be made to translate this upper limit on the fractional polarisation to an upper limit on the depolarisation ratio. Thus, for these sources we calculated the depolarisation ratio assuming $p_0=0.022$, which is the median of the Kaplan-Meier estimator of all polarised radio sources detected at $r>1.5R_\mathrm{500}$. All KM estimates were calculated using the $\mathtt{lifelines}$\footnote{\url{https://github.com/CamDavidsonPilon/lifelines/}} package \citep[][]{DavidsonPilon2019}.

\subsection{Full sample}
\begin{figure*}[tbh]
    \centering
    \includegraphics[width=1.0\textwidth]{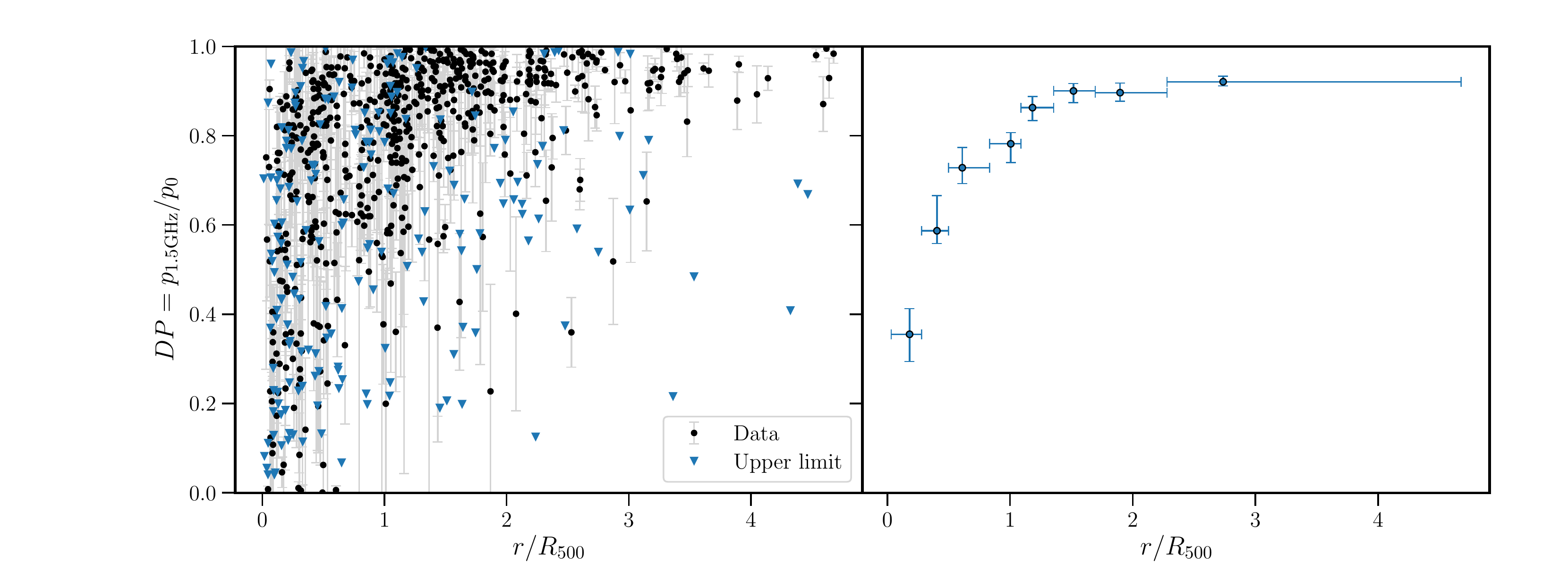}
    \caption{Depolarisation against normalised projected radius. \textit{Left panel:} Full sample of sources and relevant upper limits. The error bars reflect the 68\% confidence interval from the MCMC fitting procedure. \textit{Right panel:} median of the Kaplan-Meier estimate of the depolarisation ratio survival function in different bins of projected radius to the cluster centre. The bin width is chosen such that each bin contains an equal number of polarised sources and is denoted by the horizontal lines (i.e. 0th and 100th percentile). The points are plotted at the median radius in each bin.}
    \label{fig:KMestimate_all}
\end{figure*}

Figure \ref{fig:KMestimate_all} shows in the left panel the depolarisation fraction for all sources in our sample, including upper limits that are below $DP=1$. The right panel shows the median depolarisation fraction, calculated using the KM estimator by splitting the sample into bins of projected radius to the cluster centre. Each bin was chosen such that it contains an equal number of sources. The error bars reflect the 68\% confidence interval of the KM estimator, added in quadrature with the uncertainty introduced by the fitting procedure. The uncertainty introduced by the fitting was estimated using a Monte Carlo method. For every source, we draw 1\,000 samples from a Gaussian distribution with a standard deviation equal to the one-sigma uncertainty on the depolarisation ratio given by the MCMC chain. We note that the error is dominated by the confidence interval of the KM estimator, thus that the effect of the uncertainty in the best-fit polarisation parameters is small. This means that we are limited by the number of polarised radio sources, and not by the quality of the data. 

Figure \ref{fig:KMestimate_all} shows a clear trend of sources being more depolarised as they move towards the cluster centre, where the magnetic field strength and the line-of-sight column densities increase. The depolarisation ratio is around 0.92 beyond $2 R_{500}$, which is likely not an external, but an internal depolarisation effect because at these distances the column density and magnetic field strength of the intracluster medium would be too low to result in significant external depolarisation. 

\begin{figure*}
    \centering
    \includegraphics[width=1.0\textwidth]{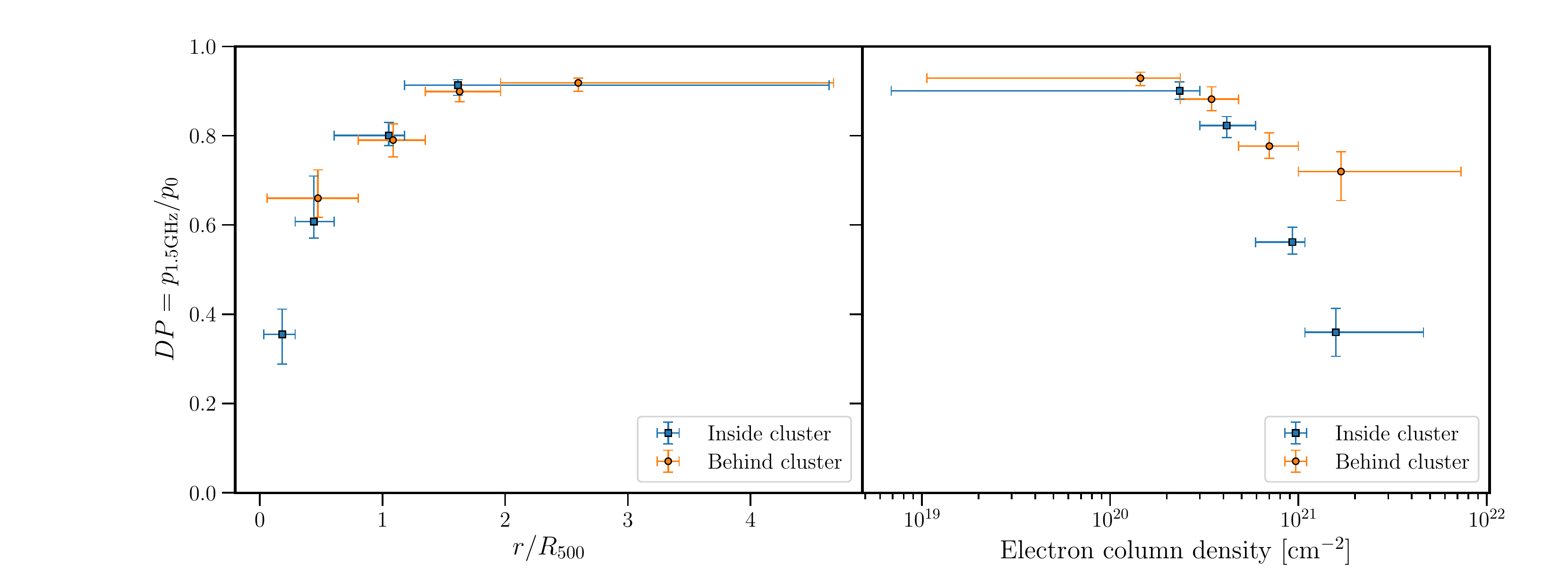}
    \caption{Median of the Kaplan-Meier estimate of the depolarisation ratio survival function in different bins of radius and electron column density. Sources inside clusters are shown in green and sources behind clusters are shown in red. The bin width is chosen such that each bin contains an equal number of polarised sources and is denoted by the horizontal lines. The points are plotted at the median radius or column density in each bin.}
    \label{fig:KMestimate_redshift}
\end{figure*}

\subsection{Background versus cluster members}\label{sec:sourcez}
To investigate whether there is a difference between depolarisation of cluster members and depolarisation of background radio sources, we classified each radio source according to the following definitions. We defined a source to be in front of a cluster if it lies at least $1\sigma_{\mathrm{z}}$ away from a chosen boundary ($\Delta z$) around the cluster redshift 
\begin{equation}
    z_\mathrm{cluster} - (z_\mathrm{source} + \sigma_{\mathrm{z}}) > \Delta z,
\end{equation}
where $z_\mathrm{cluster}$ is the cluster redshift, $z_\mathrm{source}$ the source redshift and $\sigma_{\mathrm{zsource}}$ the one-sigma uncertainty on the source redshift. The values of $z_\mathrm{source}$ and $\sigma_{\mathrm{zsource}}$ are given in Tables \ref{tab:polsources} and \ref{tab:unpolsources} in the column `$z_\mathrm{best}$' for every source.
Similarly, a source was defined to be behind a cluster if 
\begin{equation}
    (z_\mathrm{source} - \sigma_{\mathrm{z}}) - z_\mathrm{cluster} > \Delta z.
\end{equation}
All other sources were defined as inside the cluster. We have set $\Delta z=0.04(1+z)$, following the definition of cluster membership used by \citet[][]{2015ApJ...807..178W}. Sources without an optical counterpart are likely faint sources at redshifts higher than $z=0.35$, particularly because radio galaxies are often hosted by massive elliptical galaxies which should be easily detectable at $z<0.35$ at the depth of Legacy and PanSTARRS. Therefore, sources without an optical counterpart were also defined as background sources.
We verified, through a two-sample KS test, that the measured polarisation fraction of the sample of sources without an optical counterpart does not significantly differ from the sample of sources with an optical counterpart (p-value 0.14), implying that they pass similar Faraday screens.

We investigated the depolarisation effect as a function of radius and electron column density to partially split the degeneracy between magnetic field strength and electron column density, which are both a function of radius, and both increase the amount of depolarisation (Eq. \ref{eq:fdepth}). 
To determine the electron column density  for sources inside the clusters we integrated the best-fit electron density profile along the line-of-sight, from the centre of the cluster out to $R_{500}$, thus effectively integrating over half the sphere. For sources located behind the cluster, this column density was multiplied by two because we have assumed spherical symmetry in the electron profiles. 

In Figure \ref{fig:KMestimate_redshift} we show the depolarisation ratio calculated in different bins of normalised projected distance or electron column densities, for cluster members and background sources separately. Firstly, the figure shows the difficulty of detecting background sources close to the cluster centre, which means larger bins need to be used for background sources than for cluster members. For completeness, the full sample of data-points is plotted in Figure \ref{fig:DP_vs_rnorm_sourcez_scatter}.

Secondly, we detect a significant difference between cluster members and background sources in the highest column density bin (right panel). This could arise because at similar column densities background sources are projected further away from the cluster centre than cluster members and thus probe smaller magnetic field strengths, causing less depolarisation. Additionally, because cluster members are easier to detect near the cluster centre, the largest column density bin also samples preferentially higher column densities for cluster members.

Conversely, we expect that at similar radii, background sources probe higher column densities and are thus more depolarised. However, we do not significantly detect this difference given the uncertainties and the large bin size of background sources near the centre of the cluster.

\begin{figure}[tbh]
    \centering
    \includegraphics[width=1.0\columnwidth]{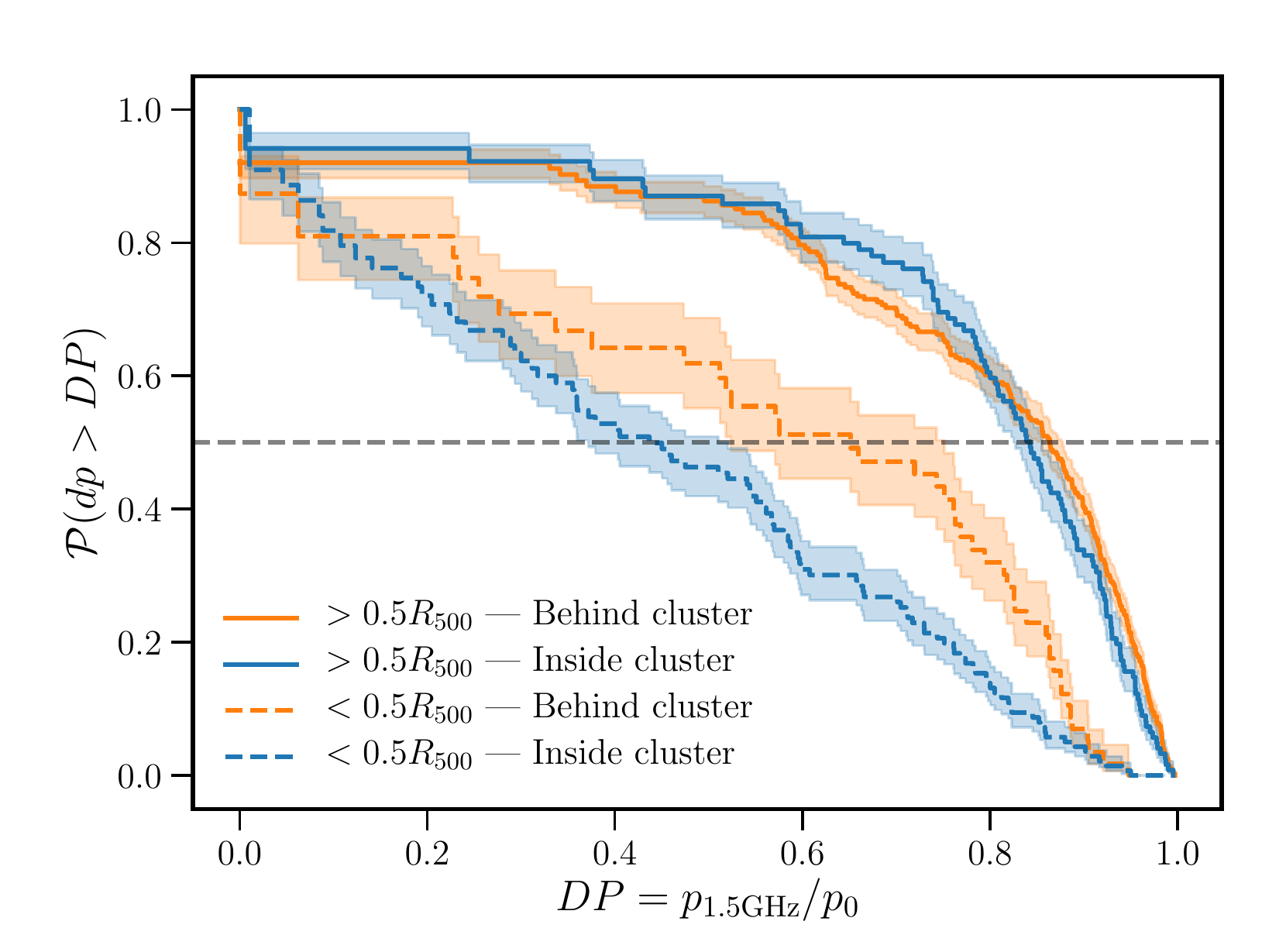}
    \caption{Survival function (i.e. 1-CDF) inferred from the Kaplan-Meier estimator for all sources located at $r/R_\mathrm{500}>0.5$, and $r/R_\mathrm{500}<0.5$. Background sources and cluster members are indicated by red and green, respectively. The grey dashed line shows the location of the 50th percentile, indicating the median for both populations.}
    \label{fig:survival_sourcez}
\end{figure}

Thirdly, at radii where we have similar sampling (i.e. $r>0.5R_\mathrm{500}$), we do not see a significant difference between cluster members and background sources. To statistically confirm this, we used the non-parametric log-rank test \citep[][]{Feigelson1985}, used frequently with other astronomical works dealing with survival analysis \citep[e.g.][]{Bonafede2011,2020ApJ...901..111K,2022A&A...661A..53V}. The resulting survival curves of cluster members and background sources are shown in Figure  \ref{fig:survival_sourcez}, and according to the log-rank test with $p$-value $0.89$ we cannot reject the null hypothesis that there is no difference between background sources and cluster members.

Lastly, at the inner radii ($r<0.5R_\mathrm{500}$) where we do not have a similar sampling of background sources and cluster members, we see a hint of more depolarisation detected in the cluster member population. However, the log-rank test returns $p=0.13$, indicating that with the current sampling, this result is not statistically significant.

\subsection{Dynamical state}\label{sec:datadynstate}

\begin{figure*}[tbh]
    \centering
    \includegraphics[width=1.0\textwidth]{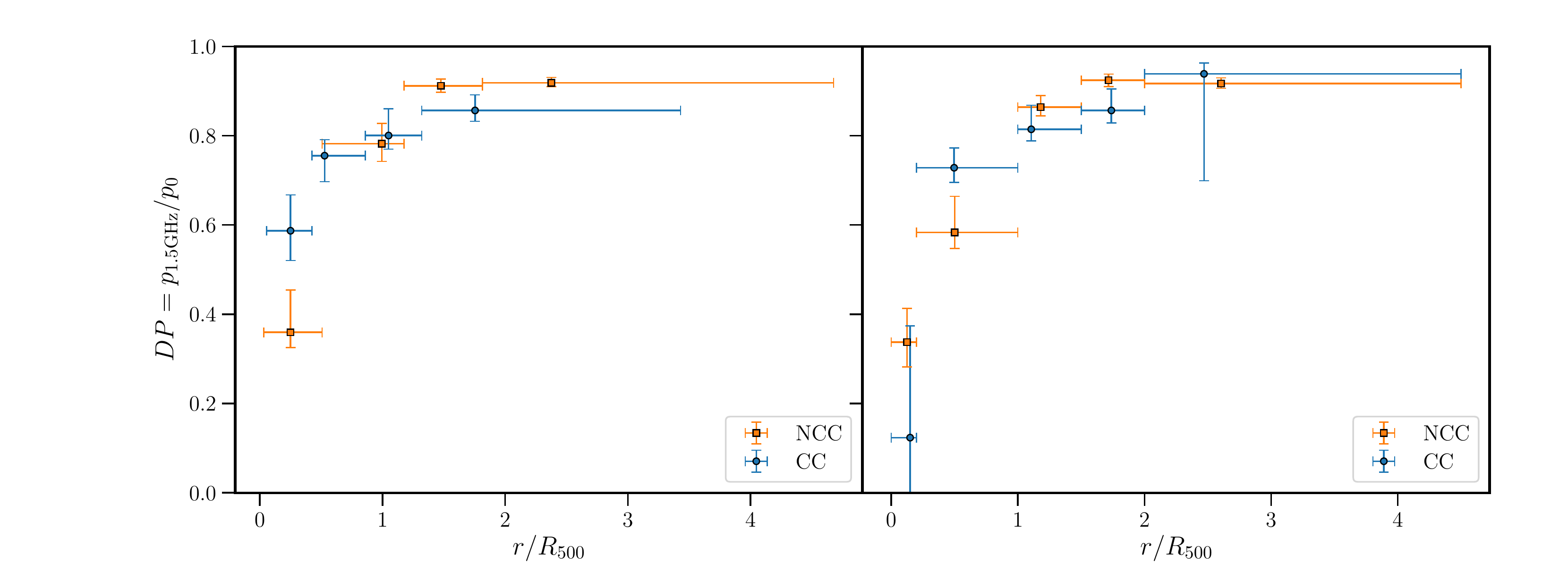}
    \caption{Median of the Kaplan-Meier estimate of the depolarisation ratio survival function in different bins of radius. Cool-core clusters are shown in blue and non-cool-core clusters are shown in orange. The left panel shows bin widths (denoted by horizontal lines) chosen such that each bin contains an equal number of sources detected in polarisation and the right panel shows manually selected bins. The points are plotted at the median radius in each bin.}
    \label{fig:p_dynstate}
\end{figure*}

The magnetic field evolution of galaxy clusters remains poorly constrained. During the lifetime of clusters, mergers with other clusters or smaller substructures can alter the structure and strength of the magnetic field significantly. This section focuses on possible differences between merging and relaxed systems.

Generally, relaxed clusters show strongly peaked, symmetrical X-ray emission that has a radiative cooling time much shorter than the Hubble time \citep[e.g. ][]{1994ARA&A..32..277F}. These clusters show the shortest cooling times in their cores and are therefore often referred to as cool-core clusters. Cluster mergers can destroy the cool core and significantly disturb the observed X-ray morphology \citep[][]{2008ApJ...675.1125B}. Thus, X-ray morphological parameters such as the concentration or cuspiness of the gas density profile can be used to determine whether a system has a cool core \citep[e.g.][]{2017ApJ...843...76A}.

We use the X-ray morphology parameters derived from the Chandra observations of 93 clusters in our sample in \citet[][]{2017ApJ...843...76A} to determine the presence or absence of a cool core.
We note that this split does not perfectly correspond to a split in the dynamical state, as there are rare examples of merging clusters that still show a cool core \citep[e.g.][]{2021ApJ...907L..12S}. However, this split is sufficient to generally divide the sample into merging and relaxed systems. Using the concentration parameter calculated in the 0.15–1.0 $R_\mathrm{500}$ range ($C_\mathrm{SB}$) by \citet[][]{2017ApJ...843...76A} to classify clusters as cool-core or non-cool-core, we found that 65\% (60/93) of the clusters in our sample are non-cool-core (NCC) and 35\% (33/93) are cool-core (CC) clusters.

Figure \ref{fig:p_dynstate} shows the depolarisation effect separately for CC and NCC clusters in equal frequency bins, the full sample is plotted in Figure \ref{fig:DP_vs_dynstate_scatter}. We see a hint in the first radius bin of detecting more depolarisation in NCC clusters than in CC clusters.

\begin{figure}[tbh]
    \centering
    \includegraphics[width=1.0\columnwidth]{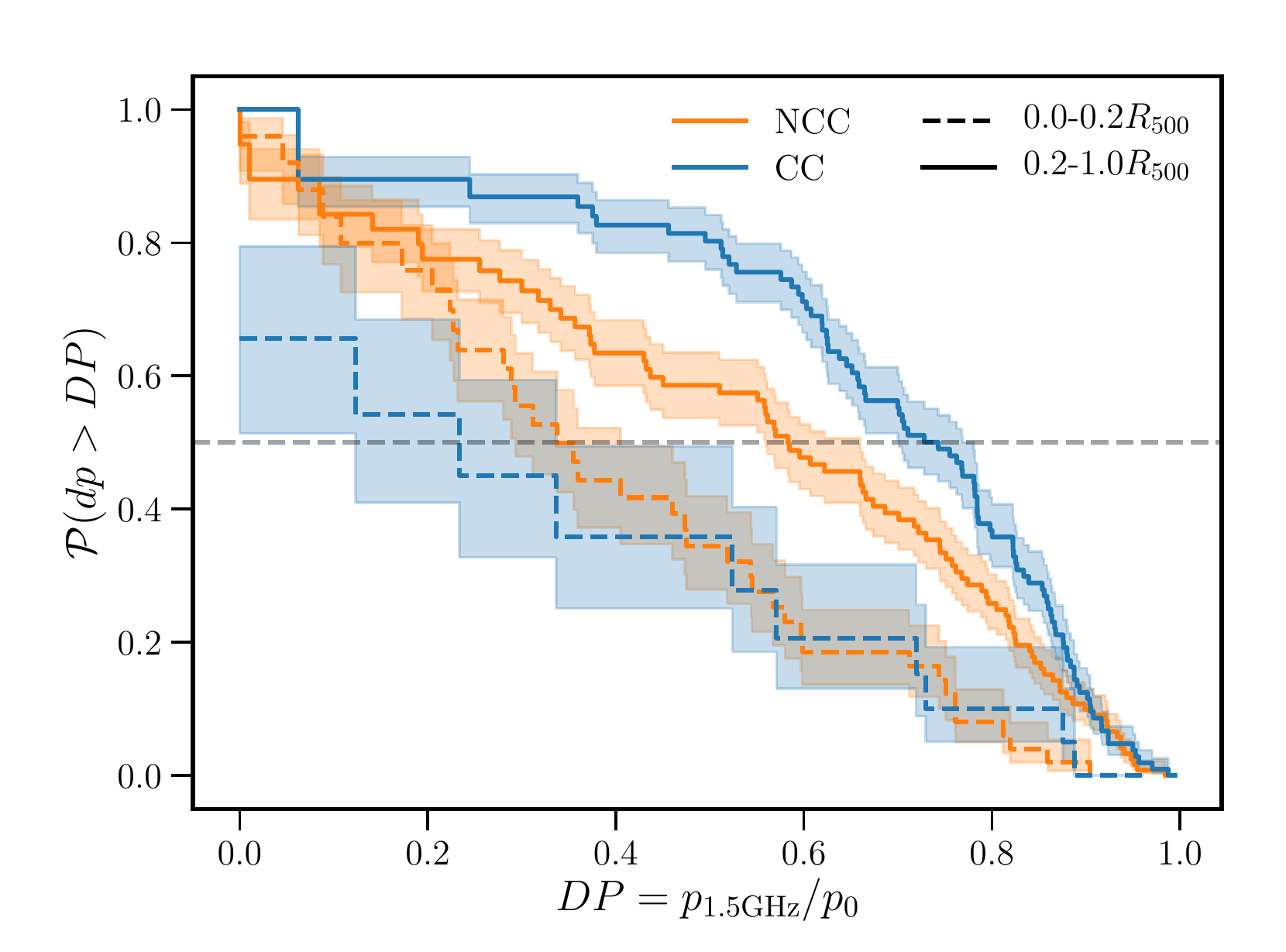}
    \caption{Survival functions (i.e. 1-CDF) inferred from the Kaplan-Meier estimator for all sources in the $0.0-0.2R_\mathrm{500}$ and $0.2-1.0R_\mathrm{500}$ bins, separately for sources detected around CC and NCC clusters. The grey dashed line shows the location of the 50th percentile, indicating the median for both populations.}
    \label{fig:survival_dynstate}
\end{figure}

To separate the effect of the central cooling core region in CC clusters, we have manually defined bins of projected radius in the right panel of Figure \ref{fig:p_dynstate}. We have chosen an inner radius bin of  $0.0-0.2R_\mathrm{500}$ because the effect of the cooling core is significant only in the inner $\sim$ 0.2$R_\mathrm{500}$ of CC clusters \citep[e.g. ][]{2006AN....327..595V,2011A&A...526A..79E}. The right panel shows that the larger depolarisation fraction in NCC is dominated by sources detected at $r>0.2R_\mathrm{500}$. In fact, sources detected at $r<0.2R_\mathrm{500}$ show a hint that there is more depolarisation in the central cooling core region of CC clusters, although the uncertainties are large due to the low number of sources detected near the centre of CC clusters. At $r<0.2R_\mathrm{200}$, we have detected only 9 sources and 16 upper limits in CC clusters, and 36 sources and 14 upper limits in NCC clusters. The significance of these results was determined by comparing the survival functions of sources detected in the $0.0-0.2R_\mathrm{500}$ and $0.2-1.0R_\mathrm{500}$ bins. The survival functions are shown in Figure \ref{fig:survival_dynstate} and the log-rank test yields $p$-values of $0.22$ and $0.001$ for the $0.0-0.2R_\mathrm{500}$ and $0.2-1.0R_\mathrm{500}$ bins, respectively. This implies that the hint is statistically significant, with less depolarisation in CC clusters than in NCC clusters outside the core region. Conversely, inside the core region we do not have enough sources to significantly detect a difference between the two samples.

\begin{figure*}[tbh]
    \centering
    \includegraphics[width=1.0\textwidth]{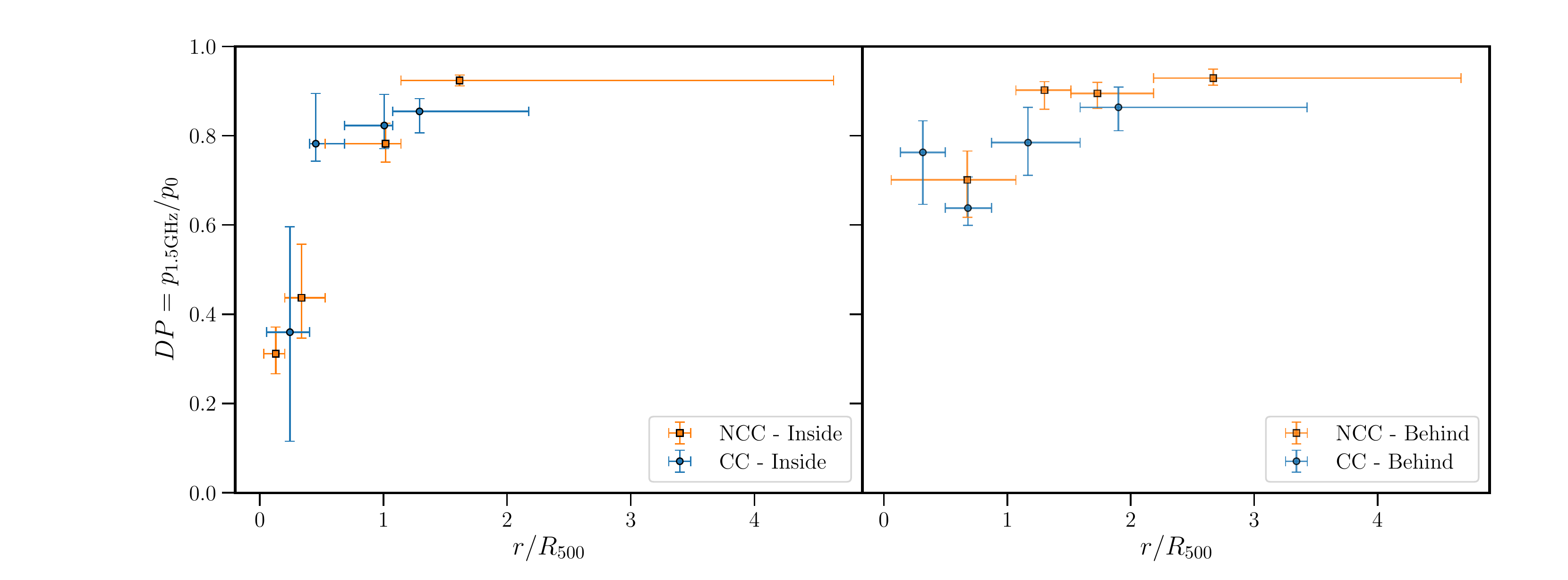}
    \caption{Same as the left panel of Figure \ref{fig:p_dynstate}, but separately for cluster members (left panel) and background sources (right panel).}
    \label{fig:p_dynstate_sourcez}
\end{figure*}

\paragraph{}
To examine to what extent the results of the CC/NCC split are affected by the position along the line-of-sight of the sources, we repeat this analysis separately for background sources and cluster members. Figure \ref{fig:p_dynstate_sourcez} shows that we do not detect a difference between NCC and CC clusters in either sub-sample. This is likely because of the low number of sources left in each sub-sample. We are mainly limited by the number of polarised sources detected near the centre of CC clusters. Comparing the survival curves of the NCC and CC sample for sources detected below $0.5R_\mathrm{500}$, the log-rank test returns $p$-values of $0.07$ and $0.24$, for the sources inside clusters and behind clusters, respectively. Thus, we cannot significantly detect differences between NCC and CC clusters when splitting the sample into background and cluster members due to the limited amount of data points. We do see that most of the depolarisation found at small radii is from cluster members, although the uncertainties become quite large due to the small sample sizes.

\subsection{Cluster mass and redshift}\label{sec:massredshiftResults}
\begin{figure}
    \centering
    \includegraphics[width=1.0\columnwidth]{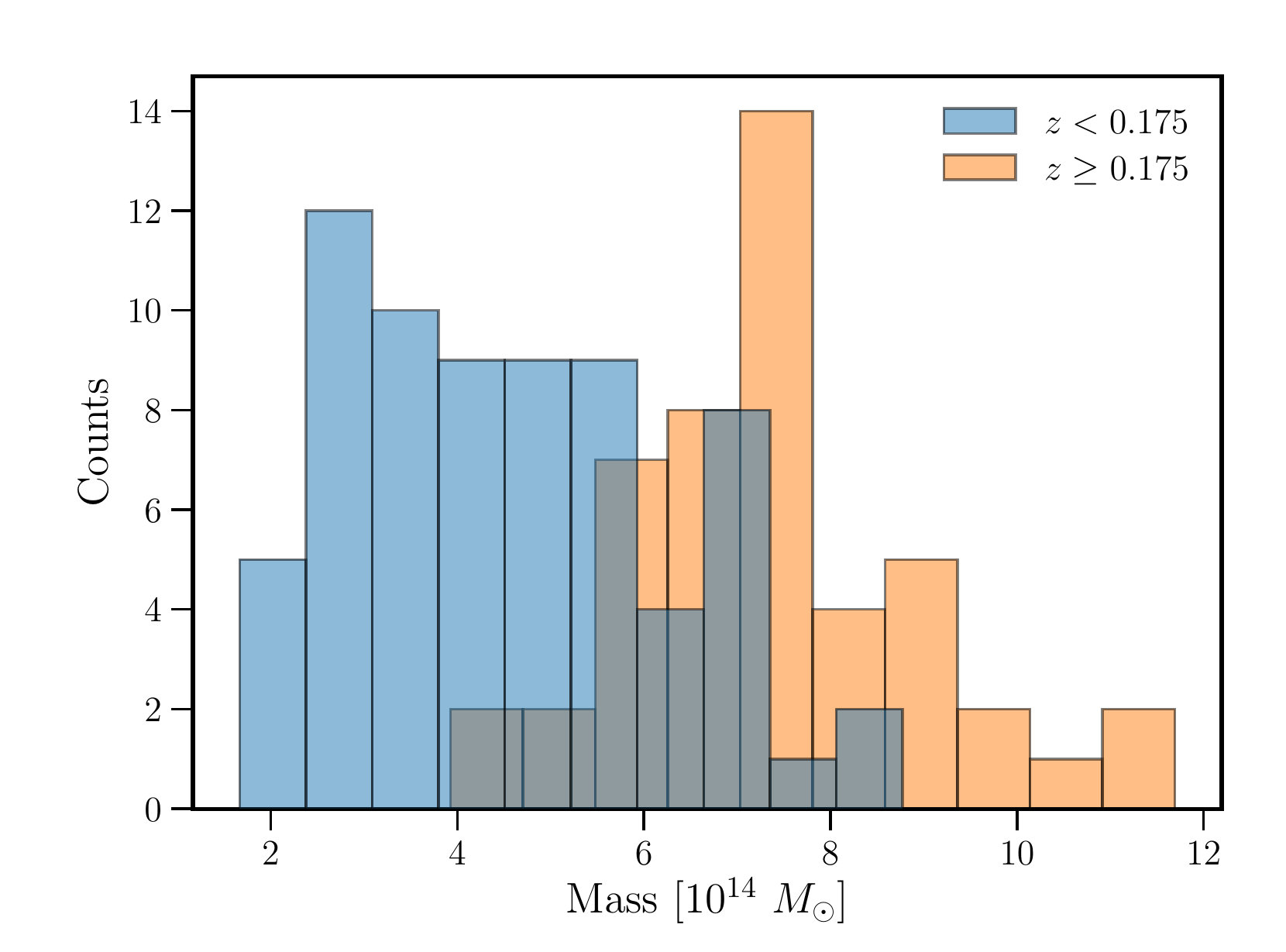}
    \caption{Distribution of cluster mass in the low- and high-redshift samples.}
    \label{fig:hist_mass_redshift}
\end{figure}
Although the sample of clusters is constrained to a relatively low redshift range ($z<0.35$), we can attempt to trace the evolution of the magnetic field of clusters, by splitting the sample based on cluster redshift. We note that the redshift of the host cluster should correlate with the amount of beam depolarisation because the same telescope resolution corresponds to larger physical areas probed at higher redshifts. This means that we effectively average over larger magnetic field scales, and thus expect more beam depolarisation.
Another effect that we have to take into account is the selection function of the Planck cluster sample. There is a strong correlation between cluster mass and redshift, with the most massive clusters preferentially being detected at high redshift \citep[see Fig. 26 in][]{2016A&A...594A..27P}. This means that a cut in redshift effectively also corresponds to a mass-cut, as shown in Figure \ref{fig:hist_mass_redshift}. As the figure shows, it is not possible to separate the effects of cluster redshift and cluster mass because there is almost no overlap in the same mass range.

\begin{figure}
    \centering
    \includegraphics[width=1.0\columnwidth]{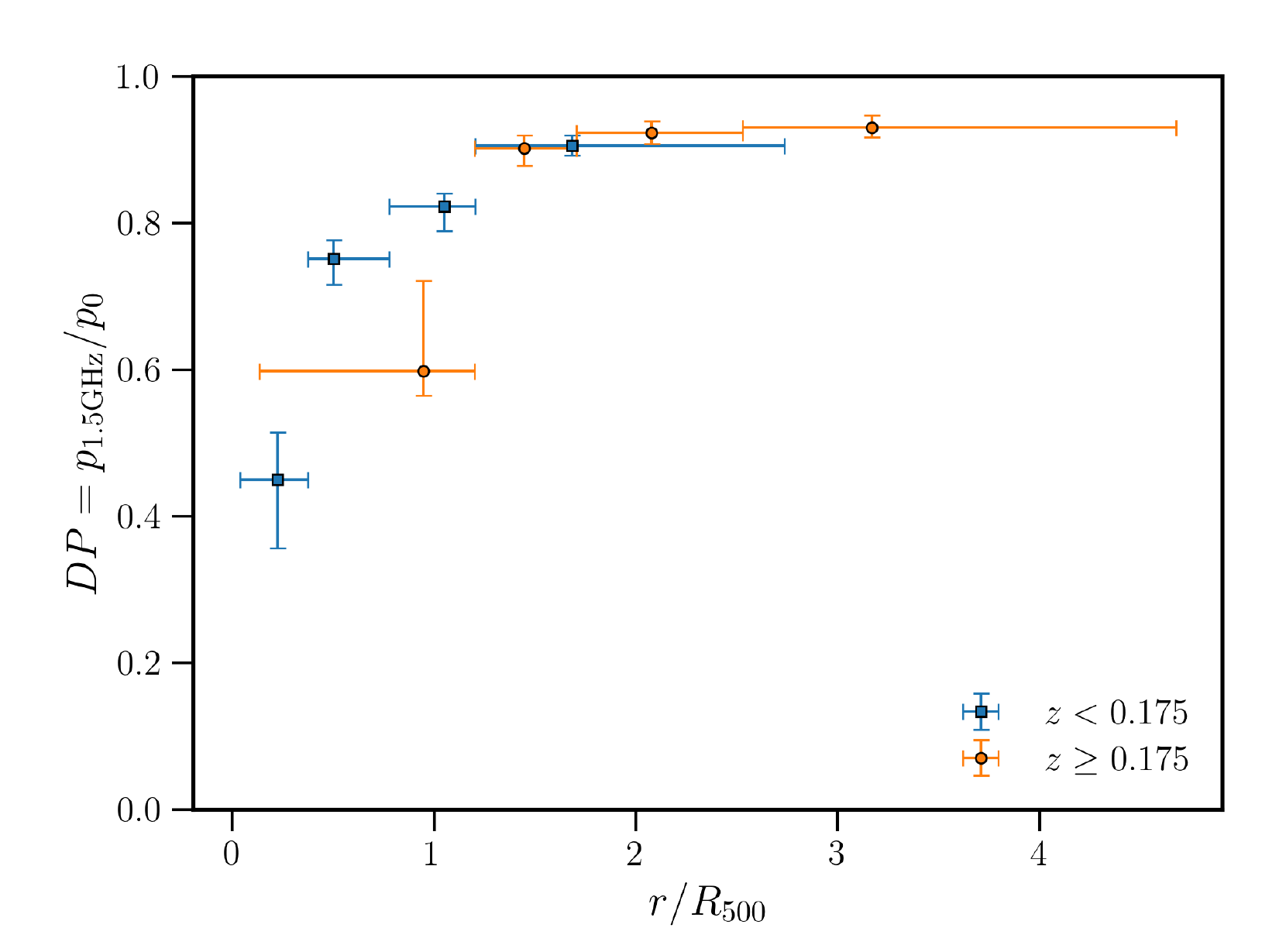}
    \caption{Median of the Kaplan-Meier estimate of the depolarisation ratio survival function in bins of projected radii and column density, for the low- and high-redshift clusters. The bin width is chosen such that each bin contains an equal number of sources detected in polarisation and is denoted by the horizontal lines. The points are plotted at the median projected radius.}
    \label{fig:redshiftbins}
\end{figure}

\begin{figure}
    \centering
    \includegraphics[width=1.0\columnwidth]{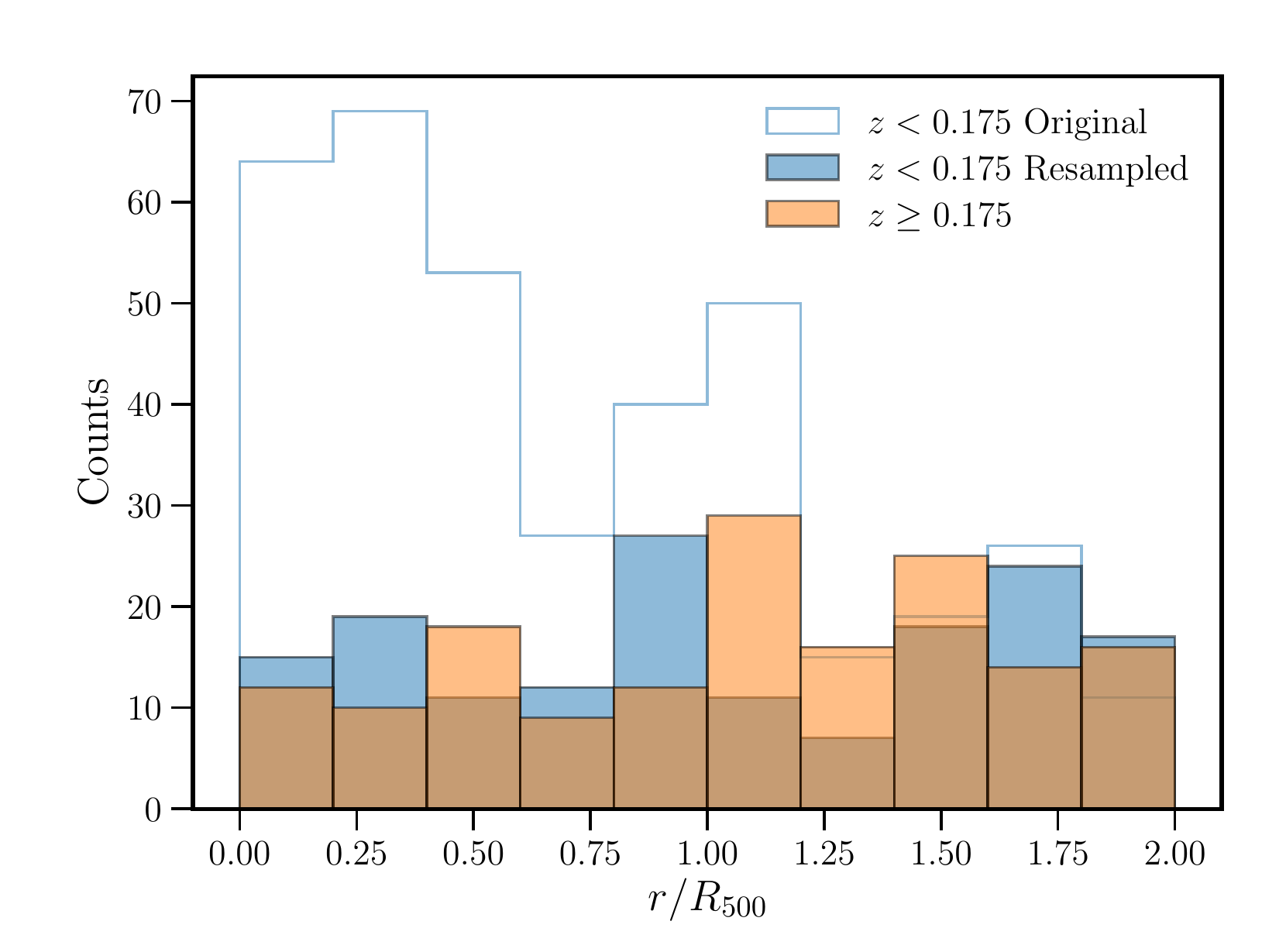}
    \caption{Distribution of the projected radius of sources detected in the low and high-redshift cluster samples. The solid blue histogram shows the result of one realisation of randomly sub-sampling sources in the low-redshift cluster sample to match the distribution of sources in the high-redshift sample.}
    \label{fig:rnorm_resample_clusterz}
\end{figure}

Figure \ref{fig:redshiftbins} shows the depolarisation trend for low- and high-redshift clusters separately, and the full sample is shown in Figure \ref{fig:DP_vs_clusterz_scatter}. The low-redshift sample contains clusters with $z<0.175$ and the high-redshift sample contains clusters with $0.175<z<0.35$. The first thing to note is that we detect significantly more sources at lower projected radii through the low-redshift clusters. Each projected radius bin has 78 detected sources through low-redshift clusters, while the high-redshift clusters have only 43 detected sources per bin. This is expected because the larger angular size of low-redshift clusters makes it easier to detect polarised sources, especially in the centre of the cluster. The low number of polarised sources detected at low projected radii in the high-redshift sample makes it difficult to compare the two populations. Therefore, we performed a bootstrap re-sampling to enforce that we have a similar sampling of radius in both low- and high-redshift clusters. This was repeated 1000 times, with one realisation of the re-sampled values shown in Figure \ref{fig:rnorm_resample_clusterz}. Out of 1000 log-rank tests comparing the high-redshift sample to the sub-sampled low-redshift sample, 4\% ($43/1000$) of the tests returned $p<0.05$, indicating that we cannot distinguish a difference in depolarisation in the low- and high-redshift sample of clusters. However, this is likely due to the low number of sources in the high-redshift sample.

\subsection{Presence of a radio halo}
There is an apparent dichotomy in clusters regarding the presence of a radio halo, where clusters that show a radio halo are almost always found to be dynamically disturbed, while clusters without a radio halo are more relaxed \citep[][]{2021A&A...647A..51C}. However, there are some cases of merging clusters without radio halos or with much fainter radio halos than usual \citep[e.g.][]{2018A&A...609A..61C,2011MNRAS.417L...1R}. While these might be special cases, it is interesting to investigate whether there are differences between the magnetic fields in merging clusters that show a radio halo and those that do not.

\begin{figure}
    \centering
    \includegraphics[width=1.0\columnwidth]{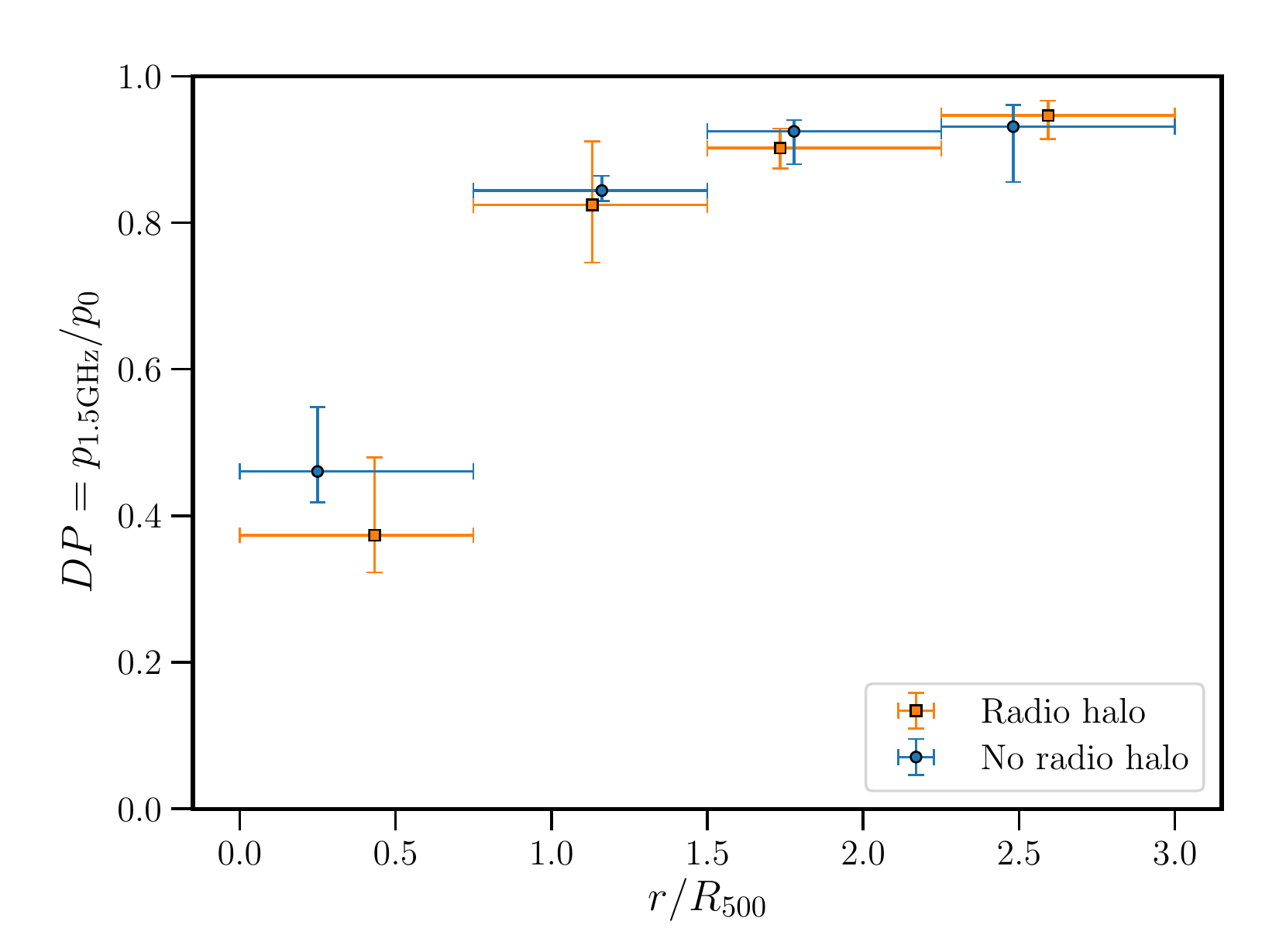}
    \caption{Comparison of the median observed depolarisation ratio against projected distance between the merging clusters with and without detected radio halos.}
    \label{fig:radiohalo}
\end{figure}

We searched the literature for every cluster in our sample and found that out of the 60 clusters classified as merging, 26 have a radio halo detection, also incorporating the results of the second Data Release of the LOFAR Two-meter Sky Survey \citep{2022A&A...660A..78B}. This thus splits the sample in about half, allowing for the same bins to be used. The resulting depolarisation curves are very similar, as shown in Figure \ref{fig:radiohalo}, and the log-rank test for similarity returned a $p$-value of $0.79$. It is thus clear that with the current sample size we see no evidence of a difference in depolarisation between clusters with radio halos and clusters without radio halos. 

\section{Results - Modelling}\label{sec:results2}
This section details the results of the modelling, and the comparison of theory with observations. We simulated magnetic fields following the approach laid out in Section \ref{sec:modelling}. We used the density profiles presented in \citet{2017ApJ...843...76A}, which were fitted to the modified double $\beta$ model shown in Equation \ref{eq:doubleBmodel}. Profiles were only available for the 102 clusters from the ESZ catalogue, so we could not model the depolarisation in the 24 new clusters from the PSZ1 \citep[][]{2015A&A...581A..14P} and PSZ2 \citep[][]{2016A&A...594A..27P} catalogues. 

\subsection{Effect of density profiles}\label{sec:effectofdensityprofiles}

\begin{figure}
    \centering
    \includegraphics[width=1.0\columnwidth]{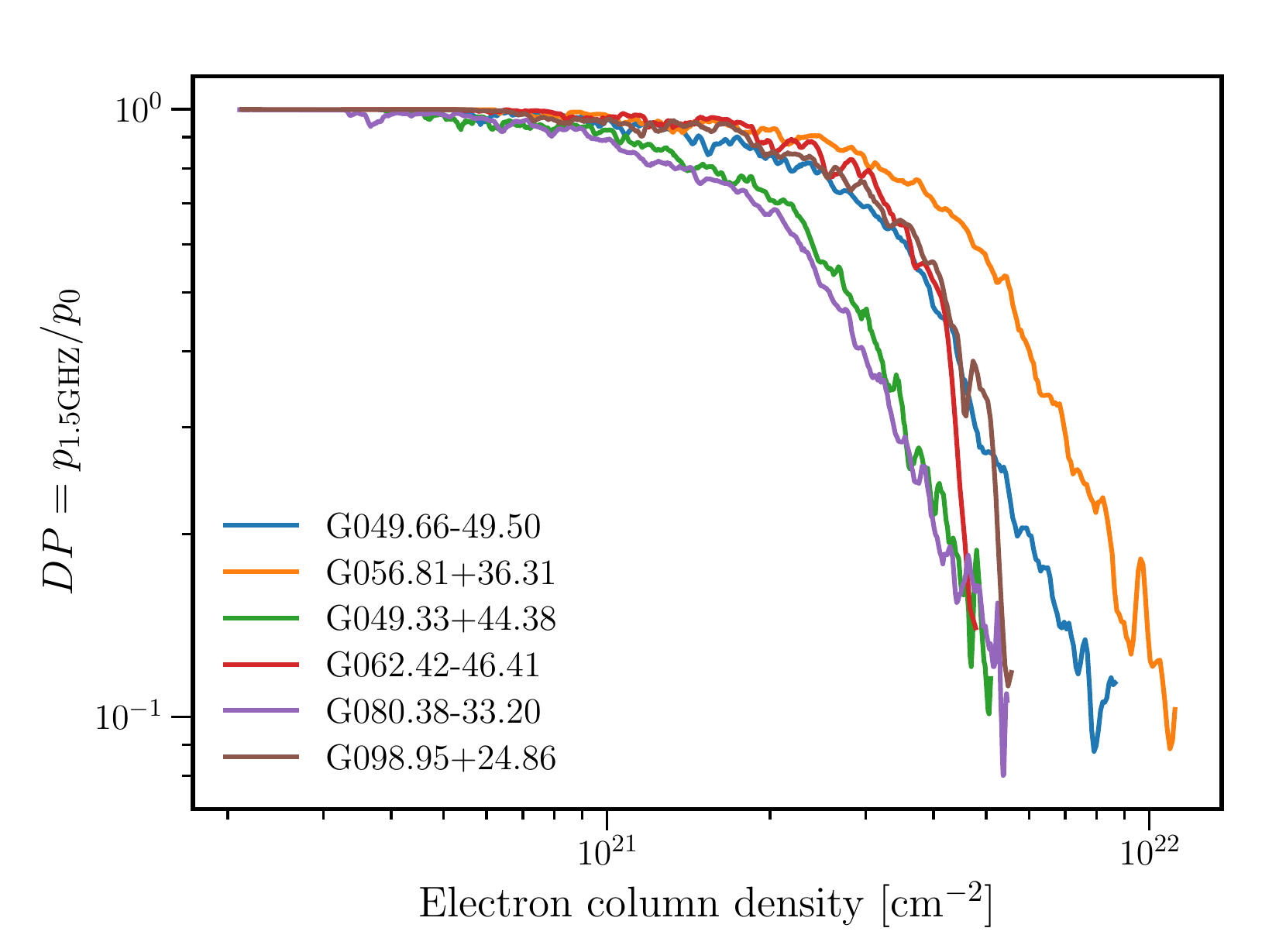}
    \caption{Modelled depolarisation ratio as a function electron column density for six different clusters around $z=0.1$. The assumed parameters for the magnetic field are $B_0=5.0\, \mu$G, $n=11/3$ and $\eta=0.5$, with $\Lambda_\mathrm{min}=4$ kpc and $\Lambda_\mathrm{max}=1024$ kpc.}
    \label{fig:simulated6}
\end{figure}

In previous works, when clusters were stacked often a mean profile was assumed \citep[e.g.][]{2004A&A...424..429M,Bonafede2011}. We first investigated the effect of the different electron density profiles. We used a subset of six arbitrarily chosen clusters around the same redshift $z=0.1$, such that we probe about the same physical scales. The modelled magnetic field parameters for this experiment are $B_0=5.0\, \mu$G, $n=11/3$ and $\eta=0.5$ with a box-size of $1024^3$ pixels, where each pixel represents $2$ kpc. The minimum magnetic field reversal scale $\Lambda_\mathrm{min}$ is thus 4 kpc, and the maximum reversal length scale, $\Lambda_\mathrm{max}=1024$ kpc. At the redshift of $0.1$, the 6$^{\prime\prime}$ beam corresponds to a physical scale of 11 kpc. 
All models start from the same random initialisation of the magnetic field vector potential $A$, meaning that the only difference between the simulated clusters is the assumed electron density profile. The properties of the clusters are given in Table \ref{tab:simulated6}.

\begin{table}[]
\caption{Properties of the six clusters that were modelled in Figure \ref{fig:simulated6}.}
\label{tab:simulated6}
\resizebox{\columnwidth}{!}{%
\begin{tabular}{@{}lllll@{}}
\toprule
Cluster       & Redshift & Dynamical state\tablefootmark{a} & \begin{tabular}[c]{@{}l@{}}Mass\\ {[$10^{14}M_\odot$]}\end{tabular} & \begin{tabular}[c]{@{}l@{}}$n_e$ ($r=10$ kpc)\tablefootmark{a}\\ {[cm$^{-3}$]}\end{tabular} \\ \midrule
G049.66-49.50 & 0.098    & CC              & 3.63$^{+0.30}_{-0.30}$                                          & 0.01                                                                   \\
G056.81+36.31 & 0.095    & CC              & 4.38$^{+0.19}_{-0.21}$                                          & 0.02                                                                   \\
G049.33+44.38 & 0.097    & NCC             & 3.67$^{+0.26}_{-0.26}$                                          & 0.004                                                                  \\
G062.42-46.41 & 0.090    & NCC             & $3.47^{+0.28}_{-0.27}$                                          & 0.02                                                                   \\
G080.38-33.20 & 0.11     & NCC             & $3.77^{+0.27}_{-0.28}$                                          & 0.004                                                                  \\
G098.95+24.86 & 0.092    & NCC             & $2.58^{+0.16}_{-0.18}$                                          & 0.01                                                                   \\ \bottomrule
\end{tabular}%
}
\tablefoot{\tablefoottext{a}{The X-ray properties are taken from \citet{2017ApJ...843...76A}.}}
\end{table}

The resulting depolarisation ratio as a function of electron column density is shown in Figure \ref{fig:simulated6}. From this figure it is clear that even at the same electron column densities, the amount of depolarisation can be quite different in different clusters, depending on the electron density profile. This is because at the same electron column densities, the local electron density profile along the line of sight can still differ quite a lot between clusters. This also influences the magnetic field strength along the line of sight because we assumed a relation between the magnetic field strength and the electron density profile. This means that it is important to take into account the different electron density profiles of the clusters, rather than define a mean electron density profile to stack the clusters. 

\subsection{Effect of the magnetic field strength and fluctuation scales}\label{sec:effectparameters}
We have chosen an arbitrary cluster, PLCKESZ G039.85-39.98, located at $z=0.176$, to investigate the qualitative effect on the depolarisation profiles of changing the scales on which the magnetic field fluctuations and the central magnetic field strength $B_0$. The effect of increasing the magnetic field is easily understood to result in more depolarisation because the scatter in RM increases \citep[e.g.][]{2004A&A...424..429M}. To understand the effects of changing the fluctuation scales, we must consider two different competing effects. First, as more power is put into larger scale fluctuations (i.e. increasing $n$), the scatter in RM over the entire cluster increases because one is integrating coherently over longer path lengths (cf. Eq. \ref{eq:fdepth}). At the same time, because the fluctuations on smaller scales are reduced, the scatter in RM over the region probed by each individual observing beam decreases. Thus, depending on the size of the observing beam, this will either increase or decrease the amount of depolarisation as $n$ changes.

\begin{figure}
    \centering
    \includegraphics[width=1.0\columnwidth]{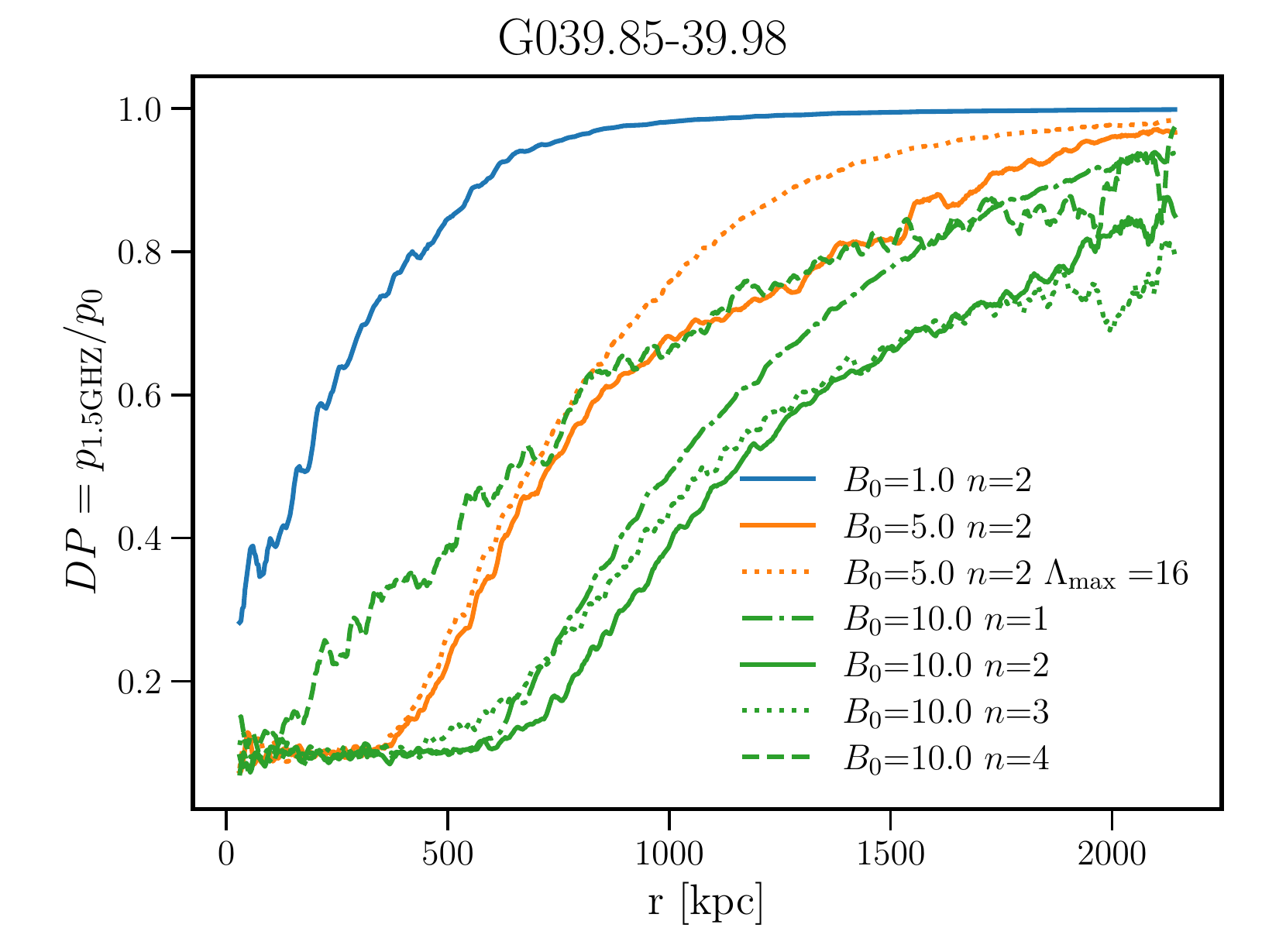}
    \caption{Effect of the magnetic field parameters on the depolarisation profile of the cluster G039.85-39.98, with $R_\mathrm{500}=1.2$ Mpc. The varied parameters are the central magnetic field strength $B_0$ in units of $\mu$G, the magnetic field power-law index $n$ and the maximum correlation scale $\Lambda_\mathrm{max}$ in kpc. For all but one model, the maximum correlation scale was set to the maximum allowed by the image.}
    \label{fig:parameters}
\end{figure}

We plot the modelled depolarisation profiles in Figure \ref{fig:parameters} as a function of different parameters. As expected, the amount of depolarisation increases with increasing magnetic field strength. When the magnetic field energy density is mostly on large scales (i.e. $n=4$), the depolarisation profile becomes quite flat as a function of projected radius because the magnetic field becomes correlated on scales larger than the beam. As the magnetic field becomes more correlated on smaller scales (i.e. from $n=4$ to $n=2$, green lines), the amount of depolarisation increases. However, as we put even more power on smaller scales (i.e. from $n=2$ to $n=1$) we reach the turn-over point where the effect of decreasing the RM scatter over the entire cluster dominates increasing the RM scatter on regions probed by the beam, resulting in less depolarisation. The exact turn-over point in the slope $n$ depends on the size of the sampling region (i.e. the observing beam size) that the simulated radio images are smoothed with and can be different for different locations in the cluster, which have different magnetic field strengths.

The degeneracy between a steep and strong magnetic field (e.g. $B_0=10\,\mu$G, $n=4$) and a shallower and weaker magnetic field (e.g. $B_0=5\,\mu$G, $n=2$) is also clear from this figure. This implies that using depolarisation alone makes it difficult to disentangle between a weaker magnetic field with a steep power-law index, or a shallower magnetic field with a flatter power-law index.  
The effect of setting a maximum fluctuation scale of $\Lambda=16$ kpc does not strongly influence the depolarisation ratio except somewhat at the cluster outskirts. This can be explained by the fact that the observing resolution (FWHM of 18 kpc at the cluster redshift) is comparable to the maximum fluctuation scale. However, the amount of depolarisation does decrease slightly because the scatter in RM over the entire cluster will be smaller due to the magnetic field being less correlated along the line of sight.

\begin{table}[]
\caption{Model parameters used for comparison to observations.}
\label{tab:modelparams}
\resizebox{\columnwidth}{!}{%
\begin{tabular}{@{}lllll@{}}
\toprule
Cluster redshift bin & \begin{tabular}[c]{@{}l@{}}Observed beam \\ FWHM (kpc)\end{tabular} & \begin{tabular}[c]{@{}l@{}}Model resolution\\ (kpc pixel$^{-1}$)\end{tabular} & \begin{tabular}[c]{@{}l@{}}Model grid size\\ (pixels)\end{tabular} & $N$\tablefootmark{a}  \\ \midrule
0.05 - 0.09          & 6-10                                                      & 1                                                                             & $2048^3$                                                           & 22 \\
0.09 - 0.16          & 10-17                                                     & 2                                                                             & $1024^3$                                                           & 23 \\
0.16 - 0.35          & 17-30                                                     & 3                                                                             & $1024^3$                                                           & 40 \\ \bottomrule
\end{tabular}%
}
\tablefoot{\tablefoottext{a}{$N$ denotes the amount of clusters with X-ray observations available such that models could be generated.}}
\end{table}

\subsection{Comparison with observations}
To compare the models to the data, we modelled the depolarisation ratio as a function of projected radius for every cluster for which a Chandra observation and thus electron density profile was available. Given the computational intensity (which scales as $N^3$ where $N$ is the number of pixels) of generating many magnetic field cubes out to typical values of $R_{500}$, we decided to simulate different clusters with different resolution depending on the cluster redshift. To simulate the depolarisation effect, the model resolution should be at least a few times the physical resolution given by the synthesised beam of the radio observations. The resolution of the synthesised beam is given in Table \ref{tab:modelparams} with the accompanying model resolution and model grid sizes used. For clusters below $z=0.05$ it was not feasible to generate simulations, since the physical resolution (FWHM) of the $6^{\prime\prime}$ synthesised beam corresponds to less than $6$ kpc, and thus the resolution of the models should be higher than $1$ kpc pixel$^{-1}$, which made the cube size unfeasibly big to generate. All modelled depolarisation profiles are shown in Figure \ref{fig:all_models} for completeness.

One final effect that we have to take into account when comparing the model to the data is internal depolarisation. Figure \ref{fig:KMestimate_redshift} showed that the depolarisation ratio is around 0.92 at the cluster outskirts. The fact that this is not 1.0 is likely caused by internal depolarisation effects. This internal depolarisation effect should not affect the trend, and therefore we multiply the simulated depolarisation ratio by the depolarisation ratio measured in the cluster outskirts to incorporate this effect. 

\subsubsection{Background versus cluster members}\label{sec:model_background}
\begin{figure}
    \centering
    \includegraphics[width=1.0\columnwidth]{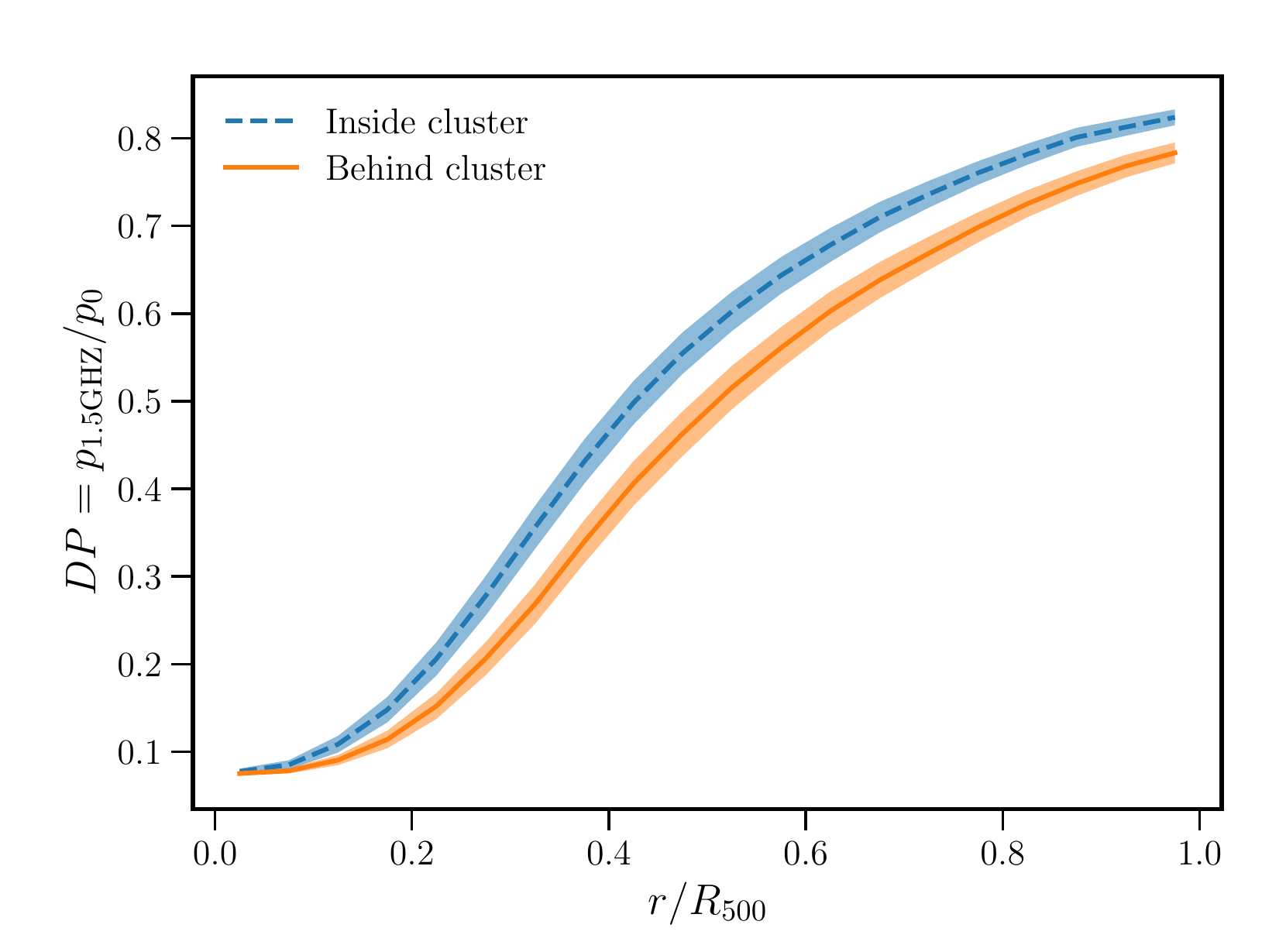}
    \caption{Average depolarisation ratio profile over all simulated clusters using a Faraday screen located behind (blue dashed line) or inside (orange solid line) the cluster. The uncertainty interval indicates the standard error on the mean of the simulated profiles.}
    \label{fig:inside_behind_models}
\end{figure}

We can model the expected difference between the depolarisation of cluster members and background sources assuming that this difference can be fully attributed to the larger path length of background sources through the cluster. We assume that the radio emission from cluster members on average intersects about half the ICM column density and that emission from background sources travels through the full column. Theoretically, this is expected to give on average a factor of two larger Faraday depth for background sources (cf. Eq. \ref{eq:fdepth}) and a factor $\sqrt{2}$ in $\sigma_\mathrm{RM}$, which theoretically should not result in more than a factor two in depolarisation (cf. Eq. \ref{eq:extdepol}) for the wavelength range that we are probing. Indeed, when modelling the depolarisation profiles occurring as a result of a Faraday screen halfway inside the cluster versus a Faraday screen behind the cluster, Figure \ref{fig:inside_behind_models} shows that the location of the Faraday screen only has a marginal effect on the resulting depolarisation. This is in agreement with the results shown in Figure \ref{fig:KMestimate_redshift}, where no clear difference between background sources and cluster members was found.

\subsubsection{Average magnetic field properties}
As shown in Section \ref{sec:sourcez}, we did not detect a significant difference in the depolarisation of cluster members and sources located behind the clusters. This is consistent with a picture where only the difference in path length between background radio emission and radio emission from the cluster medium affects the depolarisation of radio sources. To estimate the average properties of magnetic fields in galaxy clusters we can therefore compare our models with the depolarisation calculated over all sources (background and cluster members) to maximise the signal-to-noise ratio. 

Although $\eta$ is an important parameter that can influence the magnetic field estimates and the dependence on the fluctuation power-law slope $n$ \citep[][]{2020ApJ...888..101J}, we fixed $\eta=0.5$ to reduce the number of free parameters. This value is chosen such that the magnetic field energy density follows the thermal gas density \citep[as found in e.g. Coma and Abell 2382][]{2010A&A...513A..30B,2008A&A...483..699G}. We then varied $B_0=[1,5,10]\, \mu$G and $n=[1,2,3,4]$. The maximum and minimum correlation scales are also fixed to the minimum and maximum size allowed by the computational grid.

\begin{figure}
    \centering
    \includegraphics[width=1.0\columnwidth]{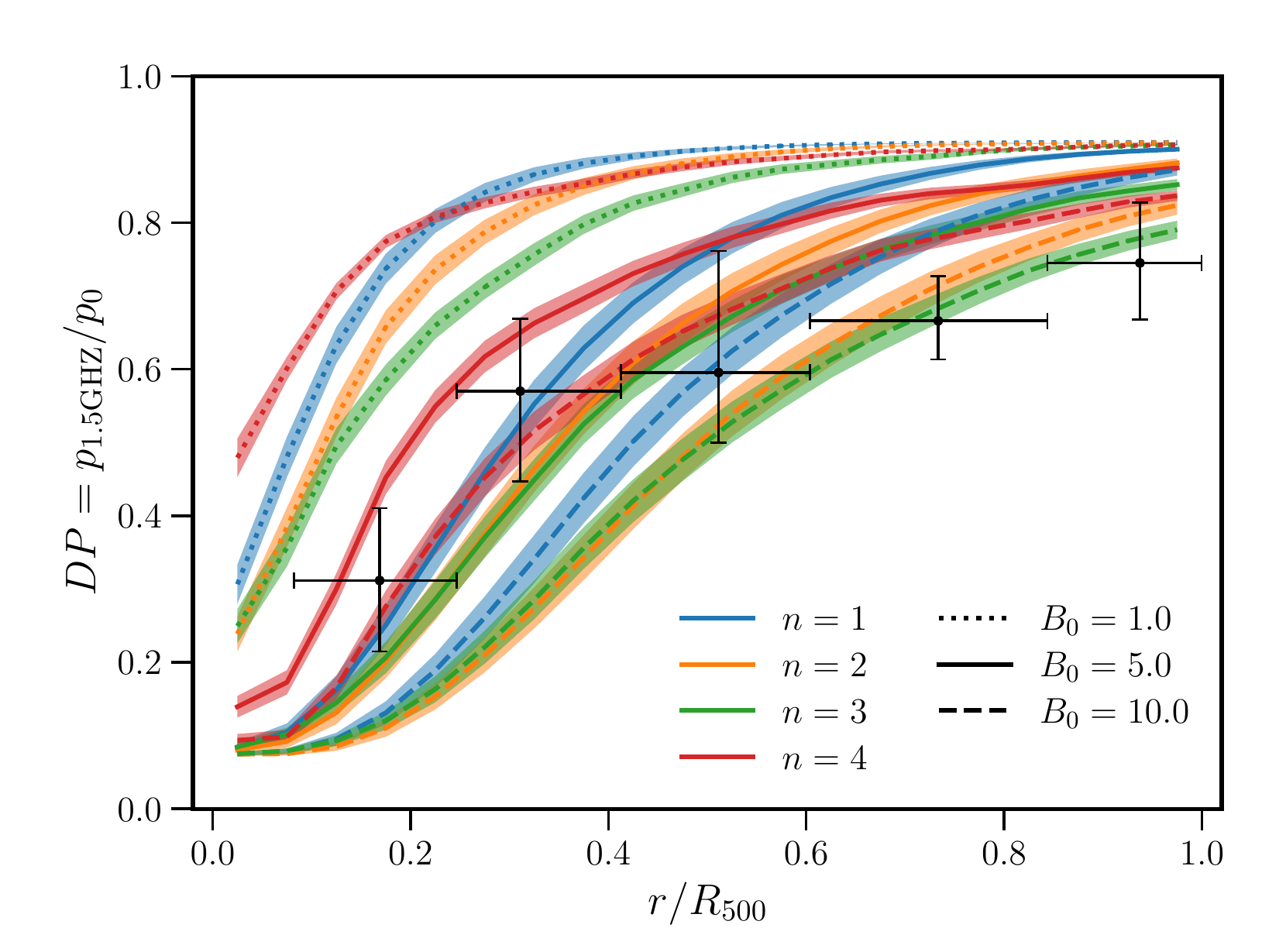}
    \caption{Comparison of the median observed depolarisation ratio in bins of projected radius, denoted by the horizontal lines, with the modelled depolarisation ratio profile for models with different magnetic field strengths (in $\mu$G) with $\eta=0.5$. The model uncertainty interval indicates the standard error on the mean of the simulated profiles.}
    \label{fig:modelall}
\end{figure}

The observed depolarisation trend was re-calculated in five equal-width bins between $0-1R_{500}$ using only sources detected in clusters that are part of the modelling, to make a fair comparison. The results of the comparison of the data with the modelled profiles are shown in Figure \ref{fig:modelall}. Because of the degeneracy between $n$ and $B_0$ (shown in Section \ref{sec:effectofdensityprofiles}) and the fact that we are averaging over many individual clusters with different electron density profiles, there is a large overlap between the different models. Still, it is clear that $B_0=1\,\mu$G does not fit the data for all values of $n$. For values of $n$ between $1$ and $4$, the best fitting average central magnetic field strength is between $5-10,\mu$G, but due to the degeneracy it is not possible to distinguish between these models. 

\subsubsection{Dynamical state}
Section \ref{sec:datadynstate} showed that NCC clusters appear to cause significantly more depolarisation than CC clusters outside 0.2$R_\mathrm{500}$. To investigate in more detail to what degree this is caused by a difference in the magnetic field properties, we average the CC and NCC clusters separately. This allows us to quantify the effect of the different electron density profiles of the two cluster samples. If the thermal gas profiles are the only cause of the discrepancy in depolarisation between NCC and CC, then the same magnetic field parameters would fit both samples. 

\begin{figure}
    \centering
    \includegraphics[width=1.0\columnwidth]{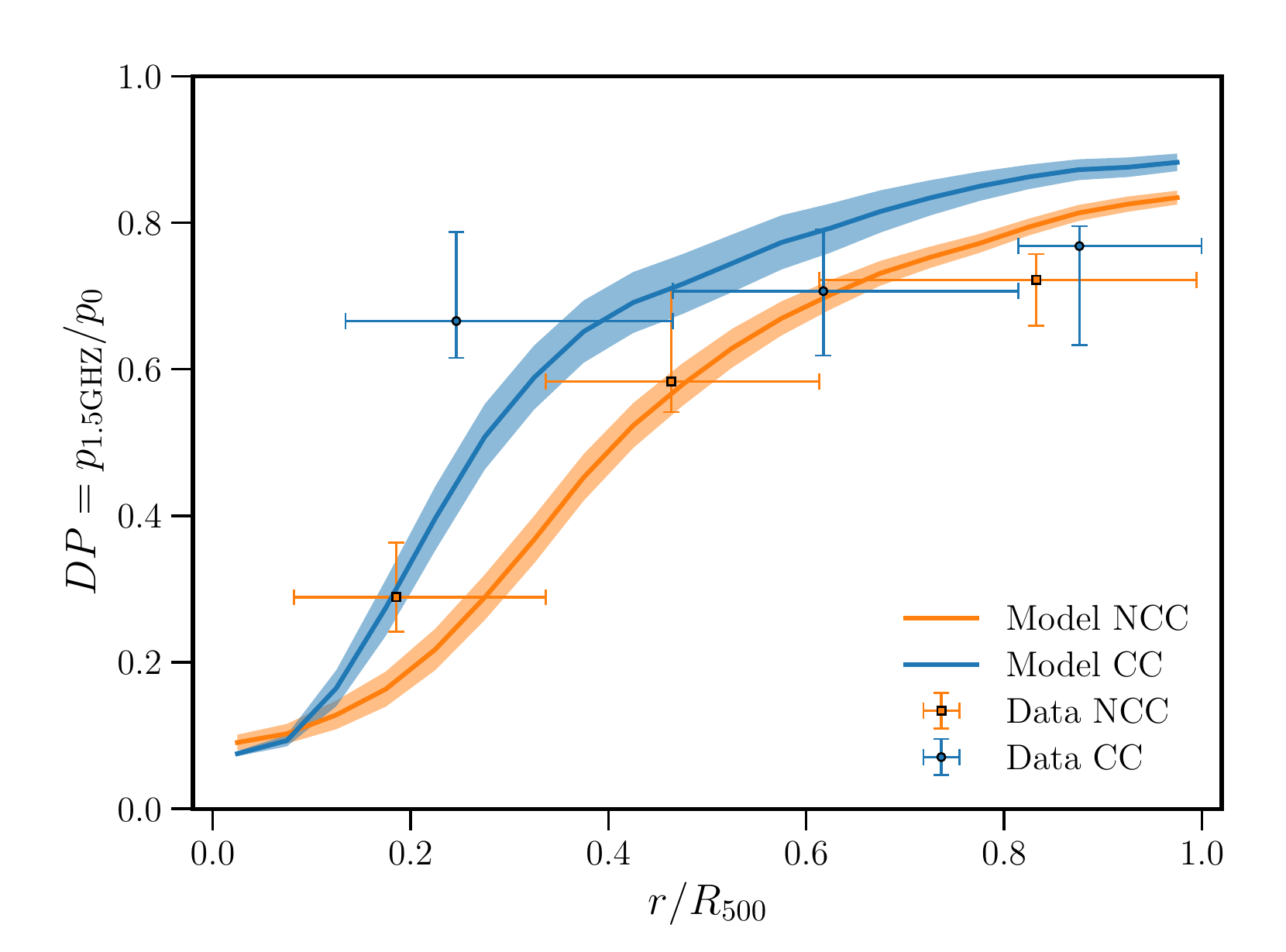}
    \caption{Comparison of the median observed depolarisation ratio with the modelled depolarisation ratio profile separately for the cool-core (CC) and non-cool-core (NCC) cluster sample. The modelled magnetic field parameters are $B_0=5 \mu$G, $\eta=0.5$ and $n=3$ for both samples. The model uncertainty interval indicates the standard error on the mean of the simulated profiles.}
    \label{fig:dynstate_model}
\end{figure}

Figure \ref{fig:dynstate_model} shows that we indeed expect more depolarisation outside the core region from NCC clusters than from CC clusters when they have the same magnetic field properties. This can be understood from the assumption that was made in Eq. \ref{eq:Bfieldprofile}, where the magnetic field energy density was assumed to follow the thermal gas density, normalised by the central electron density of the cluster. Because CC clusters generally have denser cores than NCC clusters, the magnetic field strength a few hundred kiloparsec away from the central cooling core declines faster than in NCC systems, where the denominator of Eq. \ref{eq:Bfieldprofile} is smaller. Indeed the models also show that the amount of depolarisation increases more steeply towards the centre of cooling cores than in non-cool cores, which is in line with the observations shown in the right panel of Figure \ref{fig:p_dynstate}.

\subsubsection{Mass and redshift}
Due to the low number of sources detected in the high-redshift sample, it was not possible to detect differences as a function of mass or redshift. Similar to the previous section, we can investigate to what extent we would expect a difference simply from the fact that we are probing a larger physical region at high redshift. However, in this case, the number of polarised sources detected in high-redshift clusters was already low due to the smaller angular size of the clusters and is even lower for the sample of clusters for which we also have density profiles available. 

\begin{figure}
    \centering
    \includegraphics[width=1.0\columnwidth]{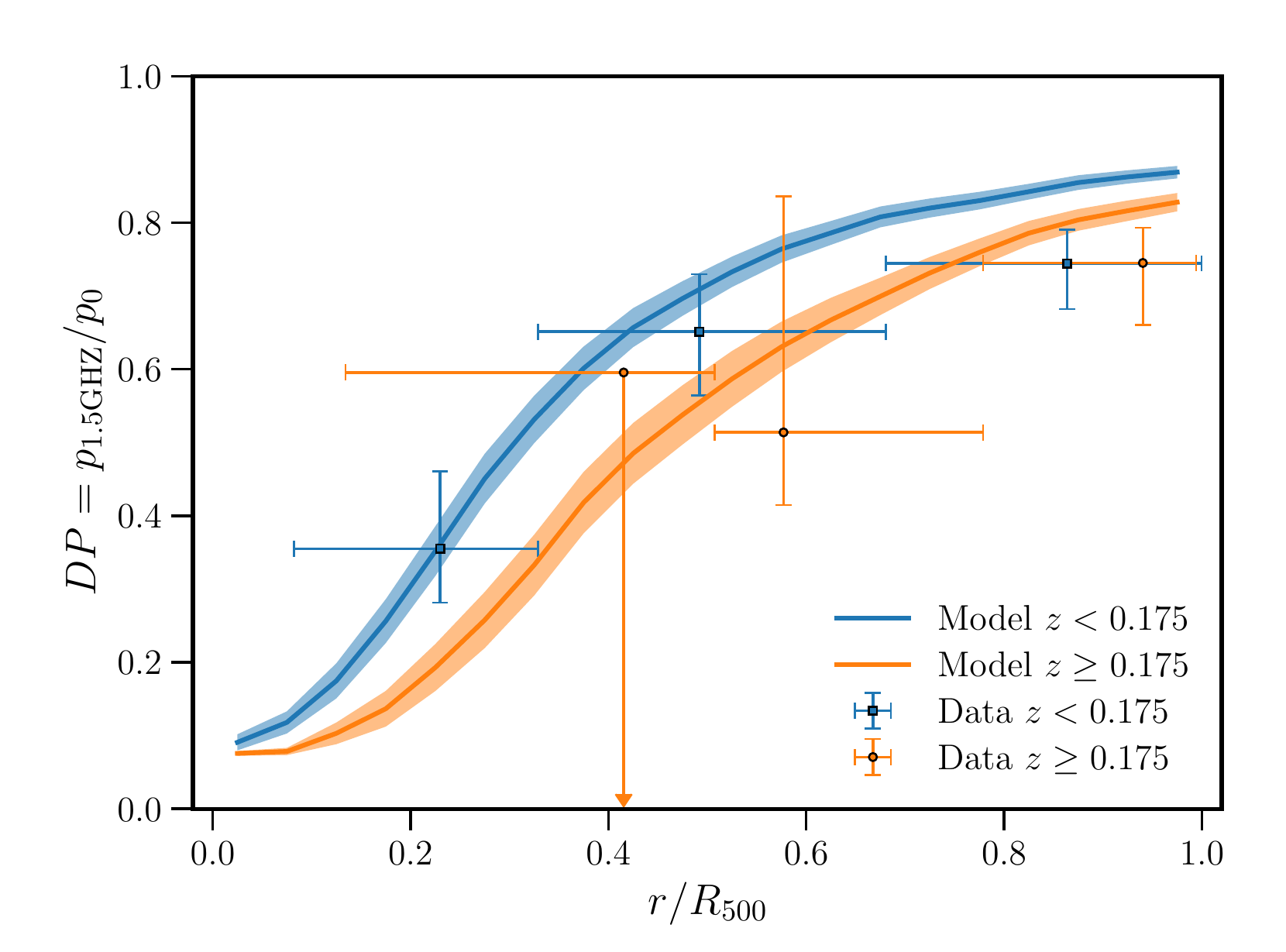}
    \caption{Comparison of the median observed depolarisation ratio with the modelled depolarisation ratio profile separately for the low- and high-redshift cluster sample. The modelled magnetic field parameters are $B_0=5 \mu$G, $\eta=0.5$ and $n=3$ for both samples. The model uncertainty interval indicates the standard error on the mean of the simulated profiles.}
    \label{fig:modeL_redshift_mass}
\end{figure}

Within $1.0R_{500}$, we have detected only 26 polarised sources in the high-redshift cluster sample, and 132 in the low-redshift sample with density profiles available. This causes large uncertainties for the high-redshift sample, particularly closest to the cluster centre. Figure \ref{fig:modeL_redshift_mass} shows that, for similar magnetic field parameters, we would expect slightly more depolarisation from the high-redshift sample than the low-redshift sample, although again this effect is not strong enough to be observed in our data. We thus do not find evidence for a difference between the magnetic field properties of the high-redshift, high-mass and low-redshift, low-mass sample of clusters. 

\section{Discussion}\label{sec:discussion}
We investigated the magnetic field properties in a sample of galaxy clusters through the effect of beam depolarisation. We confirm the hint of the depolarisation trend with cluster projected radius seen in previous studies \citep[][]{Bonafede2011,2020A&A...638A..48S} with a highly statistically significant result, as shown in Figure \ref{fig:KMestimate_all}. In this section, we discuss the implications of the results and the possible limitations of this study. 

\subsection{Cluster members versus background sources}
One of the main questions that this paper addressed is whether there is a difference between using cluster radio galaxies and background radio sources to probe the magnetic fields in galaxy clusters. One could expect such a difference because cluster radio galaxies might locally reshape the magnetic field and density structure, causing a bias in the RM and amount of depolarisation. Interactions with surrounding gas have been suggested to affect observed RM distributions in various powerful radio sources \citep[e.g. 3C75, 3C465, 3C270 and 3C353;][]{2003ApJ...588..143R,2011MNRAS.413.2525G,2012MNRAS.423.1335G}. However, this has not yet been shown in a statistical study.

Figure \ref{fig:KMestimate_redshift} and the log-rank comparison of the survival curves shown in Figure \ref{fig:survival_sourcez} demonstrated that, although near the cluster centres (i.e. $r<0.5R_\mathrm{500}$) it is difficult to have similar sampling of cluster members and background sources, at radii where we have similar sampling (i.e. $r>0.5R_\mathrm{500}$) the depolarisation of cluster members and background radio sources is similar.

In Section \ref{sec:model_background} we modelled the difference between the amount of depolarisation expected for cluster members and background sources based on the different locations of the Faraday screens. We showed that this difference is minimal, and this difference is in line with the observed depolarisation trend for background sources and cluster members. This implies that there is no significant difference between using the depolarisation properties of cluster members or background sources as a probe of the cluster magnetic field. 

Our results are in line with the findings by \citet[][]{Bonafede2011} that used source angular size as a proxy of cluster membership and by \citet[][]{2003ApJ...597..870E} that found that the biases from cluster members are not statistically significant. We note the caveat that the central region is still not well constrained with background sources.

When splitting the sample into NCC and CC clusters, there does seem to be a hint that there is a difference between background sources and cluster members near the cluster centres (Figure \ref{fig:p_dynstate_sourcez}), although only a few sources were detected near the central regions in these splits. A possible explanation for this is that there might be a pronounced effect on the cluster ICM from a select number of powerful cluster radio galaxies, which is averaged out when using a larger sample of sources. This means that when only a few cluster members are used to probe the magnetic field strength, the results may still be biased.

We thus did not find any strong differences between the depolarisation of cluster members and background sources in the full sample. However, larger samples might be able to pick up more subtle effects.

\subsection{Magnetic field parameters}
The average magnetic field properties of the cluster sample were explored by combining the depolarisation of all detected sources, irrespective of their redshift. The results in Figure \ref{fig:modelall} showed that for all power spectrum indices, a central magnetic field strength higher than $B_0=1\,\mu$G is needed to explain the observed depolarisation trend. For models with power-law indices between $n=1$ and $n=4$, an average central magnetic field strength between $5$ and $10\,\mu$G proved to be the best fit, although it was not possible to distinguish between these models. Our results agree with previous radio observations that have shown that clusters have central magnetic field strengths between $1$ and $10\,\mu$G with power spectrum indices between $n=2$ and $n=4$ \citep[][]{2004A&A...424..429M,2006A&A...460..425G,2008A&A...483..699G,2008MNRAS.391..521L,2010A&A...513A..30B,2010A&A...514A..50G,2012A&A...540A..38V,2017A&A...603A.122G,2018Galax...6..142V} and values from magneto-hydrodynamic simulations of clusters \citep[][]{2019MNRAS.486..623D}.

With larger cluster samples or deeper cluster surveys with polarisation information, such as the MeerKAT Galaxy Cluster Legacy Survey \citep[MGCLS; ][]{2022A&A...657A..56K}, it might be possible to group clusters with similar density profiles together. This would reduce the scatter in the modelled depolarisation trend and allow for a more accurate determination of magnetic field parameters. 

\subsection{Cluster properties}
We investigated possible differences in observed depolarisation as a function of various cluster properties, such as whether a cluster is undergoing a merger. The magnetic field might be altered by cluster mergers, during which a massive amount of energy is released (up to $10^{64}$ ergs on a few Gigayear timescales; \citealt{2002ASSL..272....1S}). It is expected that this energy is injected on large spatial scales and released to smaller and smaller scales through turbulent cascades \citep[][]{2018Galax...6..142V, 2019MNRAS.486..623D}.
Observations find central magnetic field strengths of around $\sim 1\,\mu$G and fluctuation scales up to a few hundreds of kpc in merging systems, while relaxed systems show higher central field strengths of around $\sim 10\,\mu$G and much smaller fluctuation scales (less than a few tens of kpc) \citep[][]{2002MNRAS.334..769T,2001ApJ...547L.111C,2018Galax...6..142V}. This implies that, theoretically, we would expect a stronger depolarisation effect in CC clusters. 

We investigated whether there are differences in the depolarisation found in CC and NCC clusters in Figure \ref{fig:p_dynstate}. Surprisingly, we found that NCC clusters show more depolarisation than CC clusters outside the cooling-core region defined as $r>0.2R_\mathrm{500}$ (Figure \ref{fig:survival_dynstate}). When modelling (Fig. \ref{fig:dynstate_model}), it was found that the same central magnetic field strength in CC clusters results in less depolarisation outside the core than in NCC clusters because the magnetic field was assumed to scale with the electron density normalised by the central electron density, which is generally higher in CC clusters than in NCC clusters. Hence, the observed differences could be explained by the same magnetic field parameters in the CC and NCC sample.

To investigate to what extent this result is dependent on the cluster classification method, we also checked different morphological parameters, splitting the sample using the cuspiness and central gas density parameters from \citet[][]{2017ApJ...843...76A}. In these splits, NCC clusters still showed more depolarisation than CC clusters outside the core region. We note that we could not use the entire sample of clusters in this analysis because only 93 out of 124 clusters are observed with Chandra in \citet[][]{2017ApJ...843...76A}. A literature search resulted in the dynamical states for 9 more galaxy clusters, which also does not change the observed depolarisation trends significantly.

Thus, we found no strong evidence that CC clusters have significantly higher magnetic field strengths or smaller fluctuations scales than NCC clusters in the central regions, although the uncertainties were large as shown in Figure \ref{fig:p_dynstate}. However, there is a hint that CC clusters indeed show more depolarisation inside the core region, as also found tentatively in \citet[][]{Bonafede2011}. Unfortunately, the typical size of the cooling cores in galaxy clusters is only 50-100 kpc, which is a region that is still poorly constrained in this study.
The potential difference between the depolarisation in CC clusters and NCC clusters both inside and outside the core region should be investigated further because the sample size is still relatively small when splitting into multiple bins. 

We also checked whether there is a correlation between magnetic field parameters and cluster mass or redshift. A positive correlation between magnetic field strength and cluster mass might be expected, as the observed radio power of giant radio halos is found to correlate with cluster mass, which can be reproduced by turbulent re-acceleration models with a positive scaling of the magnetic field strength with the cluster mass \citep[][]{2006AN....327..557C}. However, the number of polarised sources detected in the high-redshift and high-mass cluster sample was too low to investigate a trend or differences between the low-mass and high-mass samples. A deeper survey such as MGCLS might be able to overcome this problem, although the sample of clusters should be large enough or carefully selected to break the redshift-mass selection bias discussed in Section \ref{sec:massredshiftResults}. 

Finally, we checked whether the presence of a radio halo in merging systems influences the observed depolarisation trend. Models based on the turbulent re-acceleration scenario usually define a radio halo as observable if the break frequency of the radio halo spectrum is above the observing frequency. In these models, the break frequency of the spectrum depends on the magnetic field strength, the cluster mass and the merging state \citep[][]{2006MNRAS.369.1577C}. To investigate whether the magnetic field properties of clusters with a radio halo are different, we split the merging cluster sample based on the detection of a radio halo. No significant differences were observed between the depolarisation of clusters with radio halos and without radio halos, suggesting that they have similar magnetic field parameters. We checked that the cluster mass and redshift distributions are similar, (with a KS-test resulting in $p$-values of $0.2$ and $0.8$, respectively,) so the results are not biased by this. While these clusters are all classified as merging clusters according to the X-ray morphological parameters, a more in-depth study of their merging state might reveal that the clusters without radio halos are only minor merging systems where less turbulence is generated than in major mergers, which would be in line with the findings by \citet[][]{2021A&A...647A..51C}.

\subsection{Possible caveats}\label{sec:caveats}
In this section, we focus on the possible shortcomings of this work. Firstly, a single component external depolarisation model was used to fit the data. In reality, multiple interfering RM components can produce behaviour that is not proportional to $\lambda^2$ and even cause re-polarisation with decreasing frequencies \citep[e.g.][]{2016Galax...4...66P}. Observations of bright polarised sources observed at two different GHz frequencies have found that more than 25\% of sources can show re-polarisation behaviour \citep[][]{2016ApJ...829....5L}. We have found in Section \ref{sec:fracpolmeasurement} that about 25\% (193/819) polarised sources detected in this work are not well-fitted by the single component external depolarisation model. While fitting these sources with more complicated models \citep[e.g.][]{2019MNRAS.483..964B} is beyond the scope of this work, we can briefly investigate how many sources show evidence of re-polarisation by allowing $\sigma_\mathrm{RM}$ to take negative values. This test resulted in 61 sources out of 819 that show a better fit with negative values of $\sigma_\mathrm{RM}$. However, most of these sources do not show strong evidence for re-polarisation and could be fit almost equally well with a value of $\sigma_\mathrm{RM}$ that is around 0, and as such do not change the resulting depolarisation curve significantly. Additionally, the resulting median depolarisation as a function of radius is similar when incorporating the sources with bad fits, which reinforces the fact that we are not biasing the results by omitting these sources. The difference between the number of re-polarising sources found here and in the literature could be caused by the fact we measure the polarisation over the entire bandwidth, where almost always some depolarisation occurs, rather than at only two points in frequency points where multiple components might interfere and show re-polarisation. 

\begin{figure}
    \centering
    \includegraphics[width=1.0\columnwidth]{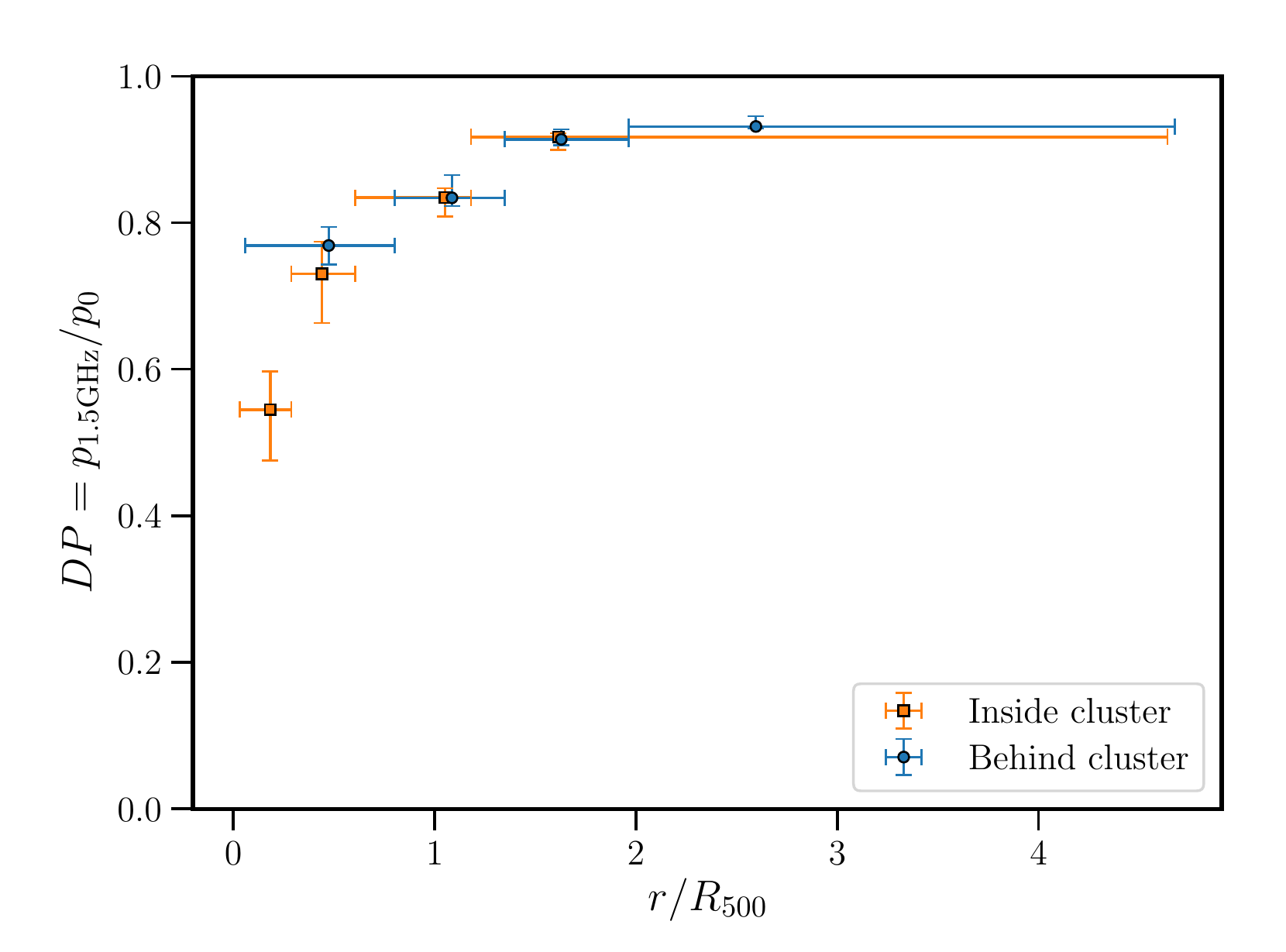}
    \caption{Median depolarisation ratio in different bins of radius without incorporating upper limits from unpolarised sources. Sources inside clusters are shown in green and sources behind clusters are shown in red. Compared to Figure \ref{fig:KMestimate_redshift}, there is significantly less depolarisation in the lowest radius bin.}
    \label{fig:KMestimate_nouplims}
\end{figure}

Secondly, to derive upper limits an assumption on the intrinsic polarisation $p_0$ had to be made. For these sources we assumed $p_0=0.022$, which was the median intrinsic polarisation fraction of sources detected at $r>1.5R_\mathrm{500}$. If this assumption was too high, we are biasing the results through the inclusion of the upper limits by calculating too much depolarisation. However, not including upper limits would cause a bias in the opposite direction by omitting preferentially the most depolarised sources. While it is impossible to determine the intrinsic polarisation value for unpolarised sources, we can show that the depolarisation curve is not dominated by upper limits. Because the upper limits were computed conservatively over the entire extent of the total intensity sources, only 196 out of the 6\,807 unpolarised sources have an upper limit on the 1.5 GHz polarisation fraction that is below the assumed $p_0=0.022$. Thus, in terms of the number of sources, the upper limits are not dominating the results. The depolarisation trend with radius when omitting upper limits entirely is shown in Figure \ref{fig:KMestimate_nouplims}. Cluster members now show less depolarisation in the centre of the cluster because the most constraining upper limits on the depolarisation fraction are found near the cluster centres, where the brightest sources are detected. However, it is clear that even without the inclusion of the upper limits, the depolarisation trend with radius is still clearly detected.  

Thirdly, for the upper limits, we computed a polarised flux threshold as described in Section \ref{sec:fracpolmeasurement}, which was dependent on the varying background noise level. This introduces a bias because all clusters are observed approximately in the pointing centre, so the upper limits are generally higher at larger projected radii. This means we are underestimating the amount of depolarisation more strongly at the edges of the field. Section \ref{sec:appendix_montecarlo} investigates trends with angular distance from the pointing centre in detail and shows that this bias is very small compared to the observed depolarisation trend. 
Because the clusters are all observed near the pointing centre, other trends with projected distance from the cluster centre could also (partially) be due to instrumental or observational trends with angular distance from the pointing centre. These biases are also investigated in detail in Appendix \ref{sec:AppendixA}, where we present that there are indeed sources of bias, but through a Monte Carlo experiment we show that the effect of these biases is minimal compared to the observed depolarisation effect. 

Fourthly, the effect of off-axis polarisation leakage can also mimic depolarisation because of the frequency dependence of the primary beam at a fixed angular distance from the pointing centre. This effect is expected to be at the order of the 1\% level for VLA L-band observations \citep[][]{2017AJ....154...56J}. To correct for this effect, full direction-dependent primary beam corrections need to be made (a-term corrections), which is possible for example with IDG \citep[][]{2018A&A...616A..27V}, but is computationally expensive for large sample sizes and beyond the scope of this paper. However, the leakage effect is in the opposite direction from the observed trend because polarisation leakage effects are stronger near the periphery of the fields, while the observed depolarisation effect is stronger near the centre of the field. Additionally, we can see in Figure \ref{fig:p0} and Figure \ref{fig:A1}c that there is no significant increase in the measured intrinsic polarisation fraction of radio sources as a function of angular separation to the pointing centre. This implies that off-axis leakage effects are negligible for this study.

\begin{figure}[tbh]
    \centering
    \includegraphics[width=\columnwidth]{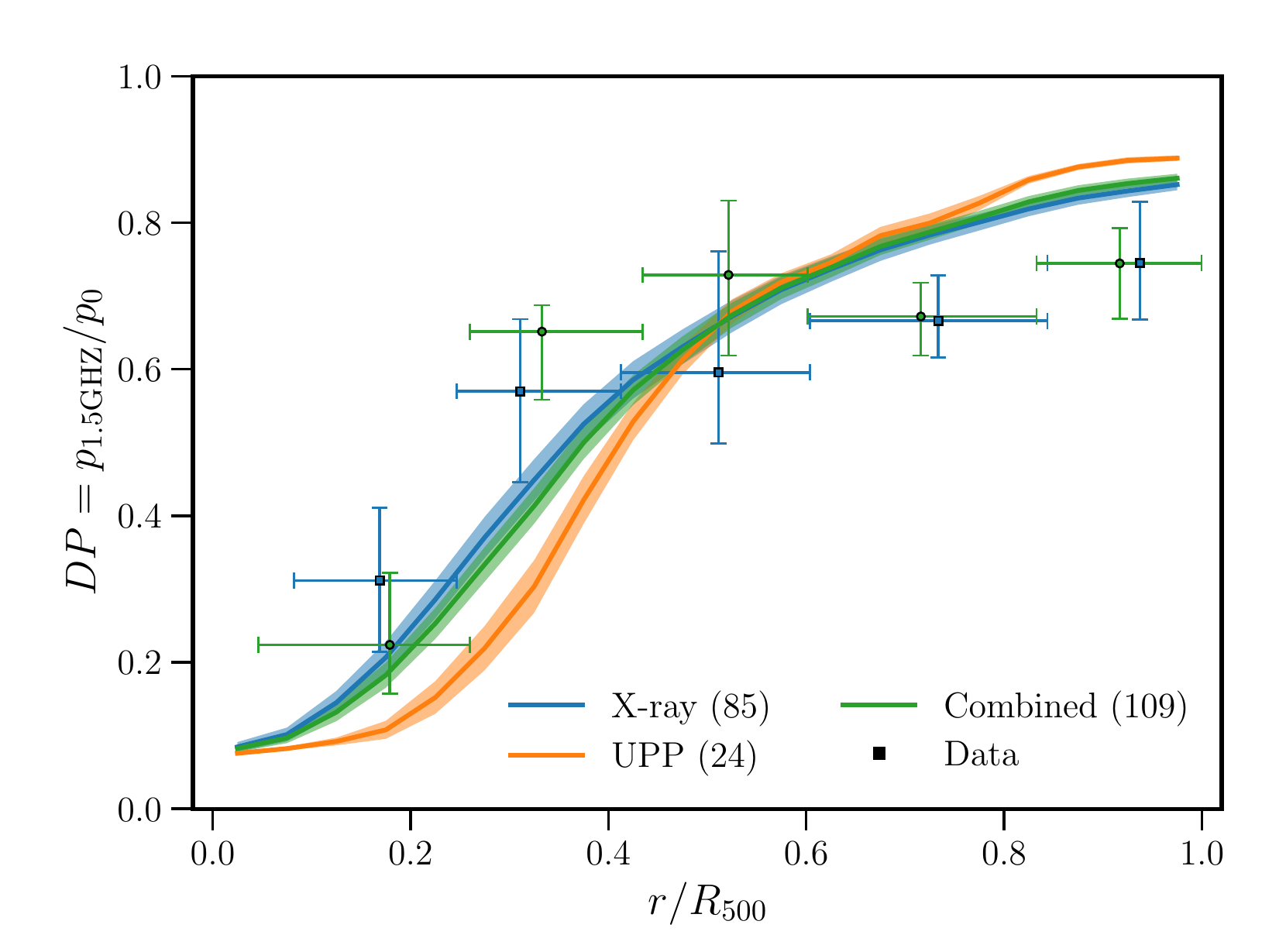}
    \caption{Comparison of the median observed depolarisation ratio with the modelled depolarisation ratio profile using only clusters that have X-ray observations (blue) and all clusters (green) by calculating electron density profiles from a universal pressure profile (orange). The model uncertainty interval indicates the standard error on the mean of the simulated profiles.} 
    \label{fig:UPP}
\end{figure}

Lastly, electron density profiles were not available for all clusters studied in this work, with the 24 new clusters from the PSZ1 and PSZ2 catalogues not having Chandra observations. However, all clusters have been observed through the Sunyaev-Zel'dovich effect, which probes the integrated pressure along the line of sight. It has been shown that the pressure profile of galaxy clusters follows a relatively universal shape, called the universal pressure profile \citep[UPP;][]{2007ApJ...668....1N,2010A&A...517A..92A}. This profile scales in terms of the cluster properties $M_\mathrm{500},R_\mathrm{500}$ and $P_\mathrm{500}$, where $P_\mathrm{500}$ is the characteristic pressure at an overdensity of $500$. With an assumption on the cluster temperature, we can thus calculate a general electron density profile from the UPP to include these 24 clusters in our modelling. To derive the electron density profiles, we use the best-fit parameters for the UPP fit on Planck ESZ clusters from \citet[][]{2013A&A...550A.131P} combined with the best-fit mass-temperature relation on ESZ clusters from \citet[][]{2020ApJ...892..102L}. Including these 24 additional clusters in our modelling results in slightly more average depolarisation as a function of radius, but the final results do not change significantly, as shown in Figure \ref{fig:UPP}. Thus, assuming a universal pressure profile might be useful for future studies of larger, or higher redshift samples of clusters where X-ray observations are not available. 

\section{Conclusion}\label{sec:conclusion}
In this work, we have utilised VLA L-band polarisation observations of a sample of 124 clusters from the Chandra-Planck Legacy Program for Massive Clusters of Galaxies to measure the depolarisation properties of radio sources inside and behind clusters. The main aims of the paper were to use the depolarisation ratio to i) determine the average magnetic field properties in clusters, ii) investigate whether there is a difference between using cluster members and background sources as probes and iii) quantify the dependence of the magnetic field with cluster properties such as mass and dynamical state. We compared the data with modelled depolarisation trends by assuming the magnetic field is a Gaussian random field that follows the thermal electron density profile of the cluster. For the first time in a statistical polarisation study, we took into account the individual electron density profiles of different clusters when modelling the depolarisation ratio. We showed that the depolarisation ratio is a good probe of the magnetic fields in galaxy clusters. Our main results can be summarised as follows:

\begin{enumerate}
    \item We clearly detect a trend of radio sources becoming more depolarised as they move (in projection) towards the cluster centre. This trend can be explained by models with a central magnetic field strength of $5-10\,\mu$G with power-law indices between $n=1$ and $n=4$ and cannot be easily attributed to observational or other systematic biases in the analysis.
    
    \item The individual thermal electron density profiles of the clusters should be taken into account when modelling multiple clusters, as the theoretical depolarisation in separate clusters can be significantly different, even at similar electron column densities. This scatter might be overcome with larger samples when clusters with similar density profiles are grouped together.
    
    \item The relation between the simulated beam depolarisation and fluctuation scale spectral slope is not monotonic. We found that simulated beam depolarisation can increase or decrease with increasing fluctuation scale spectral slope $n$ depending on the size of the observing beam and the location in the cluster.
    
    \item We found no statistically significant difference between the depolarisation properties of background and cluster sources, although background sources were rare to detect near the cluster centre, where cluster members were most often detected. The fact that we see no strong difference implies that the interaction between the radio sources in clusters and their local surrounding medium generally does not strongly influence their polarisation properties. Thus, in statistical studies, both in-cluster and background sources can be used as a probe of the magnetic fields.

    \item Disturbed (non-cool-core) clusters showed more depolarisation in the $0.2-1.0R_\mathrm{500}$ region than cool-core clusters. After modelling, this effect was not strong enough to warrant different magnetic field parameters for disturbed or relaxed systems. While literature suggests that cool-core clusters have stronger magnetic fields inside the core region and should thus show more depolarisation, we did not significantly detect different polarisation fractions inside $0.2R\mathrm{500}$ in cool-core and non-cool-core clusters. However, the uncertainties were large due to the low number of sources, and the most central ($\sim$100 kpc) cooling core region is even more unconstrained in this study and should be investigated further.
    
    \item The observed depolarisation in merging clusters that show a radio halo and merging galaxy clusters that do not show a radio halo is similar. This implies that the presence or absence of a radio halo in merging clusters is likely not dominated by the cluster magnetic field properties.
\end{enumerate}

The biggest limitation of the study of magnetic fields in clusters through depolarisation is currently the number of polarised sources that are detected. With deeper cluster surveys and the advent of the SKA, depolarisation of radio sources will be a promising tool to study cluster magnetic fields.

\paragraph{}

\begin{acknowledgements}
The authors would like to thank the anonymous referee for the comments and suggestions that improved the quality of the manuscript.
      EO and RJvW acknowledge support from the VIDI research programme with project number 639.042.729, which is financed by the Netherlands Organisation for Scientific Research (NWO). 
      LR acknowledges partial support from U.S. National Science Foundation grant AST17-14205 to the University of Minnesota.
AB acknowledges support from the ERC Starting Grant `DRANOEL', number 714245 and from the from Italian MIUR grant FARE `SMS'. 
      KJD acknowledges funding from the European Union’s Horizon 2020 research and innovation programme under the Marie Sk\l{}odowska-Curie grant agreement No. 892117 (HIZRAD). 
Basic research in radio astronomy at the Naval Research Laboratory is supported by 6.1 Base funding.
      This work was performed using the compute resources from the Academic Leiden Interdisciplinary Cluster Environment (ALICE) provided by Leiden University. 
  This paper has made use of observational material taken with an NRAO instrument. The National Radio Astronomy Observatory is a facility of the National Science Foundation operated under cooperative agreement by Associated Universities, Inc. 
The Pan-STARRS1 Surveys (PS1) and the PS1 public science archive have been made possible through contributions by the Institute for Astronomy, the University of Hawaii, the Pan-STARRS Project Office, the Max-Planck Society and its participating institutes, the Max Planck Institute for Astronomy, Heidelberg and the Max Planck Institute for Extraterrestrial Physics, Garching, The Johns Hopkins University, Durham University, the University of Edinburgh, the Queen's University Belfast, the Harvard-Smithsonian Center for Astrophysics, the Las Cumbres Observatory Global Telescope Network Incorporated, the National Central University of Taiwan, the Space Telescope Science Institute, the National Aeronautics and Space Administration under Grant No. NNX08AR22G issued through the Planetary Science Division of the NASA Science Mission Directorate, the National Science Foundation Grant No. AST-1238877, the University of Maryland, Eotvos Lorand University (ELTE), the Los Alamos National Laboratory, and the Gordon and Betty Moore Foundation. 
    The Legacy Surveys consist of three individual and complementary projects: the Dark Energy Camera Legacy Survey (DECaLS; Proposal ID 2014B-0404; PIs: David Schlegel and Arjun Dey), the Beijing-Arizona Sky Survey (BASS; NOAO Prop. ID 2015A-0801; PIs: Zhou Xu and Xiaohui Fan), and the Mayall z-band Legacy Survey (MzLS; Prop. ID 2016A-0453; PI: Arjun Dey). DECaLS, BASS and MzLS together include data obtained, respectively, at the Blanco telescope, Cerro Tololo Inter-American Observatory, NSF’s NOIRLab; the Bok telescope, Steward Observatory, University of Arizona; and the Mayall telescope, Kitt Peak National Observatory, NOIRLab. The Legacy Surveys project is honoured to be permitted to conduct astronomical research on Iolkam Du’ag (Kitt Peak), a mountain with particular significance to the Tohono O’odham Nation. 
SDSS-IV is managed by the Astrophysical Research Consortium for the Participating Institutions of the SDSS Collaboration including the Brazilian Participation Group, the Carnegie Institution for Science, Carnegie Mellon University, Center for Astrophysics | Harvard \& Smithsonian, the Chilean Participation Group, the French Participation Group, Instituto de Astrof\'isica de Canarias, The Johns Hopkins University, Kavli Institute for the Physics and Mathematics of the Universe (IPMU) / University of Tokyo, the Korean Participation Group, Lawrence Berkeley National Laboratory, Leibniz Institut f\"ur Astrophysik Potsdam (AIP),  Max-Planck-Institut f\"ur Astronomie (MPIA Heidelberg), Max-Planck-Institut f\"ur Astrophysik (MPA Garching), Max-Planck-Institut f\"ur Extraterrestrische Physik (MPE), National Astronomical Observatories of China, New Mexico State University, New York University, University of Notre Dame, Observat\'ario Nacional / MCTI, The Ohio State University, Pennsylvania State University, Shanghai Astronomical Observatory, United Kingdom Participation Group, Universidad Nacional Aut\'onoma de M\'exico, University of Arizona, University of Colorado Boulder, University of Oxford, University of Portsmouth, University of Utah, University of Virginia, University of Washington, University of Wisconsin, Vanderbilt University, and Yale University
This research has made use of the NASA/IPAC Extragalactic Database (NED), which is operated by the Jet Propulsion Laboratory, California Institute of Technology, under contract with the National Aeronautics and Space Administration. 
This research has made use of NASA's Astrophysics Data System (ADS).

\end{acknowledgements}

\bibliographystyle{aa}
\bibliography{firstbib.bib}

\begin{thebibliography}{100}
\expandafter\ifx\csname natexlab\endcsname\relax\def\natexlab#1{#1}\fi

\bibitem[{{Ahumada} {et~al.}(2020){Ahumada}, {Prieto}, {Almeida}, {Anders},
  {Anderson}, {Andrews}, {Anguiano}, {Arcodia}, {Armengaud}, {Aubert}, {Avila},
  {Avila-Reese}, {Badenes}, {Balland}, {Barger}, {Barrera-Ballesteros}, {Basu},
  {Bautista}, {Beaton}, {Beers}, {Benavides}, {Bender}, {Bernardi}, {Bershady},
  {Beutler}, {Bidin}, {Bird}, {Bizyaev}, {Blanc}, {Blanton}, {Boquien},
  {Borissova}, {Bovy}, {Brandt}, {Brinkmann}, {Brownstein}, {Bundy}, {Bureau},
  {Burgasser}, {Burtin}, {Cano-D{\'\i}az}, {Capasso}, {Cappellari}, {Carrera},
  {Chabanier}, {Chaplin}, {Chapman}, {Cherinka}, {Chiappini}, {Doohyun Choi},
  {Chojnowski}, {Chung}, {Clerc}, {Coffey}, {Comerford}, {Comparat}, {da
  Costa}, {Cousinou}, {Covey}, {Crane}, {Cunha}, {Ilha}, {Dai}, {Damsted},
  {Darling}, {Davidson}, {Davies}, {Dawson}, {De}, {de la Macorra}, {De Lee},
  {Queiroz}, {Deconto Machado}, {de la Torre}, {Dell'Agli}, {du Mas des
  Bourboux}, {Diamond-Stanic}, {Dillon}, {Donor}, {Drory}, {Duckworth},
  {Dwelly}, {Ebelke}, {Eftekharzadeh}, {Davis Eigenbrot}, {Elsworth},
  {Eracleous}, {Erfanianfar}, {Escoffier}, {Fan}, {Farr},
  {Fern{\'a}ndez-Trincado}, {Feuillet}, {Finoguenov}, {Fofie},
  {Fraser-McKelvie}, {Frinchaboy}, {Fromenteau}, {Fu}, {Galbany}, {Garcia},
  {Garc{\'\i}a-Hern{\'a}ndez}, {Oehmichen}, {Ge}, {Maia}, {Geisler}, {Gelfand},
  {Goddy}, {Gonzalez-Perez}, {Grabowski}, {Green}, {Grier}, {Guo}, {Guy},
  {Harding}, {Hasselquist}, {Hawken}, {Hayes}, {Hearty}, {Hekker}, {Hogg},
  {Holtzman}, {Horta}, {Hou}, {Hsieh}, {Huber}, {Hunt}, {Chitham}, {Imig},
  {Jaber}, {Angel}, {Johnson}, {Jones}, {J{\"o}nsson}, {Jullo}, {Kim},
  {Kinemuchi}, {Kirkpatrick}, {Kite}, {Klaene}, {Kneib}, {Kollmeier}, {Kong},
  {Kounkel}, {Krishnarao}, {Lacerna}, {Lan}, {Lane}, {Law}, {Le Goff}, {Leung},
  {Lewis}, {Li}, {Lian}, {Lin}, {Long}, {Longa-Pe{\~n}a}, {Lundgren}, {Lyke},
  {Ted Mackereth}, {MacLeod}, {Majewski}, {Manchado}, {Maraston}, {Martini},
  {Masseron}, {Masters}, {Mathur}, {McDermid}, {Merloni}, {Merrifield},
  {M{\'e}sz{\'a}ros}, {Miglio}, {Minniti}, {Minsley}, {Miyaji}, {Mohammad},
  {Mosser}, {Mueller}, {Muna}, {Mu{\~n}oz-Guti{\'e}rrez}, {Myers}, {Nadathur},
  {Nair}, {Nandra}, {do Nascimento}, {Nevin}, {Newman}, {Nidever}, {Nitschelm},
  {Noterdaeme}, {O'Connell}, {Olmstead}, {Oravetz}, {Oravetz}, {Osorio},
  {Pace}, {Padilla}, {Palanque-Delabrouille}, {Palicio}, {Pan}, {Pan},
  {Parker}, {Paviot}, {Peirani}, {Ram{\'r}ez}, {Penny}, {Percival},
  {Perez-Fournon}, {P{\'e}rez-R{\`a}fols}, {Petitjean}, {Pieri},
  {Pinsonneault}, {Poovelil}, {Povick}, {Prakash}, {Price-Whelan}, {Raddick},
  {Raichoor}, {Ray}, {Rembold}, {Rezaie}, {Riffel}, {Riffel}, {Rix}, {Robin},
  {Roman-Lopes}, {Rom{\'a}n-Z{\'u}{\~n}iga}, {Rose}, {Ross}, {Rossi},
  {Rowlands}, {Rubin}, {Salvato}, {S{\'a}nchez}, {S{\'a}nchez-Menguiano},
  {S{\'a}nchez-Gallego}, {Sayres}, {Schaefer}, {Schiavon}, {Schimoia},
  {Schlafly}, {Schlegel}, {Schneider}, {Schultheis}, {Schwope}, {Seo},
  {Serenelli}, {Shafieloo}, {Shamsi}, {Shao}, {Shen}, {Shetrone}, {Shirley},
  {Aguirre}, {Simon}, {Skrutskie}, {Slosar}, {Smethurst}, {Sobeck}, {Sodi},
  {Souto}, {Stark}, {Stassun}, {Steinmetz}, {Stello}, {Stermer},
  {Storchi-Bergmann}, {Streblyanska}, {Stringfellow}, {Stutz}, {Su{\'a}rez},
  {Sun}, {Taghizadeh-Popp}, {Talbot}, {Tayar}, {Thakar}, {Theriault}, {Thomas},
  {Thomas}, {Tinker}, {Tojeiro}, {Toledo}, {Tremonti}, {Troup}, {Tuttle},
  {Unda-Sanzana}, {Valentini}, {Vargas-Gonz{\'a}lez}, {Vargas-Maga{\~n}a},
  {V{\'a}zquez-Mata}, {Vivek}, {Wake}, {Wang}, {Weaver}, {Weijmans}, {Wild},
  {Wilson}, {Wilson}, {Wolthuis}, {Wood-Vasey}, {Yan}, {Yang}, {Y{\`e}che},
  {Zamora}, {Zarrouk}, {Zasowski}, {Zhang}, {Zhao}, {Zhao}, {Zheng}, {Zheng},
  {Zhu}, \& {Zou}}]{2020ApJS..249....3A}
{Ahumada}, R., {Prieto}, C.~A., {Almeida}, A., {et~al.} 2020, \apjs, 249, 3

\bibitem[{{Akahori} {et~al.}(2018){Akahori}, {Nakanishi}, {Sofue}, {Fujita},
  {Ichiki}, {Ideguchi}, {Kameya}, {Kudoh}, {Kudoh}, {Machida}, {Miyashita},
  {Ohno}, {Ozawa}, {Takahashi}, {Takizawa}, \&
  {Yamazaki}}]{2018PASJ...70R...2A}
{Akahori}, T., {Nakanishi}, H., {Sofue}, Y., {et~al.} 2018, \pasj, 70, R2

\bibitem[{{Andrade-Santos} {et~al.}(2017){Andrade-Santos}, {Jones}, {Forman},
  {Lovisari}, {Vikhlinin}, {van Weeren}, {Murray}, {Arnaud}, {Pratt},
  {D{\'e}mocl{\`e}s}, {Kraft}, {Mazzotta}, {B{\"o}hringer}, {Chon},
  {Giacintucci}, {Clarke}, {Borgani}, {David}, {Douspis}, {Pointecouteau},
  {Dahle}, {Brown}, {Aghanim}, \& {Rasia}}]{2017ApJ...843...76A}
{Andrade-Santos}, F., {Jones}, C., {Forman}, W.~R., {et~al.} 2017, \apj, 843,
  76

\bibitem[{{Andrade-Santos} {et~al.}(2021){Andrade-Santos}, {Pratt}, {Melin},
  {Arnaud}, {Jones}, {Forman}, {Pointecouteau}, {Bartalucci}, {Vikhlinin},
  {Murray}, {Mazzotta}, {Borgani}, {Lovisari}, {van Weeren}, {Kraft}, {David},
  \& {Giacintucci}}]{2021ApJ...914...58A}
{Andrade-Santos}, F., {Pratt}, G.~W., {Melin}, J.-B., {et~al.} 2021, \apj, 914,
  58

\bibitem[{{Arnaud} {et~al.}(2010){Arnaud}, {Pratt}, {Piffaretti},
  {B{\"o}hringer}, {Croston}, \& {Pointecouteau}}]{2010A&A...517A..92A}
{Arnaud}, M., {Pratt}, G.~W., {Piffaretti}, R., {et~al.} 2010, \aap, 517, A92

\bibitem[{{Arshakian} \& {Beck}(2011)}]{2011MNRAS.418.2336A}
{Arshakian}, T.~G. \& {Beck}, R. 2011, \mnras, 418, 2336

\bibitem[{{Beck} \& {Krause}(2005)}]{2005AN....326..414B}
{Beck}, R. \& {Krause}, M. 2005, Astronomische Nachrichten, 326, 414

\bibitem[{{B{\"o}hringer} {et~al.}(2016){B{\"o}hringer}, {Chon}, \&
  {Kronberg}}]{2016A&A...596A..22B}
{B{\"o}hringer}, H., {Chon}, G., \& {Kronberg}, P.~P. 2016, \aap, 596, A22

\bibitem[{{Bonafede} {et~al.}(2010){Bonafede}, {Feretti}, {Murgia}, {Govoni},
  {Giovannini}, {Dallacasa}, {Dolag}, \& {Taylor}}]{2010A&A...513A..30B}
{Bonafede}, A., {Feretti}, L., {Murgia}, M., {et~al.} 2010, \aap, 513, A30

\bibitem[{{Bonafede} {et~al.}(2011){Bonafede}, {Govoni}, {Feretti}, {Murgia},
  {Giovannini}, \& {Br{\"u}ggen}}]{Bonafede2011}
{Bonafede}, A., {Govoni}, F., {Feretti}, L., {et~al.} 2011, \aap, 530, A24

\bibitem[{{Bonafede} {et~al.}(2013){Bonafede}, {Vazza}, {Br{\"u}ggen},
  {Murgia}, {Govoni}, {Feretti}, {Giovannini}, \&
  {Ogrean}}]{2013MNRAS.433.3208B}
{Bonafede}, A., {Vazza}, F., {Br{\"u}ggen}, M., {et~al.} 2013, \mnras, 433,
  3208

\bibitem[{{Botteon} {et~al.}(2022){Botteon}, {Shimwell}, {Cassano}, {Cuciti},
  {Zhang}, {Bruno}, {Camillini}, {Natale}, {Jones}, {Gastaldello},
  {Simionescu}, {Rossetti}, {Akamatsu}, {van Weeren}, {Brunetti},
  {Br{\"u}ggen}, {Groeneveld}, {Hoang}, {Hardcastle}, {Ignesti}, {Di Gennaro},
  {Bonafede}, {Drabent}, {R{\"o}ttgering}, {Hoeft}, \& {de
  Gasperin}}]{2022A&A...660A..78B}
{Botteon}, A., {Shimwell}, T.~W., {Cassano}, R., {et~al.} 2022, \aap, 660, A78

\bibitem[{{Brentjens} \& {de Bruyn}(2005)}]{BrentjensBruyn}
{Brentjens}, M.~A. \& {de Bruyn}, A.~G. 2005, \aap, 441, 1217

\bibitem[{{Brown} {et~al.}(2019){Brown}, {Bergerud}, {Costa}, {Gaensler},
  {Isbell}, {LaRocca}, {Norris}, {Purcell}, {Rudnick}, \&
  {Sun}}]{2019MNRAS.483..964B}
{Brown}, S., {Bergerud}, B., {Costa}, A., {et~al.} 2019, \mnras, 483, 964

\bibitem[{{Burn}(1966)}]{1966MNRAS.133...67B}
{Burn}, B.~J. 1966, \mnras, 133, 67

\bibitem[{{Burns} {et~al.}(2008){Burns}, {Hallman}, {Gantner}, {Motl}, \&
  {Norman}}]{2008ApJ...675.1125B}
{Burns}, J.~O., {Hallman}, E.~J., {Gantner}, B., {Motl}, P.~M., \& {Norman},
  M.~L. 2008, \apj, 675, 1125

\bibitem[{{Carilli} \& {Taylor}(2002)}]{2002ARA&A..40..319C}
{Carilli}, C.~L. \& {Taylor}, G.~B. 2002, \araa, 40, 319

\bibitem[{{Cassano} {et~al.}(2006{\natexlab{a}}){Cassano}, {Brunetti}, \&
  {Setti}}]{2006AN....327..557C}
{Cassano}, R., {Brunetti}, G., \& {Setti}, G. 2006{\natexlab{a}}, Astronomische
  Nachrichten, 327, 557

\bibitem[{{Cassano} {et~al.}(2006{\natexlab{b}}){Cassano}, {Brunetti}, \&
  {Setti}}]{2006MNRAS.369.1577C}
{Cassano}, R., {Brunetti}, G., \& {Setti}, G. 2006{\natexlab{b}}, \mnras, 369,
  1577

\bibitem[{{Chambers} {et~al.}(2016){Chambers}, {Magnier}, {Metcalfe},
  {Flewelling}, {Huber}, {Waters}, {Denneau}, {Draper}, {Farrow}, {Finkbeiner},
  {Holmberg}, {Koppenhoefer}, {Price}, {Rest}, {Saglia}, {Schlafly}, {Smartt},
  {Sweeney}, {Wainscoat}, {Burgett}, {Chastel}, {Grav}, {Heasley}, {Hodapp},
  {Jedicke}, {Kaiser}, {Kudritzki}, {Luppino}, {Lupton}, {Monet}, {Morgan},
  {Onaka}, {Shiao}, {Stubbs}, {Tonry}, {White}, {Ba{\~n}ados}, {Bell},
  {Bender}, {Bernard}, {Boegner}, {Boffi}, {Botticella}, {Calamida},
  {Casertano}, {Chen}, {Chen}, {Cole}, {Deacon}, {Frenk}, {Fitzsimmons},
  {Gezari}, {Gibbs}, {Goessl}, {Goggia}, {Gourgue}, {Goldman}, {Grant},
  {Grebel}, {Hambly}, {Hasinger}, {Heavens}, {Heckman}, {Henderson}, {Henning},
  {Holman}, {Hopp}, {Ip}, {Isani}, {Jackson}, {Keyes}, {Koekemoer}, {Kotak},
  {Le}, {Liska}, {Long}, {Lucey}, {Liu}, {Martin}, {Masci}, {McLean}, {Mindel},
  {Misra}, {Morganson}, {Murphy}, {Obaika}, {Narayan}, {Nieto-Santisteban},
  {Norberg}, {Peacock}, {Pier}, {Postman}, {Primak}, {Rae}, {Rai}, {Riess},
  {Riffeser}, {Rix}, {R{\"o}ser}, {Russel}, {Rutz}, {Schilbach}, {Schultz},
  {Scolnic}, {Strolger}, {Szalay}, {Seitz}, {Small}, {Smith}, {Soderblom},
  {Taylor}, {Thomson}, {Taylor}, {Thakar}, {Thiel}, {Thilker}, {Unger},
  {Urata}, {Valenti}, {Wagner}, {Walder}, {Walter}, {Watters}, {Werner},
  {Wood-Vasey}, \& {Wyse}}]{2016arXiv161205560C}
{Chambers}, K.~C., {Magnier}, E.~A., {Metcalfe}, N., {et~al.} 2016, arXiv
  e-prints, arXiv:1612.05560

\bibitem[{{Clarke} {et~al.}(2001){Clarke}, {Kronberg}, \&
  {B{\"o}hringer}}]{2001ApJ...547L.111C}
{Clarke}, T.~E., {Kronberg}, P.~P., \& {B{\"o}hringer}, H. 2001, \apjl, 547,
  L111

\bibitem[{{Cuciti} {et~al.}(2018){Cuciti}, {Brunetti}, {van Weeren},
  {Bonafede}, {Dallacasa}, {Cassano}, {Venturi}, \&
  {Kale}}]{2018A&A...609A..61C}
{Cuciti}, V., {Brunetti}, G., {van Weeren}, R., {et~al.} 2018, \aap, 609, A61

\bibitem[{{Cuciti} {et~al.}(2021){Cuciti}, {Cassano}, {Brunetti}, {Dallacasa},
  {de Gasperin}, {Ettori}, {Giacintucci}, {Kale}, {Pratt}, {van Weeren}, \&
  {Venturi}}]{2021A&A...647A..51C}
{Cuciti}, V., {Cassano}, R., {Brunetti}, G., {et~al.} 2021, \aap, 647, A51

\bibitem[{{Dahlen} {et~al.}(2013){Dahlen}, {Mobasher}, {Faber}, {Ferguson},
  {Barro}, {Finkelstein}, {Finlator}, {Fontana}, {Gruetzbauch}, {Johnson},
  {Pforr}, {Salvato}, {Wiklind}, {Wuyts}, {Acquaviva}, {Dickinson}, {Guo},
  {Huang}, {Huang}, {Newman}, {Bell}, {Conselice}, {Galametz}, {Gawiser},
  {Giavalisco}, {Grogin}, {Hathi}, {Kocevski}, {Koekemoer}, {Koo}, {Lee},
  {McGrath}, {Papovich}, {Peth}, {Ryan}, {Somerville}, {Weiner}, \&
  {Wilson}}]{2013ApJ...775...93D}
{Dahlen}, T., {Mobasher}, B., {Faber}, S.~M., {et~al.} 2013, \apj, 775, 93

\bibitem[{Davidson-Pilon(2019)}]{DavidsonPilon2019}
Davidson-Pilon, C. 2019, Journal of Open Source Software, 4, 1317

\bibitem[{{de Gasperin} {et~al.}(2022){de Gasperin}, {Rudnick}, {Finoguenov},
  {Wittor}, {Akamatsu}, {Br{\"u}ggen}, {Chibueze}, {Clarke}, {Cotton},
  {Cuciti}, {Dom{\'\i}nguez-Fern{\'a}ndez}, {Knowles}, {O'Sullivan}, \&
  {Sebokolodi}}]{2022A&A...659A.146D}
{de Gasperin}, F., {Rudnick}, L., {Finoguenov}, A., {et~al.} 2022, \aap, 659,
  A146

\bibitem[{{Dey} {et~al.}(2019){Dey}, {Schlegel}, {Lang}, {Blum}, {Burleigh},
  {Fan}, {Findlay}, {Finkbeiner}, {Herrera}, {Juneau}, {Landriau}, {Levi},
  {McGreer}, {Meisner}, {Myers}, {Moustakas}, {Nugent}, {Patej}, {Schlafly},
  {Walker}, {Valdes}, {Weaver}, {Y{\`e}che}, {Zou}, {Zhou}, {Abareshi},
  {Abbott}, {Abolfathi}, {Aguilera}, {Alam}, {Allen}, {Alvarez}, {Annis},
  {Ansarinejad}, {Aubert}, {Beechert}, {Bell}, {BenZvi}, {Beutler}, {Bielby},
  {Bolton}, {Brice{\~n}o}, {Buckley-Geer}, {Butler}, {Calamida}, {Carlberg},
  {Carter}, {Casas}, {Castander}, {Choi}, {Comparat}, {Cukanovaite}, {Delubac},
  {DeVries}, {Dey}, {Dhungana}, {Dickinson}, {Ding}, {Donaldson}, {Duan},
  {Duckworth}, {Eftekharzadeh}, {Eisenstein}, {Etourneau}, {Fagrelius},
  {Farihi}, {Fitzpatrick}, {Font-Ribera}, {Fulmer}, {G{\"a}nsicke},
  {Gaztanaga}, {George}, {Gerdes}, {Gontcho}, {Gorgoni}, {Green}, {Guy},
  {Harmer}, {Hernandez}, {Honscheid}, {Huang}, {James}, {Jannuzi}, {Jiang},
  {Joyce}, {Karcher}, {Karkar}, {Kehoe}, {Kneib}, {Kueter-Young}, {Lan},
  {Lauer}, {Le Guillou}, {Le Van Suu}, {Lee}, {Lesser}, {Perreault Levasseur},
  {Li}, {Mann}, {Marshall}, {Mart{\'\i}nez-V{\'a}zquez}, {Martini}, {du Mas des
  Bourboux}, {McManus}, {Meier}, {M{\'e}nard}, {Metcalfe},
  {Mu{\~n}oz-Guti{\'e}rrez}, {Najita}, {Napier}, {Narayan}, {Newman}, {Nie},
  {Nord}, {Norman}, {Olsen}, {Paat}, {Palanque-Delabrouille}, {Peng},
  {Poppett}, {Poremba}, {Prakash}, {Rabinowitz}, {Raichoor}, {Rezaie},
  {Robertson}, {Roe}, {Ross}, {Ross}, {Rudnick}, {Safonova}, {Saha},
  {S{\'a}nchez}, {Savary}, {Schweiker}, {Scott}, {Seo}, {Shan}, {Silva},
  {Slepian}, {Soto}, {Sprayberry}, {Staten}, {Stillman}, {Stupak}, {Summers},
  {Sien Tie}, {Tirado}, {Vargas-Maga{\~n}a}, {Vivas}, {Wechsler}, {Williams},
  {Yang}, {Yang}, {Yapici}, {Zaritsky}, {Zenteno}, {Zhang}, {Zhang}, {Zhou}, \&
  {Zhou}}]{2019AJ....157..168D}
{Dey}, A., {Schlegel}, D.~J., {Lang}, D., {et~al.} 2019, \aj, 157, 168

\bibitem[{{Di Gennaro} {et~al.}(2021{\natexlab{a}}){Di Gennaro}, {van Weeren},
  {Brunetti}, {Cassano}, {Br{\"u}ggen}, {Hoeft}, {Shimwell}, {R{\"o}ttgering},
  {Bonafede}, {Botteon}, {Cuciti}, {Dallacasa}, {de Gasperin},
  {Dom{\'\i}nguez-Fern{\'a}ndez}, {En{\ss}lin}, {Gastaldello}, {Mandal},
  {Rossetti}, \& {Simionescu}}]{2021NatAs...5..268D}
{Di Gennaro}, G., {van Weeren}, R.~J., {Brunetti}, G., {et~al.}
  2021{\natexlab{a}}, Nature Astronomy, 5, 268

\bibitem[{{Di Gennaro} {et~al.}(2021{\natexlab{b}}){Di Gennaro}, {van Weeren},
  {Rudnick}, {Hoeft}, {Br{\"u}ggen}, {Ryu}, {R{\"o}ttgering}, {Forman},
  {Stroe}, {Shimwell}, {Kraft}, {Jones}, \& {Hoang}}]{2021ApJ...911....3D}
{Di Gennaro}, G., {van Weeren}, R.~J., {Rudnick}, L., {et~al.}
  2021{\natexlab{b}}, \apj, 911, 3

\bibitem[{{Dolag} {et~al.}(2008){Dolag}, {Bykov}, \&
  {Diaferio}}]{2008SSRv..134..311D}
{Dolag}, K., {Bykov}, A.~M., \& {Diaferio}, A. 2008, \ssr, 134, 311

\bibitem[{{Dom{\'\i}nguez-Fern{\'a}ndez}
  {et~al.}(2019){Dom{\'\i}nguez-Fern{\'a}ndez}, {Vazza}, {Br{\"u}ggen}, \&
  {Brunetti}}]{2019MNRAS.486..623D}
{Dom{\'\i}nguez-Fern{\'a}ndez}, P., {Vazza}, F., {Br{\"u}ggen}, M., \&
  {Brunetti}, G. 2019, \mnras, 486, 623

\bibitem[{{Donnert} {et~al.}(2018){Donnert}, {Vazza}, {Br{\"u}ggen}, \&
  {ZuHone}}]{2018SSRv..214..122D}
{Donnert}, J., {Vazza}, F., {Br{\"u}ggen}, M., \& {ZuHone}, J. 2018, \ssr, 214,
  122

\bibitem[{{Duncan}(2022)}]{2022MNRAS.512.3662D}
{Duncan}, K.~J. 2022, \mnras, 512, 3662

\bibitem[{{Eckert} {et~al.}(2011){Eckert}, {Molendi}, \&
  {Paltani}}]{2011A&A...526A..79E}
{Eckert}, D., {Molendi}, S., \& {Paltani}, S. 2011, \aap, 526, A79

\bibitem[{{Ensslin} {et~al.}(2003){Ensslin}, {Vogt}, {Clarke}, \&
  {Taylor}}]{2003ApJ...597..870E}
{Ensslin}, T.~A., {Vogt}, C., {Clarke}, T.~E., \& {Taylor}, G.~B. 2003, \apj,
  597, 870

\bibitem[{{Fabian}(1994)}]{1994ARA&A..32..277F}
{Fabian}, A.~C. 1994, \araa, 32, 277

\bibitem[{{Farnes} {et~al.}(2018){Farnes}, {Heald}, {Junklewitz}, {Mulcahy},
  {Haverkorn}, {Van Eck}, {Riseley}, {Brentjens}, {Horellou}, {Vacca}, {Jones},
  {Horneffer}, \& {Paladino}}]{2018MNRAS.474.3280F}
{Farnes}, J.~S., {Heald}, G., {Junklewitz}, H., {et~al.} 2018, \mnras, 474,
  3280

\bibitem[{{Feigelson} \& {Nelson}(1985)}]{Feigelson1985}
{Feigelson}, E.~D. \& {Nelson}, P.~I. 1985, \apj, 293, 192

\bibitem[{{Feretti} {et~al.}(1998){Feretti}, {Giovannini}, {Klein}, {Mack},
  {Sijbring}, \& {Zech}}]{1998A&A...331..475F}
{Feretti}, L., {Giovannini}, G., {Klein}, U., {et~al.} 1998, \aap, 331, 475

\bibitem[{{Ferrari} {et~al.}(2008){Ferrari}, {Govoni}, {Schindler}, {Bykov}, \&
  {Rephaeli}}]{2008SSRv..134...93F}
{Ferrari}, C., {Govoni}, F., {Schindler}, S., {Bykov}, A.~M., \& {Rephaeli}, Y.
  2008, \ssr, 134, 93

\bibitem[{{Govoni} \& {Feretti}(2004)}]{2004IJMPD..13.1549G}
{Govoni}, F. \& {Feretti}, L. 2004, International Journal of Modern Physics D,
  13, 1549

\bibitem[{{Govoni} {et~al.}(2006){Govoni}, {Murgia}, {Feretti}, {Giovannini},
  {Dolag}, \& {Taylor}}]{2006A&A...460..425G}
{Govoni}, F., {Murgia}, M., {Feretti}, L., {et~al.} 2006, \aap, 460, 425

\bibitem[{{Govoni} {et~al.}(2017){Govoni}, {Murgia}, {Vacca}, {Loi}, {Girardi},
  {Gastaldello}, {Giovannini}, {Feretti}, {Paladino}, {Carretti}, {Concu},
  {Melis}, {Poppi}, {Valente}, {Bernardi}, {Bonafede}, {Boschin}, {Brienza},
  {Clarke}, {Colafrancesco}, {de Gasperin}, {Eckert}, {En{\ss}lin}, {Ferrari},
  {Gregorini}, {Johnston-Hollitt}, {Junklewitz}, {Orr{\`u}}, {Parma}, {Perley},
  {Rossetti}, {B Taylor}, \& {Vazza}}]{2017A&A...603A.122G}
{Govoni}, F., {Murgia}, M., {Vacca}, V., {et~al.} 2017, \aap, 603, A122

\bibitem[{{Guidetti} {et~al.}(2011){Guidetti}, {Laing}, {Bridle}, {Parma}, \&
  {Gregorini}}]{2011MNRAS.413.2525G}
{Guidetti}, D., {Laing}, R.~A., {Bridle}, A.~H., {Parma}, P., \& {Gregorini},
  L. 2011, \mnras, 413, 2525

\bibitem[{{Guidetti} {et~al.}(2012){Guidetti}, {Laing}, {Croston}, {Bridle}, \&
  {Parma}}]{2012MNRAS.423.1335G}
{Guidetti}, D., {Laing}, R.~A., {Croston}, J.~H., {Bridle}, A.~H., \& {Parma},
  P. 2012, \mnras, 423, 1335

\bibitem[{{Guidetti} {et~al.}(2010){Guidetti}, {Laing}, {Murgia}, {Govoni},
  {Gregorini}, \& {Parma}}]{2010A&A...514A..50G}
{Guidetti}, D., {Laing}, R.~A., {Murgia}, M., {et~al.} 2010, \aap, 514, A50

\bibitem[{{Guidetti} {et~al.}(2008){Guidetti}, {Murgia}, {Govoni}, {Parma},
  {Gregorini}, {de Ruiter}, {Cameron}, \& {Fanti}}]{2008A&A...483..699G}
{Guidetti}, D., {Murgia}, M., {Govoni}, F., {et~al.} 2008, \aap, 483, 699

\bibitem[{{Hardcastle} {et~al.}(2019){Hardcastle}, {Williams}, {Best},
  {Croston}, {Duncan}, {R{\"o}ttgering}, {Sabater}, {Shimwell}, {Tasse},
  {Callingham}, {Cochrane}, {de Gasperin}, {G{\"u}rkan}, {Jarvis}, {Mahatma},
  {Miley}, {Mingo}, {Mooney}, {Morabito}, {O'Sullivan}, {Prandoni},
  {Shulevski}, \& {Smith}}]{2019A&A...622A..12H}
{Hardcastle}, M.~J., {Williams}, W.~L., {Best}, P.~N., {et~al.} 2019, \aap,
  622, A12

\bibitem[{{Jagannathan} {et~al.}(2017){Jagannathan}, {Bhatnagar}, {Rau}, \&
  {Taylor}}]{2017AJ....154...56J}
{Jagannathan}, P., {Bhatnagar}, S., {Rau}, U., \& {Taylor}, A.~R. 2017, \aj,
  154, 56

\bibitem[{{Johnson} {et~al.}(2020){Johnson}, {Rudnick}, {Jones}, {Mendygral},
  \& {Dolag}}]{2020ApJ...888..101J}
{Johnson}, A.~R., {Rudnick}, L., {Jones}, T.~W., {Mendygral}, P.~J., \&
  {Dolag}, K. 2020, \apj, 888, 101

\bibitem[{{Kang} {et~al.}(2020){Kang}, {Wang}, \& {Kang}}]{2020ApJ...901..111K}
{Kang}, J., {Wang}, J., \& {Kang}, W. 2020, \apj, 901, 111

\bibitem[{{Knowles} {et~al.}(2022){Knowles}, {Cotton}, {Rudnick}, {Camilo},
  {Goedhart}, {Deane}, {Ramatsoku}, {Bietenholz}, {Br{\"u}ggen}, {Button},
  {Chen}, {Chibueze}, {Clarke}, {de Gasperin}, {Ianjamasimanana}, {J{\'o}zsa},
  {Hilton}, {Kesebonye}, {Kolokythas}, {Kraan-Korteweg}, {Lawrie}, {Lochner},
  {Loubser}, {Marchegiani}, {Mhlahlo}, {Moodley}, {Murphy}, {Namumba},
  {Oozeer}, {Parekh}, {Pillay}, {Passmoor}, {Ramaila}, {Ranchod},
  {Retana-Montenegro}, {Sebokolodi}, {Sikhosana}, {Smirnov}, {Thorat},
  {Venturi}, {Abbott}, {Adam}, {Adams}, {Aldera}, {Bauermeister}, {Bennett},
  {Bode}, {Botha}, {Botha}, {Brederode}, {Buchner}, {Burger}, {Cheetham}, {de
  Villiers}, {Dikgale-Mahlakoana}, {du Toit}, {Esterhuyse}, {Fadana},
  {Fanaroff}, {Fataar}, {Foley}, {Fourie}, {Frank}, {Gamatham}, {Gatsi},
  {Geyer}, {Gouws}, {Gumede}, {Heywood}, {Hlakola}, {Hokwana}, {Hoosen},
  {Horn}, {Horrell}, {Hugo}, {Isaacson}, {Jonas}, {Jordaan}, {Joubert},
  {Julie}, {Kapp}, {Kasper}, {Kenyon}, {Kotz{\'e}}, {Kotze}, {Kriek}, {Kriel},
  {Krishnan}, {Kusel}, {Legodi}, {Lehmensiek}, {Liebenberg}, {Lord}, {Lunsky},
  {Madisa}, {Magnus}, {Main}, {Makhaba}, {Makhathini}, {Malan}, {Manley},
  {Marais}, {Maree}, {Martens}, {Mauch}, {McAlpine}, {Merry}, {Millenaar},
  {Mokone}, {Monama}, {Mphego}, {New}, {Ngcebetsha}, {Ngoasheng}, {Ockards},
  {Otto}, {Patel}, {Peens-Hough}, {Perkins}, {Ramanujam}, {Ramudzuli},
  {Ratcliffe}, {Renil}, {Robyntjies}, {Rust}, {Salie}, {Sambu}, {Schollar},
  {Schwardt}, {Schwartz}, {Serylak}, {Siebrits}, {Sirothia}, {Slabber},
  {Sofeya}, {Taljaard}, {Tasse}, {Tiplady}, {Toruvanda}, {Twum}, {van Balla},
  {van der Byl}, {van der Merwe}, {van Dyk}, {Van Tonder}, {Van Wyk}, {Venter},
  {Venter}, {Welz}, {Williams}, \& {Xaia}}]{2022A&A...657A..56K}
{Knowles}, K., {Cotton}, W.~D., {Rudnick}, L., {et~al.} 2022, \aap, 657, A56

\bibitem[{{Kron}(1980)}]{1980ApJS...43..305K}
{Kron}, R.~G. 1980, \apjs, 43, 305

\bibitem[{{Laing} {et~al.}(2008){Laing}, {Bridle}, {Parma}, \&
  {Murgia}}]{2008MNRAS.391..521L}
{Laing}, R.~A., {Bridle}, A.~H., {Parma}, P., \& {Murgia}, M. 2008, \mnras,
  391, 521

\bibitem[{{Lamee} {et~al.}(2016){Lamee}, {Rudnick}, {Farnes}, {Carretti},
  {Gaensler}, {Haverkorn}, \& {Poppi}}]{2016ApJ...829....5L}
{Lamee}, M., {Rudnick}, L., {Farnes}, J.~S., {et~al.} 2016, \apj, 829, 5

\bibitem[{{Lovisari} {et~al.}(2020){Lovisari}, {Schellenberger}, {Sereno},
  {Ettori}, {Pratt}, {Forman}, {Jones}, {Andrade-Santos}, {Randall}, \&
  {Kraft}}]{2020ApJ...892..102L}
{Lovisari}, L., {Schellenberger}, G., {Sereno}, M., {et~al.} 2020, \apj, 892,
  102

\bibitem[{{McMullin} {et~al.}(2007){McMullin}, {Waters}, {Schiebel}, {Young},
  \& {Golap}}]{2007ASPC..376..127M}
{McMullin}, J.~P., {Waters}, B., {Schiebel}, D., {Young}, W., \& {Golap}, K.
  2007, in Astronomical Society of the Pacific Conference Series, Vol. 376,
  Astronomical Data Analysis Software and Systems XVI, ed. R.~A. {Shaw},
  F.~{Hill}, \& D.~J. {Bell}, 127

\bibitem[{{Mohan} \& {Rafferty}(2015)}]{2015ascl.soft02007M}
{Mohan}, N. \& {Rafferty}, D. 2015, {PyBDSF: Python Blob Detection and Source
  Finder}

\bibitem[{{Murgia} {et~al.}(2004){Murgia}, {Govoni}, {Feretti}, {Giovannini},
  {Dallacasa}, {Fanti}, {Taylor}, \& {Dolag}}]{2004A&A...424..429M}
{Murgia}, M., {Govoni}, F., {Feretti}, L., {et~al.} 2004, \aap, 424, 429

\bibitem[{{Nagai} {et~al.}(2007){Nagai}, {Kravtsov}, \&
  {Vikhlinin}}]{2007ApJ...668....1N}
{Nagai}, D., {Kravtsov}, A.~V., \& {Vikhlinin}, A. 2007, \apj, 668, 1

\bibitem[{{Offringa} {et~al.}(2014){Offringa}, {McKinley}, {Hurley-Walker},
  {Briggs}, {Wayth}, {Kaplan}, {Bell}, {Feng}, {Neben}, {Hughes}, {Rhee},
  {Murphy}, {Bhat}, {Bernardi}, {Bowman}, {Cappallo}, {Corey}, {Deshpand e},
  {Emrich}, {Ewall-Wice}, {Gaensler}, {Goeke}, {Greenhill}, {Hazelton},
  {Hindson}, {Johnston-Hollitt}, {Jacobs}, {Kasper}, {Kratzenberg}, {Lenc},
  {Lonsdale}, {Lynch}, {McWhirter}, {Mitchell}, {Morales}, {Morgan},
  {Kudryavtseva}, {Oberoi}, {Ord}, {Pindor}, {Procopio}, {Prabu}, {Riding},
  {Roshi}, {Shankar}, {Srivani}, {Subrahmanyan}, {Tingay}, {Waterson},
  {Webster}, {Whitney}, {Williams}, \& {Williams}}]{2014MNRAS.444..606O}
{Offringa}, A.~R., {McKinley}, B., {Hurley-Walker}, N., {et~al.} 2014, \mnras,
  444, 606

\bibitem[{{Offringa} {et~al.}(2012){Offringa}, {van de Gronde}, \&
  {Roerdink}}]{2012A&A...539A..95O}
{Offringa}, A.~R., {van de Gronde}, J.~J., \& {Roerdink}, J.~B.~T.~M. 2012,
  \aap, 539, A95

\bibitem[{{O'Sullivan} {et~al.}(2019){O'Sullivan}, {Machalski}, {Van Eck},
  {Heald}, {Br{\"u}ggen}, {Fynbo}, {Heintz}, {Lara-Lopez}, {Vacca},
  {Hardcastle}, {Shimwell}, {Tasse}, {Vazza}, {Andernach}, {Birkinshaw},
  {Haverkorn}, {Horellou}, {Williams}, {Harwood}, {Brunetti}, {Anderson},
  {Mao}, {Nikiel-Wroczy{\'n}ski}, {Takahashi}, {Carretti}, {Vernstrom}, {van
  Weeren}, {Orr{\'u}}, {Morabito}, \& {Callingham}}]{2019A&A...622A..16O}
{O'Sullivan}, S.~P., {Machalski}, J., {Van Eck}, C.~L., {et~al.} 2019, \aap,
  622, A16

\bibitem[{{Pasetto} {et~al.}(2016){Pasetto}, {Carrasco-Gonz{\'a}lez}, {Bruni},
  {Basu}, {O'Sullivan}, {Kraus}, \& {Mack}}]{2016Galax...4...66P}
{Pasetto}, A., {Carrasco-Gonz{\'a}lez}, C., {Bruni}, G., {et~al.} 2016,
  Galaxies, 4, 66

\bibitem[{{Pearson}(1895)}]{1895RSPS...58..240P}
{Pearson}, K. 1895, Proceedings of the Royal Society of London Series I, 58,
  240

\bibitem[{{Perley} \& {Butler}(2017)}]{PerleyButler2017}
{Perley}, R.~A. \& {Butler}, B.~J. 2017, \apjs, 230, 7

\bibitem[{{Planck Collaboration} {et~al.}(2015){Planck Collaboration}, {Ade},
  {Aghanim}, {Armitage-Caplan}, {Arnaud}, {Ashdown}, {Atrio-Barandela},
  {Aumont}, {Aussel}, {Baccigalupi}, {Banday}, {Barreiro}, {Barrena},
  {Bartelmann}, {Bartlett}, {Battaner}, {Benabed}, {Beno{\^\i}t},
  {Benoit-L{\'e}vy}, {Bernard}, {Bersanelli}, {Bielewicz}, {Bikmaev}, {Bobin},
  {Bock}, {B{\"o}hringer}, {Bonaldi}, {Bond}, {Borrill}, {Bouchet}, {Bridges},
  {Bucher}, {Burenin}, {Burigana}, {Butler}, {Cardoso}, {Carvalho}, {Catalano},
  {Challinor}, {Chamballu}, {Chary}, {Chen}, {Chiang}, {Chiang}, {Chon},
  {Christensen}, {Churazov}, {Church}, {Clements}, {Colombi}, {Colombo},
  {Comis}, {Couchot}, {Coulais}, {Crill}, {Curto}, {Cuttaia}, {Da Silva},
  {Dahle}, {Danese}, {Davies}, {Davis}, {de Bernardis}, {de Rosa}, {de Zotti},
  {Delabrouille}, {Delouis}, {D{\'e}mocl{\`e}s}, {D{\'e}sert}, {Dickinson},
  {Diego}, {Dolag}, {Dole}, {Donzelli}, {Dor{\'e}}, {Douspis}, {Dupac},
  {Efstathiou}, {En{\ss}lin}, {Eriksen}, {Feroz}, {Ferragamo}, {Finelli},
  {Flores-Cacho}, {Forni}, {Frailis}, {Franceschi}, {Fromenteau}, {Galeotta},
  {Ganga}, {G{\'e}nova-Santos}, {Giard}, {Giardino}, {Gilfanov},
  {Giraud-H{\'e}raud}, {Gonz{\'a}lez-Nuevo}, {G{\'o}rski}, {Grainge},
  {Gratton}, {Gregorio}, {Groeneboom}, {Gruppuso}, {Hansen}, {Hanson},
  {Harrison}, {Hempel}, {Henrot-Versill{\'e}}, {Hern{\'a}ndez-Monteagudo},
  {Herranz}, {Hildebrandt}, {Hivon}, {Hobson}, {Holmes}, {Hornstrup}, {Hovest},
  {Huffenberger}, {Hurier}, {Hurley-Walker}, {Jaffe}, {Jaffe}, {Jones},
  {Juvela}, {Keih{\"a}nen}, {Keskitalo}, {Khamitov}, {Kisner}, {Kneissl},
  {Knoche}, {Knox}, {Kunz}, {Kurki-Suonio}, {Lagache}, {L{\"a}hteenm{\"a}ki},
  {Lamarre}, {Lasenby}, {Laureijs}, {Lawrence}, {Leahy}, {Leonardi},
  {Le{\'o}n-Tavares}, {Lesgourgues}, {Li}, {Liddle}, {Liguori}, {Lilje},
  {Linden-V{\o}rnle}, {L{\'o}pez-Caniego}, {Lubin}, {Mac{\'\i}as-P{\'e}rez},
  {MacTavish}, {Maffei}, {Maino}, {Mandolesi}, {Maris}, {Marshall}, {Martin},
  {Mart{\'\i}nez-Gonz{\'a}lez}, {Masi}, {Massardi}, {Matarrese}, {Matthai},
  {Mazzotta}, {Mei}, {Meinhold}, {Melchiorri}, {Melin}, {Mendes}, {Mennella},
  {Migliaccio}, {Mikkelsen}, {Mitra}, {Miville-Desch{\^e}nes}, {Moneti},
  {Montier}, {Morgante}, {Mortlock}, {Munshi}, {Murphy}, {Naselsky}, {Nastasi},
  {Nati}, {Natoli}, {Nesvadba}, {Netterfield}, {N{\o}rgaard-Nielsen},
  {Noviello}, {Novikov}, {Novikov}, {O'Dwyer}, {Olamaie}, {Osborne},
  {Oxborrow}, {Paci}, {Pagano}, {Pajot}, {Paoletti}, {Pasian}, {Patanchon},
  {Pearson}, {Perdereau}, {Perotto}, {Perrott}, {Perrotta}, {Piacentini},
  {Piat}, {Pierpaoli}, {Pietrobon}, {Plaszczynski}, {Pointecouteau}, {Polenta},
  {Ponthieu}, {Popa}, {Poutanen}, {Pratt}, {Pr{\'e}zeau}, {Prunet}, {Puget},
  {Rachen}, {Reach}, {Rebolo}, {Reinecke}, {Remazeilles}, {Renault},
  {Ricciardi}, {Riller}, {Ristorcelli}, {Rocha}, {Rosset}, {Roudier},
  {Rowan-Robinson}, {Rubi{\~n}o-Mart{\'\i}n}, {Rumsey}, {Rusholme}, {Sandri},
  {Santos}, {Saunders}, {Savini}, {Schammel}, {Scott}, {Seiffert}, {Shellard},
  {Shimwell}, {Spencer}, {Starck}, {Stolyarov}, {Stompor}, {Streblyanska},
  {Sudiwala}, {Sunyaev}, {Sureau}, {Sutton}, {Suur-Uski}, {Sygnet}, {Tauber},
  {Tavagnacco}, {Terenzi}, {Toffolatti}, {Tomasi}, {Tramonte}, {Tristram},
  {Tucci}, {Tuovinen}, {T{\"u}rler}, {Umana}, {Valenziano}, {Valiviita}, {Van
  Tent}, {Vibert}, {Vielva}, {Villa}, {Vittorio}, {Wade}, {Wandelt}, {White},
  {White}, {Yvon}, {Zacchei}, \& {Zonca}}]{2015A&A...581A..14P}
{Planck Collaboration}, {Ade}, P.~A.~R., {Aghanim}, N., {et~al.} 2015, \aap,
  581, A14

\bibitem[{{Planck Collaboration} {et~al.}(2013){Planck Collaboration}, {Ade},
  {Aghanim}, {Arnaud}, {Ashdown}, {Atrio-Barandela}, {Aumont}, {Baccigalupi},
  {Balbi}, {Banday}, {Barreiro}, {Bartlett}, {Battaner}, {Benabed},
  {Beno{\^\i}t}, {Bernard}, {Bersanelli}, {Bhatia}, {Bikmaev}, {Bobin},
  {B{\"o}hringer}, {Bonaldi}, {Bond}, {Borgani}, {Borrill}, {Bouchet},
  {Bourdin}, {Brown}, {Burenin}, {Burigana}, {Cabella}, {Cardoso}, {Carvalho},
  {Castex}, {Catalano}, {Cay{\'o}n}, {Chamballu}, {Chiang}, {Chon},
  {Christensen}, {Churazov}, {Clements}, {Colafrancesco}, {Colombi}, {Colombo},
  {Comis}, {Coulais}, {Crill}, {Cuttaia}, {Da Silva}, {Dahle}, {Danese},
  {Davis}, {de Bernardis}, {de Gasperis}, {de Zotti}, {Delabrouille},
  {D{\'e}mocl{\`e}s}, {D{\'e}sert}, {Diego}, {Dolag}, {Dole}, {Donzelli},
  {Dor{\'e}}, {D{\"o}rl}, {Douspis}, {Dupac}, {Efstathiou}, {En{\ss}lin},
  {Eriksen}, {Finelli}, {Flores-Cacho}, {Forni}, {Fosalba}, {Frailis},
  {Franceschi}, {Frommert}, {Galeotta}, {Ganga}, {G{\'e}nova-Santos}, {Giard},
  {Giraud-H{\'e}raud}, {Gonz{\'a}lez-Nuevo}, {G{\'o}rski}, {Gregorio},
  {Gruppuso}, {Hansen}, {Harrison}, {Hempel}, {Henrot-Versill{\'e}},
  {Hern{\'a}ndez-Monteagudo}, {Herranz}, {Hildebrandt}, {Hivon}, {Hobson},
  {Holmes}, {Hurier}, {Jaffe}, {Jaffe}, {Jagemann}, {Jones}, {Juvela},
  {Keih{\"a}nen}, {Khamitov}, {Kisner}, {Kneissl}, {Knoche}, {Knox}, {Kunz},
  {Kurki-Suonio}, {Lagache}, {L{\"a}hteenm{\"a}ki}, {Lamarre}, {Lasenby},
  {Lawrence}, {Le Jeune}, {Leonardi}, {Liddle}, {Lilje}, {L{\'o}pez-Caniego},
  {Luzzi}, {Mac{\'\i}as-P{\'e}rez}, {Maino}, {Mandolesi}, {Maris}, {Marleau},
  {Marshall}, {Mart{\'\i}nez-Gonz{\'a}lez}, {Masi}, {Massardi}, {Matarrese},
  {Mazzotta}, {Mei}, {Melchiorri}, {Melin}, {Mendes}, {Mennella}, {Mitra},
  {Miville-Desch{\^e}nes}, {Moneti}, {Montier}, {Morgante}, {Mortlock},
  {Munshi}, {Murphy}, {Naselsky}, {Nati}, {Natoli}, {N{\o}rgaard-Nielsen},
  {Noviello}, {Novikov}, {Novikov}, {Osborne}, {Pajot}, {Paoletti}, {Pasian},
  {Patanchon}, {Perdereau}, {Perotto}, {Perrotta}, {Piacentini}, {Piat},
  {Pierpaoli}, {Piffaretti}, {Plaszczynski}, {Pointecouteau}, {Polenta},
  {Ponthieu}, {Popa}, {Poutanen}, {Pratt}, {Prunet}, {Puget}, {Rachen},
  {Reach}, {Rebolo}, {Reinecke}, {Remazeilles}, {Renault}, {Ricciardi},
  {Riller}, {Ristorcelli}, {Rocha}, {Roman}, {Rosset}, {Rossetti},
  {Rubi{\~n}o-Mart{\'\i}n}, {Rusholme}, {Sandri}, {Savini}, {Scott}, {Smoot},
  {Starck}, {Sudiwala}, {Sunyaev}, {Sutton}, {Suur-Uski}, {Sygnet}, {Tauber},
  {Terenzi}, {Toffolatti}, {Tomasi}, {Tristram}, {Tuovinen}, {Valenziano}, {Van
  Tent}, {Varis}, {Vielva}, {Villa}, {Vittorio}, {Wade}, {Wandelt}, {Welikala},
  {White}, {White}, {Yvon}, {Zacchei}, \& {Zonca}}]{2013A&A...550A.131P}
{Planck Collaboration}, {Ade}, P.~A.~R., {Aghanim}, N., {et~al.} 2013, \aap,
  550, A131

\bibitem[{{Planck Collaboration} {et~al.}(2011){Planck Collaboration}, {Ade},
  {Aghanim}, {Arnaud}, {Ashdown}, {Aumont}, {Baccigalupi}, {Balbi}, {Banday},
  {Barreiro}, {Bartelmann}, {Bartlett}, {Battaner}, {Battye}, {Benabed},
  {Beno{\^\i}t}, {Bernard}, {Bersanelli}, {Bhatia}, {Bock}, {Bonaldi}, {Bond},
  {Borrill}, {Bouchet}, {Brown}, {Bucher}, {Burigana}, {Cabella}, {Cantalupo},
  {Cardoso}, {Carvalho}, {Catalano}, {Cay{\'o}n}, {Challinor}, {Chamballu},
  {Chary}, {Chiang}, {Chiang}, {Chon}, {Christensen}, {Churazov}, {Clements},
  {Colafrancesco}, {Colombi}, {Couchot}, {Coulais}, {Crill}, {Cuttaia}, {da
  Silva}, {Dahle}, {Danese}, {Davis}, {de Bernardis}, {de Gasperis}, {de Rosa},
  {de Zotti}, {Delabrouille}, {Delouis}, {D{\'e}sert}, {Dickinson}, {Diego},
  {Dolag}, {Dole}, {Donzelli}, {Dor{\'e}}, {D{\"o}rl}, {Douspis}, {Dupac},
  {Efstathiou}, {Eisenhardt}, {En{\ss}lin}, {Feroz}, {Finelli}, {Flores-Cacho},
  {Forni}, {Fosalba}, {Frailis}, {Franceschi}, {Fromenteau}, {Galeotta},
  {Ganga}, {G{\'e}nova-Santos}, {Giard}, {Giardino}, {Giraud-H{\'e}raud},
  {Gonz{\'a}lez-Nuevo}, {Gonz{\'a}lez-Riestra}, {G{\'o}rski}, {Grainge},
  {Gratton}, {Gregorio}, {Gruppuso}, {Harrison}, {Hein{\"a}m{\"a}ki},
  {Henrot-Versill{\'e}}, {Hern{\'a}ndez-Monteagudo}, {Herranz}, {Hildebrandt},
  {Hivon}, {Hobson}, {Holmes}, {Hovest}, {Hoyland}, {Huffenberger}, {Hurier},
  {Hurley-Walker}, {Jaffe}, {Jones}, {Juvela}, {Keih{\"a}nen}, {Keskitalo},
  {Kisner}, {Kneissl}, {Knox}, {Kurki-Suonio}, {Lagache}, {Lamarre}, {Lasenby},
  {Laureijs}, {Lawrence}, {Le Jeune}, {Leach}, {Leonardi}, {Li}, {Liddle},
  {Lilje}, {Linden-V{\o}rnle}, {L{\'o}pez-Caniego}, {Lubin},
  {Mac{\'\i}as-P{\'e}rez}, {MacTavish}, {Maffei}, {Maino}, {Mandolesi}, {Mann},
  {Maris}, {Marleau}, {Mart{\'\i}nez-Gonz{\'a}lez}, {Masi}, {Matarrese},
  {Matthai}, {Mazzotta}, {Mei}, {Meinhold}, {Melchiorri}, {Melin}, {Mendes},
  {Mennella}, {Mitra}, {Miville-Desch{\^e}nes}, {Moneti}, {Montier},
  {Morgante}, {Mortlock}, {Munshi}, {Murphy}, {Naselsky}, {Nati}, {Natoli},
  {Netterfield}, {N{\o}rgaard-Nielsen}, {Noviello}, {Novikov}, {Novikov},
  {Olamaie}, {Osborne}, {Pajot}, {Pasian}, {Patanchon}, {Pearson}, {Perdereau},
  {Perotto}, {Perrotta}, {Piacentini}, {Piat}, {Pierpaoli}, {Piffaretti},
  {Plaszczynski}, {Pointecouteau}, {Polenta}, {Ponthieu}, {Poutanen}, {Pratt},
  {Pr{\'e}zeau}, {Prunet}, {Puget}, {Rachen}, {Reach}, {Rebolo}, {Reinecke},
  {Renault}, {Ricciardi}, {Riller}, {Ristorcelli}, {Rocha}, {Rosset},
  {Rubi{\~n}o-Mart{\'\i}n}, {Rusholme}, {Saar}, {Sandri}, {Santos}, {Saunders},
  {Savini}, {Schaefer}, {Scott}, {Seiffert}, {Shellard}, {Smoot}, {Stanford},
  {Starck}, {Stivoli}, {Stolyarov}, {Stompor}, {Sudiwala}, {Sunyaev}, {Sutton},
  {Sygnet}, {Taburet}, {Tauber}, {Terenzi}, {Toffolatti}, {Tomasi}, {Torre},
  {Tristram}, {Tuovinen}, {Valenziano}, {Vibert}, {Vielva}, {Villa},
  {Vittorio}, {Wade}, {Wandelt}, {Weller}, {White}, {White}, {Yvon}, {Zacchei},
  \& {Zonca}}]{2011A&A...536A...8P}
{Planck Collaboration}, {Ade}, P.~A.~R., {Aghanim}, N., {et~al.} 2011, \aap,
  536, A8

\bibitem[{{Planck Collaboration} {et~al.}(2016){Planck Collaboration}, {Ade},
  {Aghanim}, {Arnaud}, {Ashdown}, {Aumont}, {Baccigalupi}, {Banday},
  {Barreiro}, {Barrena}, {Bartlett}, {Bartolo}, {Battaner}, {Battye},
  {Benabed}, {Beno{\^\i}t}, {Benoit-L{\'e}vy}, {Bernard}, {Bersanelli},
  {Bielewicz}, {Bikmaev}, {B{\"o}hringer}, {Bonaldi}, {Bonavera}, {Bond},
  {Borrill}, {Bouchet}, {Bucher}, {Burenin}, {Burigana}, {Butler}, {Calabrese},
  {Cardoso}, {Carvalho}, {Catalano}, {Challinor}, {Chamballu}, {Chary},
  {Chiang}, {Chon}, {Christensen}, {Clements}, {Colombi}, {Colombo}, {Combet},
  {Comis}, {Couchot}, {Coulais}, {Crill}, {Curto}, {Cuttaia}, {Dahle},
  {Danese}, {Davies}, {Davis}, {de Bernardis}, {de Rosa}, {de Zotti},
  {Delabrouille}, {D{\'e}sert}, {Dickinson}, {Diego}, {Dolag}, {Dole},
  {Donzelli}, {Dor{\'e}}, {Douspis}, {Ducout}, {Dupac}, {Efstathiou},
  {Eisenhardt}, {Elsner}, {En{\ss}lin}, {Eriksen}, {Falgarone}, {Fergusson},
  {Feroz}, {Ferragamo}, {Finelli}, {Forni}, {Frailis}, {Fraisse}, {Franceschi},
  {Frejsel}, {Galeotta}, {Galli}, {Ganga}, {G{\'e}nova-Santos}, {Giard},
  {Giraud-H{\'e}raud}, {Gjerl{\o}w}, {Gonz{\'a}lez-Nuevo}, {G{\'o}rski},
  {Grainge}, {Gratton}, {Gregorio}, {Gruppuso}, {Gudmundsson}, {Hansen},
  {Hanson}, {Harrison}, {Hempel}, {Henrot-Versill{\'e}},
  {Hern{\'a}ndez-Monteagudo}, {Herranz}, {Hildebrandt}, {Hivon}, {Hobson},
  {Holmes}, {Hornstrup}, {Hovest}, {Huffenberger}, {Hurier}, {Jaffe}, {Jaffe},
  {Jin}, {Jones}, {Juvela}, {Keih{\"a}nen}, {Keskitalo}, {Khamitov}, {Kisner},
  {Kneissl}, {Knoche}, {Kunz}, {Kurki-Suonio}, {Lagache}, {Lamarre}, {Lasenby},
  {Lattanzi}, {Lawrence}, {Leonardi}, {Lesgourgues}, {Levrier}, {Liguori},
  {Lilje}, {Linden-V{\o}rnle}, {L{\'o}pez-Caniego}, {Lubin},
  {Mac{\'\i}as-P{\'e}rez}, {Maggio}, {Maino}, {Mak}, {Mandolesi}, {Mangilli},
  {Martin}, {Mart{\'\i}nez-Gonz{\'a}lez}, {Masi}, {Matarrese}, {Mazzotta},
  {McGehee}, {Mei}, {Melchiorri}, {Melin}, {Mendes}, {Mennella}, {Migliaccio},
  {Mitra}, {Miville-Desch{\^e}nes}, {Moneti}, {Montier}, {Morgante},
  {Mortlock}, {Moss}, {Munshi}, {Murphy}, {Naselsky}, {Nastasi}, {Nati},
  {Natoli}, {Netterfield}, {N{\o}rgaard-Nielsen}, {Noviello}, {Novikov},
  {Novikov}, {Olamaie}, {Oxborrow}, {Paci}, {Pagano}, {Pajot}, {Paoletti},
  {Pasian}, {Patanchon}, {Pearson}, {Perdereau}, {Perotto}, {Perrott},
  {Perrotta}, {Pettorino}, {Piacentini}, {Piat}, {Pierpaoli}, {Pietrobon},
  {Plaszczynski}, {Pointecouteau}, {Polenta}, {Pratt}, {Pr{\'e}zeau}, {Prunet},
  {Puget}, {Rachen}, {Reach}, {Rebolo}, {Reinecke}, {Remazeilles}, {Renault},
  {Renzi}, {Ristorcelli}, {Rocha}, {Rosset}, {Rossetti}, {Roudier}, {Rozo},
  {Rubi{\~n}o-Mart{\'\i}n}, {Rumsey}, {Rusholme}, {Rykoff}, {Sandri}, {Santos},
  {Saunders}, {Savelainen}, {Savini}, {Schammel}, {Scott}, {Seiffert},
  {Shellard}, {Shimwell}, {Spencer}, {Stanford}, {Stern}, {Stolyarov},
  {Stompor}, {Streblyanska}, {Sudiwala}, {Sunyaev}, {Sutton}, {Suur-Uski},
  {Sygnet}, {Tauber}, {Terenzi}, {Toffolatti}, {Tomasi}, {Tramonte},
  {Tristram}, {Tucci}, {Tuovinen}, {Umana}, {Valenziano}, {Valiviita}, {Van
  Tent}, {Vielva}, {Villa}, {Wade}, {Wandelt}, {Wehus}, {White}, {Wright},
  {Yvon}, {Zacchei}, \& {Zonca}}]{2016A&A...594A..27P}
{Planck Collaboration}, {Ade}, P.~A.~R., {Aghanim}, N., {et~al.} 2016, \aap,
  594, A27

\bibitem[{{Rajpurohit} {et~al.}(2022){Rajpurohit}, {Hoeft}, {Wittor}, {van
  Weeren}, {Vazza}, {Rudnick}, {Rajpurohit}, {Forman}, {Riseley}, {Brienza},
  {Bonafede}, {Rajpurohit}, {Dom{\'\i}nguez-Fern{\'a}ndez}, {Eilek},
  {Bonnassieux}, {Br{\"u}ggen}, {Loi}, {R{\"o}ttgering}, {Drabent},
  {Locatelli}, {Botteon}, {Brunetti}, \& {Clarke}}]{2022A&A...657A...2R}
{Rajpurohit}, K., {Hoeft}, M., {Wittor}, D., {et~al.} 2022, \aap, 657, A2

\bibitem[{{Rudnick} \& {Blundell}(2003)}]{2003ApJ...588..143R}
{Rudnick}, L. \& {Blundell}, K.~M. 2003, \apj, 588, 143

\bibitem[{{Rudnick} \& {Owen}(2014)}]{2014ApJ...785...45R}
{Rudnick}, L. \& {Owen}, F.~N. 2014, \apj, 785, 45

\bibitem[{{Russell} {et~al.}(2011){Russell}, {van Weeren}, {Edge}, {McNamara},
  {Sanders}, {Fabian}, {Baum}, {Canning}, {Donahue}, \&
  {O'Dea}}]{2011MNRAS.417L...1R}
{Russell}, H.~R., {van Weeren}, R.~J., {Edge}, A.~C., {et~al.} 2011, \mnras,
  417, L1

\bibitem[{{Sarazin}(2002)}]{2002ASSL..272....1S}
{Sarazin}, C.~L. 2002, in Astrophysics and Space Science Library, Vol. 272,
  Merging Processes in Galaxy Clusters, ed. L.~{Feretti}, I.~M. {Gioia}, \&
  G.~{Giovannini}, 1--38

\bibitem[{{Schlegel} {et~al.}(1998){Schlegel}, {Finkbeiner}, \&
  {Davis}}]{1998ApJ...500..525S}
{Schlegel}, D.~J., {Finkbeiner}, D.~P., \& {Davis}, M. 1998, \apj, 500, 525

\bibitem[{{Sebokolodi} {et~al.}(2020){Sebokolodi}, {Perley}, {Eilek},
  {Carilli}, {Smirnov}, {Laing}, {Greisen}, \& {Wise}}]{2020ApJ...903...36S}
{Sebokolodi}, M. L.~L., {Perley}, R., {Eilek}, J., {et~al.} 2020, \apj, 903, 36

\bibitem[{{Sokoloff} {et~al.}(1998){Sokoloff}, {Bykov}, {Shukurov},
  {Berkhuijsen}, {Beck}, \& {Poezd}}]{1998MNRAS.299..189S}
{Sokoloff}, D.~D., {Bykov}, A.~A., {Shukurov}, A., {et~al.} 1998, \mnras, 299,
  189

\bibitem[{{Somboonpanyakul} {et~al.}(2021){Somboonpanyakul}, {McDonald},
  {Bayliss}, {Voit}, {Donahue}, {Gaspari}, {Dahle}, {Rivera-Thorsen}, \&
  {Stark}}]{2021ApJ...907L..12S}
{Somboonpanyakul}, T., {McDonald}, M., {Bayliss}, M., {et~al.} 2021, \apjl,
  907, L12

\bibitem[{Spearman(1904)}]{10.2307/1412159}
Spearman, C. 1904, The American Journal of Psychology, 15, 72

\bibitem[{{Stuardi} {et~al.}(2021){Stuardi}, {Bonafede}, {Lovisari},
  {Dom{\'\i}nguez-Fern{\'a}ndez}, {Vazza}, {Br{\"u}ggen}, {van Weeren}, \& {de
  Gasperin}}]{2021MNRAS.502.2518S}
{Stuardi}, C., {Bonafede}, A., {Lovisari}, L., {et~al.} 2021, \mnras, 502, 2518

\bibitem[{{Stuardi} {et~al.}(2020){Stuardi}, {O'Sullivan}, {Bonafede},
  {Br{\"u}ggen}, {Dabhade}, {Horellou}, {Morganti}, {Carretti}, {Heald},
  {Iacobelli}, \& {Vacca}}]{2020A&A...638A..48S}
{Stuardi}, C., {O'Sullivan}, S.~P., {Bonafede}, A., {et~al.} 2020, \aap, 638,
  A48

\bibitem[{{Sunyaev} \& {Zeldovich}(1970)}]{1970CoASP...2...66S}
{Sunyaev}, R.~A. \& {Zeldovich}, Y.~B. 1970, Comments on Astrophysics and Space
  Physics, 2, 66

\bibitem[{{Sunyaev} \& {Zeldovich}(1972)}]{1972CoASP...4..173S}
{Sunyaev}, R.~A. \& {Zeldovich}, Y.~B. 1972, Comments on Astrophysics and Space
  Physics, 4, 173

\bibitem[{{Tarr{\'\i}o} \& {Zarattini}(2020)}]{2020A&A...642A.102T}
{Tarr{\'\i}o}, P. \& {Zarattini}, S. 2020, \aap, 642, A102

\bibitem[{{Tasse}(2014)}]{2014A&A...566A.127T}
{Tasse}, C. 2014, \aap, 566, A127

\bibitem[{{Taylor} {et~al.}(2002){Taylor}, {Fabian}, \&
  {Allen}}]{2002MNRAS.334..769T}
{Taylor}, G.~B., {Fabian}, A.~C., \& {Allen}, S.~W. 2002, \mnras, 334, 769

\bibitem[{{Taylor} {et~al.}(2006){Taylor}, {Gugliucci}, {Fabian}, {Sanders},
  {Gentile}, \& {Allen}}]{2006MNRAS.368.1500T}
{Taylor}, G.~B., {Gugliucci}, N.~E., {Fabian}, A.~C., {et~al.} 2006, \mnras,
  368, 1500

\bibitem[{{Taylor} {et~al.}(1990){Taylor}, {Perley}, {Inoue}, {Kato}, {Tabara},
  \& {Aizu}}]{1990ApJ...360...41T}
{Taylor}, G.~B., {Perley}, R.~A., {Inoue}, M., {et~al.} 1990, \apj, 360, 41

\bibitem[{{Tribble}(1991)}]{1991MNRAS.253..147T}
{Tribble}, P.~C. 1991, \mnras, 253, 147

\bibitem[{{Uson} \& {Cotton}(2008)}]{2008A&A...486..647U}
{Uson}, J.~M. \& {Cotton}, W.~D. 2008, \aap, 486, 647

\bibitem[{{Vacca} {et~al.}(2018){Vacca}, {Murgia}, {Govoni}, {En{\ss}lin},
  {Oppermann}, {Feretti}, {Giovannini}, \& {Loi}}]{2018Galax...6..142V}
{Vacca}, V., {Murgia}, M., {Govoni}, F., {et~al.} 2018, Galaxies, 6, 142

\bibitem[{{Vacca} {et~al.}(2012){Vacca}, {Murgia}, {Govoni}, {Feretti},
  {Giovannini}, {Perley}, \& {Taylor}}]{2012A&A...540A..38V}
{Vacca}, V., {Murgia}, M., {Govoni}, F., {et~al.} 2012, \aap, 540, A38

\bibitem[{{van der Tol} {et~al.}(2018){van der Tol}, {Veenboer}, \&
  {Offringa}}]{2018A&A...616A..27V}
{van der Tol}, S., {Veenboer}, B., \& {Offringa}, A.~R. 2018, \aap, 616, A27

\bibitem[{{van Terwisga} {et~al.}(2022){van Terwisga}, {Hacar}, {van Dishoeck},
  {Oonk}, \& {Portegies Zwart}}]{2022A&A...661A..53V}
{van Terwisga}, S.~E., {Hacar}, A., {van Dishoeck}, E.~F., {Oonk}, R., \&
  {Portegies Zwart}, S. 2022, \aap, 661, A53

\bibitem[{{van Weeren} {et~al.}(2019){van Weeren}, {de Gasperin}, {Akamatsu},
  {Br{\"u}ggen}, {Feretti}, {Kang}, {Stroe}, \&
  {Zandanel}}]{2019SSRv..215...16V}
{van Weeren}, R.~J., {de Gasperin}, F., {Akamatsu}, H., {et~al.} 2019, \ssr,
  215, 16

\bibitem[{{Vikhlinin} {et~al.}(2006){Vikhlinin}, {Kravtsov}, {Forman}, {Jones},
  {Markevitch}, {Murray}, \& {Van Speybroeck}}]{2006ApJ...640..691V}
{Vikhlinin}, A., {Kravtsov}, A., {Forman}, W., {et~al.} 2006, \apj, 640, 691

\bibitem[{{Vogt} \& {En{\ss}lin}(2006)}]{2006AN....327..595V}
{Vogt}, C. \& {En{\ss}lin}, T.~A. 2006, Astronomische Nachrichten, 327, 595

\bibitem[{{Wen} \& {Han}(2015)}]{2015ApJ...807..178W}
{Wen}, Z.~L. \& {Han}, J.~L. 2015, \apj, 807, 178

\bibitem[{{Williams} {et~al.}(2019){Williams}, {Hardcastle}, {Best}, {Sabater},
  {Croston}, {Duncan}, {Shimwell}, {R{\"o}ttgering}, {Nisbet}, {G{\"u}rkan},
  {Alegre}, {Cochrane}, {Goyal}, {Hale}, {Jackson}, {Jamrozy}, {Kondapally},
  {Kunert-Bajraszewska}, {Mahatma}, {Mingo}, {Morabito}, {Prandoni},
  {Roskowinski}, {Shulevski}, {Smith}, {Tasse}, {Urquhart}, {Webster}, {White},
  {Beswick}, {Callingham}, {Chy{\.z}y}, {de Gasperin}, {Harwood}, {Hoeft},
  {Iacobelli}, {McKean}, {Mechev}, {Miley}, {Schwarz}, \& {van
  Weeren}}]{2019A&A...622A...2W}
{Williams}, W.~L., {Hardcastle}, M.~J., {Best}, P.~N., {et~al.} 2019, \aap,
  622, A2

\end{thebibliography}

\clearpage
\newpage

\begin{appendix}

\section{Possible biases}\label{sec:AppendixA}
This appendix aims to investigate possible biases that might affect the analysis. The polarisation properties of radio sources might be influenced by other source properties such as size, total intensity, and whether the source is a single or multi-component source. If, for example, multi-component sources are easier to detect near the pointing centre and have different polarisation properties than single-component sources, the observed depolarisation effect with projected radius might be biased. This is because all clusters have been observed approximately in the pointing centre. As mentioned in Section \ref{sec:caveats}, this also introduces a bias through the inclusion of the upper limits because the upper limits on the fractional polarisation are generally higher at larger projected radii due to the primary beam response. In the first section of this appendix, we check whether biases because of source properties are present, and in the second section we quantify the biases through a Monte Carlo experiment.

\subsection{Source properties}
To investigate whether there is a dependence between observed source properties and polarisation, taking into account the distance to the pointing centre, we plot running medians of various quantities versus the angular distance to the pointing centre. The uncertainty, $\sigma_{\pm}$, on the running median $M$ is calculated as in \citet[][]{2016ApJ...829....5L},
\begin{equation}
    \sigma_{\pm} = |M - [p_{16},p_{84}]/\sqrt{N},
\end{equation}
where $p_x$ denotes the $x$-th percentile of the distribution, and $N$ is the number of points in a bin. The amount of correlation is quantified by the Pearson \citep{1895RSPS...58..240P} and Spearman \citep[][]{10.2307/1412159} coefficients, shown in Table \ref{tab:properties1}. We define weak correlation for values of Pearson $|r| \leq 0.3$, moderate correlation for $ 0.3 < |r| \leq 0.7$ and strong correlation for $r > 0.7$, using the common cutoff $p$-value of 0.05 for statistical significance. Because the Pearson coefficient only measures the linear relationship we also report the Spearman coefficient. The monotonicity of the correlation is given by the Spearman coefficient, where an absolute value of $1$ indicates a perfectly monotonic relationship. 

\begin{figure*}
    \centering
    \begin{subfigure}[b]{0.475\textwidth}
        \centering
        \includegraphics[width=\textwidth]{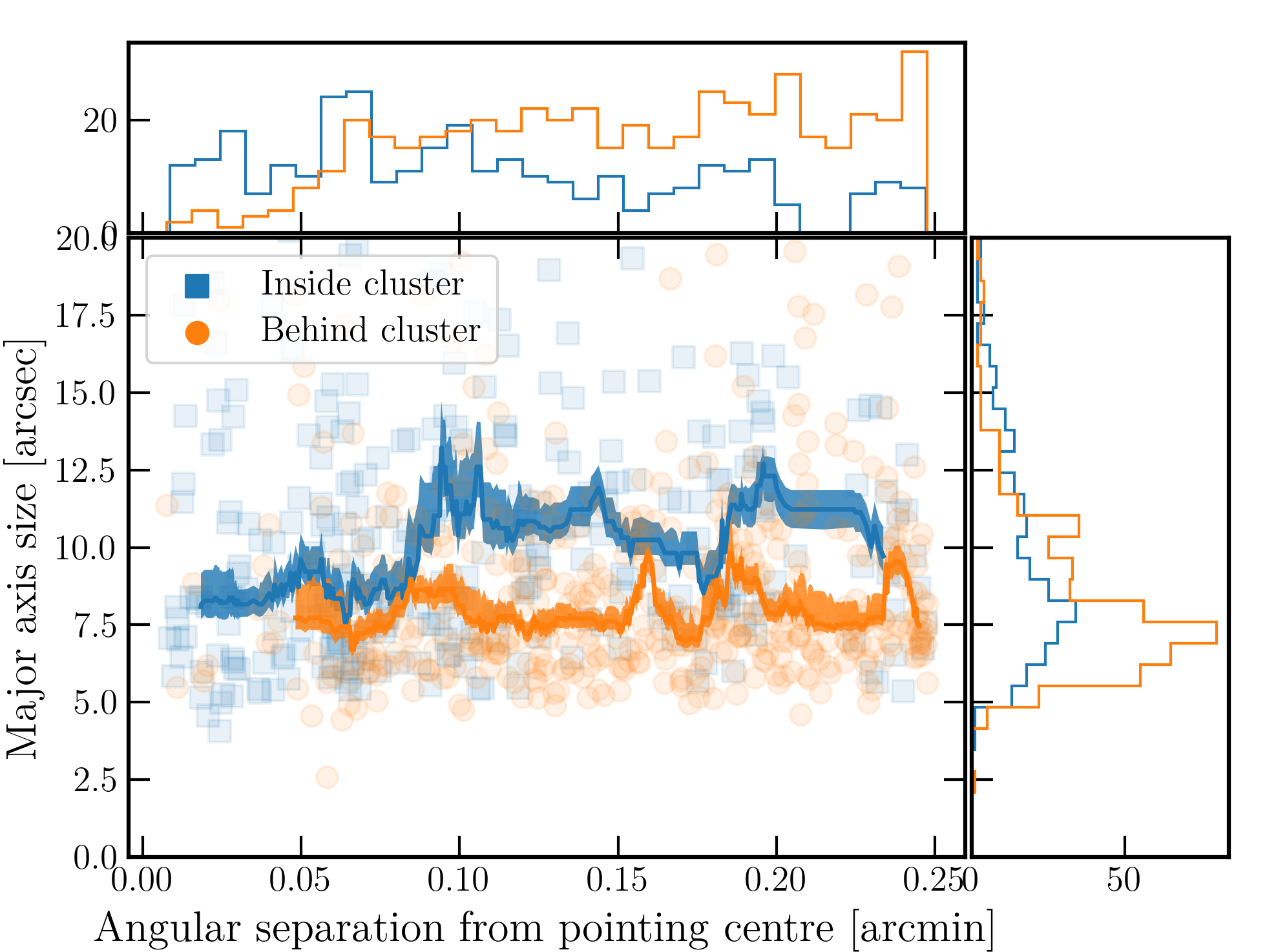}
        \caption[]%
        {{\small Polarised source sizes}}    
        \label{fig:size_thetap}
    \end{subfigure}
    \hfill
    \begin{subfigure}[b]{0.475\textwidth}  
        \centering 
        \includegraphics[width=\textwidth]{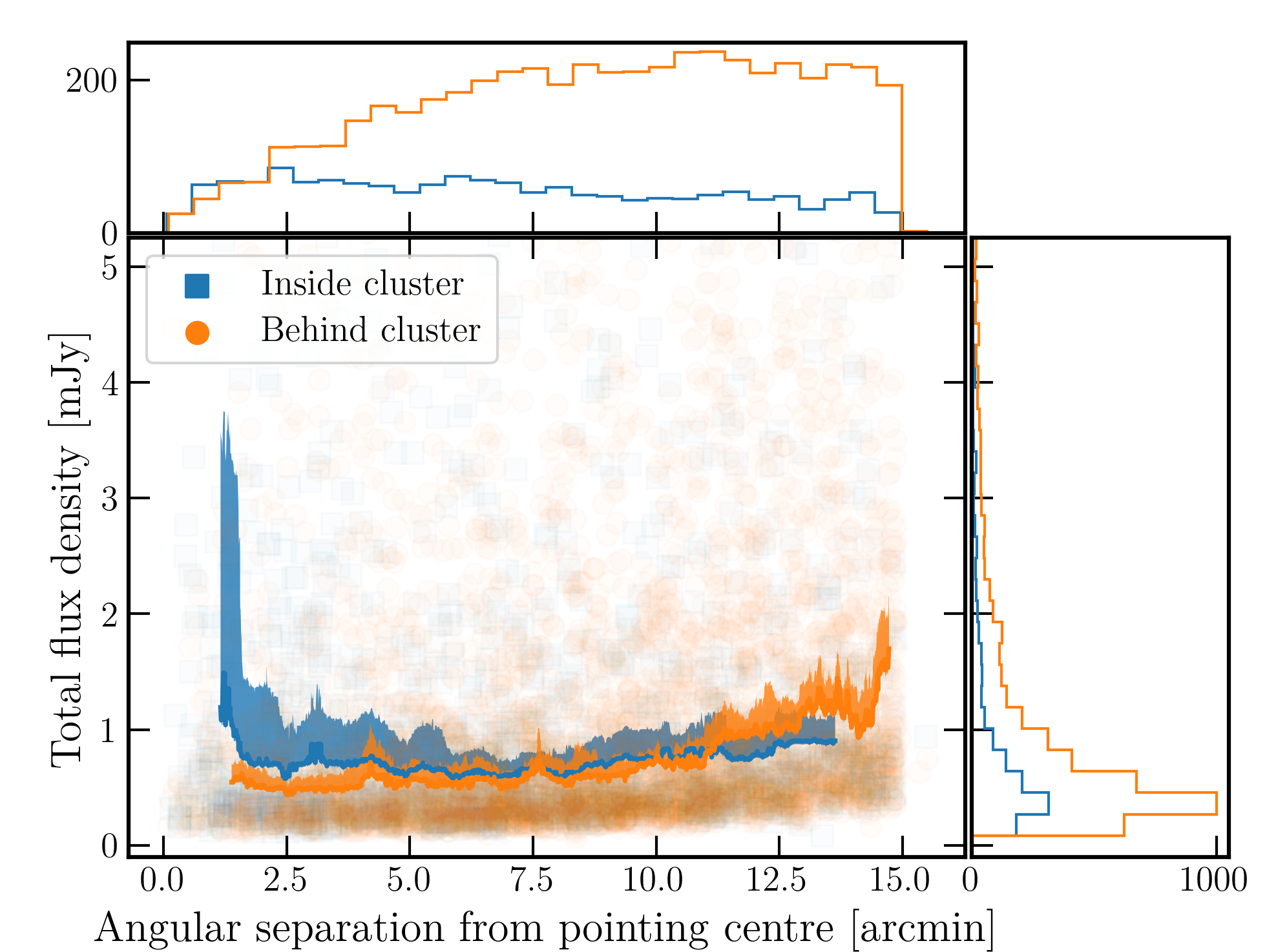}
        \caption[]%
        {{\small Total Stokes I flux density}}    
        \label{fig:flux_thetap}
    \end{subfigure}
    \vskip\baselineskip
    \begin{subfigure}[b]{0.475\textwidth}   
        \centering 
        \includegraphics[width=\textwidth]{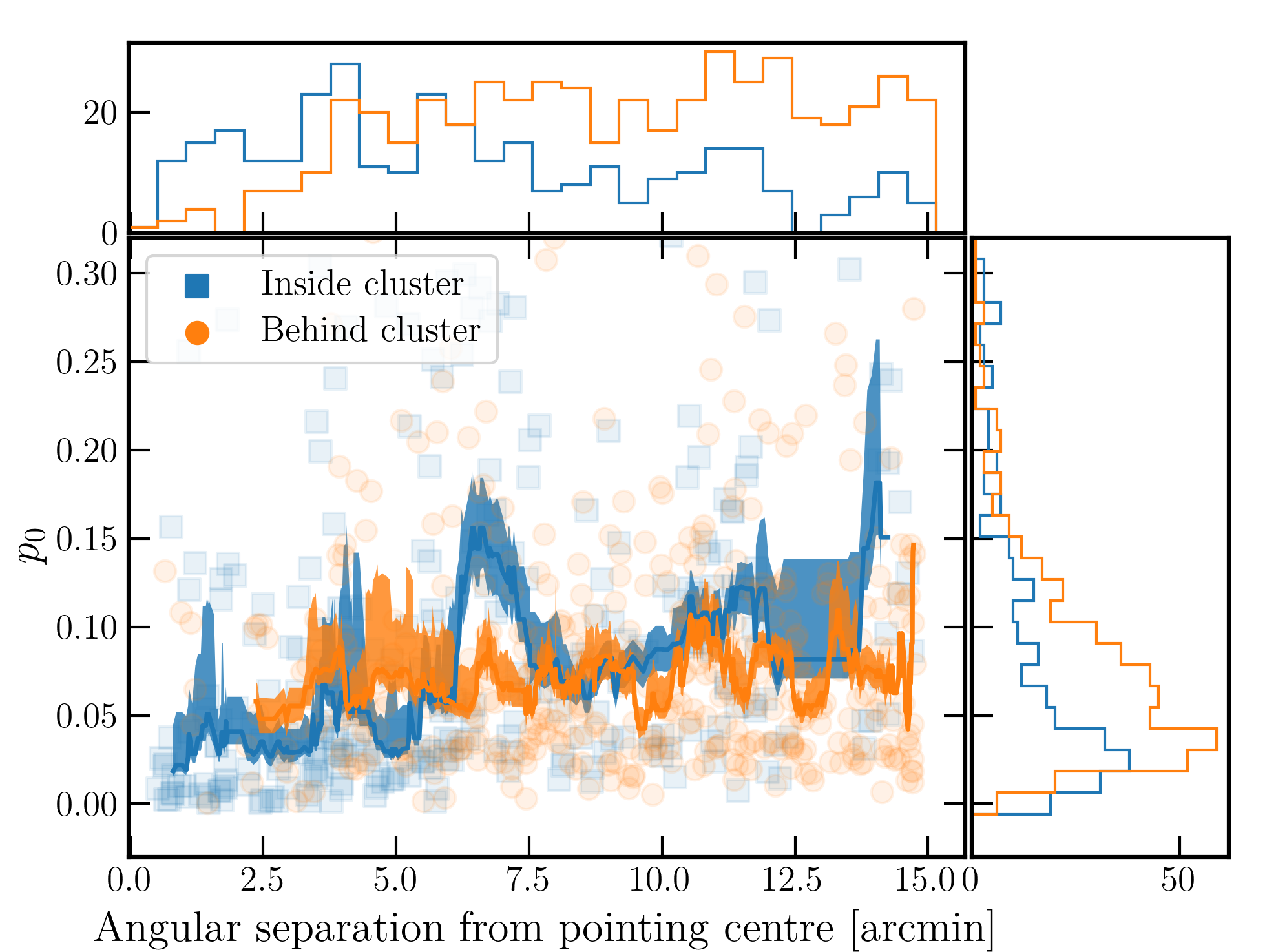}
        \caption[]%
        {{\small Intrinsic polarisation fraction}}    
        \label{fig:p0_thetap}
    \end{subfigure}
    \hfill
    \begin{subfigure}[b]{0.475\textwidth}   
        \centering 
        \includegraphics[width=\textwidth]{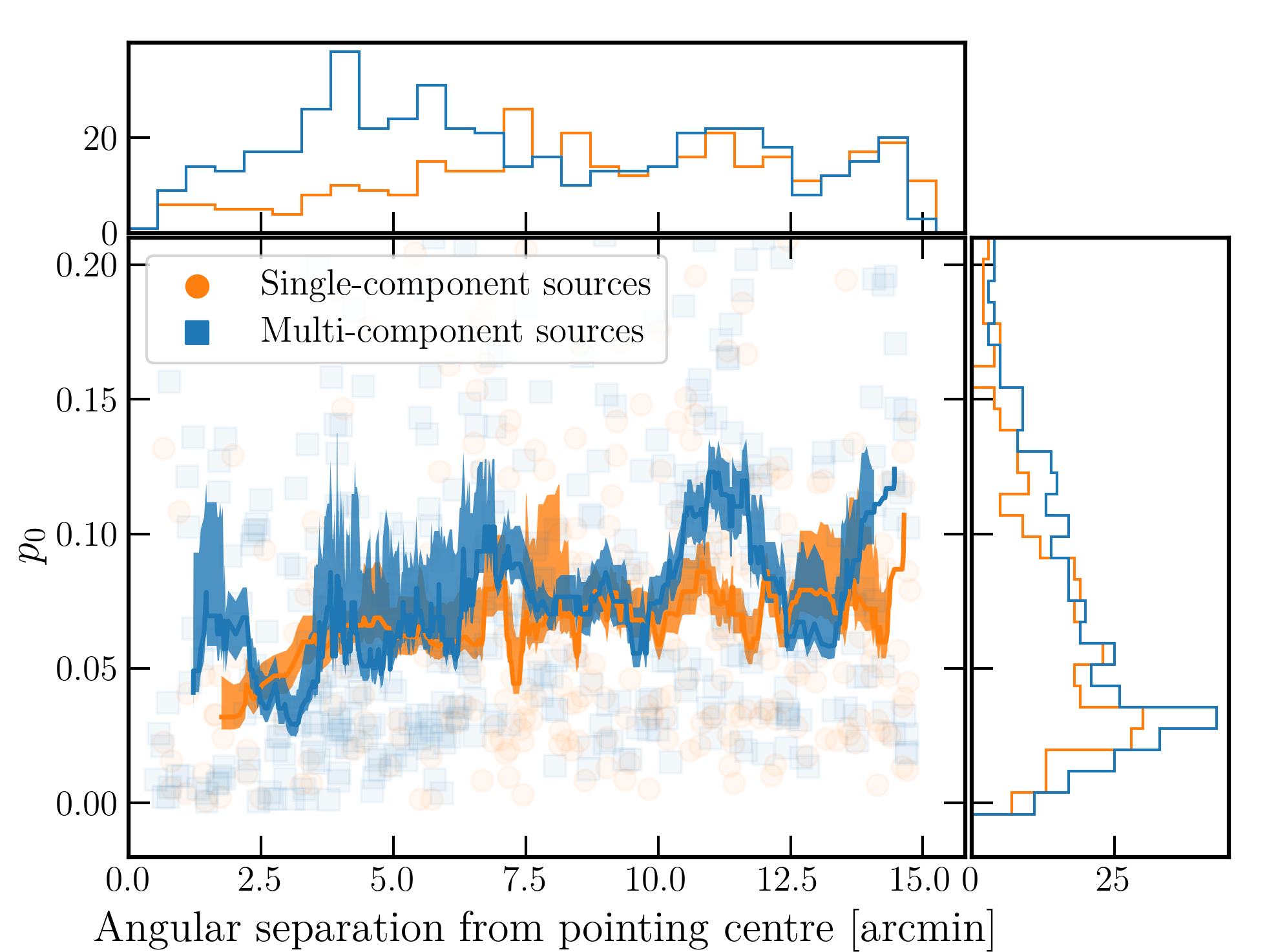}
        \caption[]%
        {{\small Intrinsic polarisation fraction for single- and multi-component sources}}    
        \label{fig:p0_thetap_multi}
    \end{subfigure}
    \caption[ The average and standard deviation of critical parameters ]
    {\small Various quantities against angular separation separately for cluster members and background sources. The running median is shown with uncertainties in the shaded region. Histograms show the projected distributions along the axes.} 
    \label{fig:A1}
\end{figure*}

We plot in Figure \ref{fig:size_thetap} the running median of the observed polarised source major axis versus the angular distance to the pointing centre. As the figure shows, there is some dependence of polarised source size on angular separation. Cluster members show an increase in median source size around 5 arcminutes from the pointing centre, while for background sources there is no significant correlation. The median source size is larger for cluster members than for background sources, which is expected simply because they are nearby sources. 

The total flux density versus angular separation is shown in Figure \ref{fig:flux_thetap}. The background population clearly shows the effect of the primary beam response. Cluster members are less affected by this, possibly because they are all low redshift ($z<0.35$) sources for which we are already sensitive enough to probe the majority of the cluster population. There is an excess of bright cluster members in the centre of the image, indicated by the peak at low angular separation. This means that we are detecting the brightest cluster members preferentially in the centre of the images. This is not unexpected because the centres of the clusters lie near the centres of the images, but it might bias the results if the total flux density is correlated with polarisation properties.

Although we are mainly interested in the depolarisation of radio sources, it is important to investigate the best-fit intrinsic polarisation fraction as a function of angular distance. Because of the chosen model and finite amount of bandwidth, there is a degeneracy between $p_0$ and $\sigma_\mathrm{RM}$ (see Fig. \ref{fig:examplefit}). While this degeneracy is simply a result of the fitting, real correlations between $p_0$ and $\sigma_\mathrm{RM}$ have been claimed before in the literature \citep[e.g.][]{2016ApJ...829....5L}. It is therefore important to inspect trends of $p_0$ with angular separation because that could create biases in the depolarisation trend. In Figure \ref{fig:p0_thetap} we plot the best-fit intrinsic polarisation fraction. Here we see that there is a weak correlation with angular distance for cluster members and no significant correlation for background sources. The fact that the median $p_0$ is lower near the centre of the image is likely (at least partly) caused by the fact that the brightest cluster members often lie in the centre of the cluster (Fig.\ref{fig:flux_thetap}), where it is thus possible to detect smaller polarisation fractions. 

Lastly, there might be a difference in polarisation properties of single and multi-component sources, so we also plot this separation in Figure \ref{fig:p0_thetap_multi}. We note that multi-component sources are often, but not always, cluster members. Both populations show no evidence for a strong correlation of the intrinsic polarisation with angular separation, and both populations have similar distributions of intrinsic polarisation, so it is unlikely that this is biasing the results significantly. 

\begin{figure*}
\centering
\begin{minipage}{.5\textwidth}
  \centering
  \includegraphics[width=1.0\linewidth]{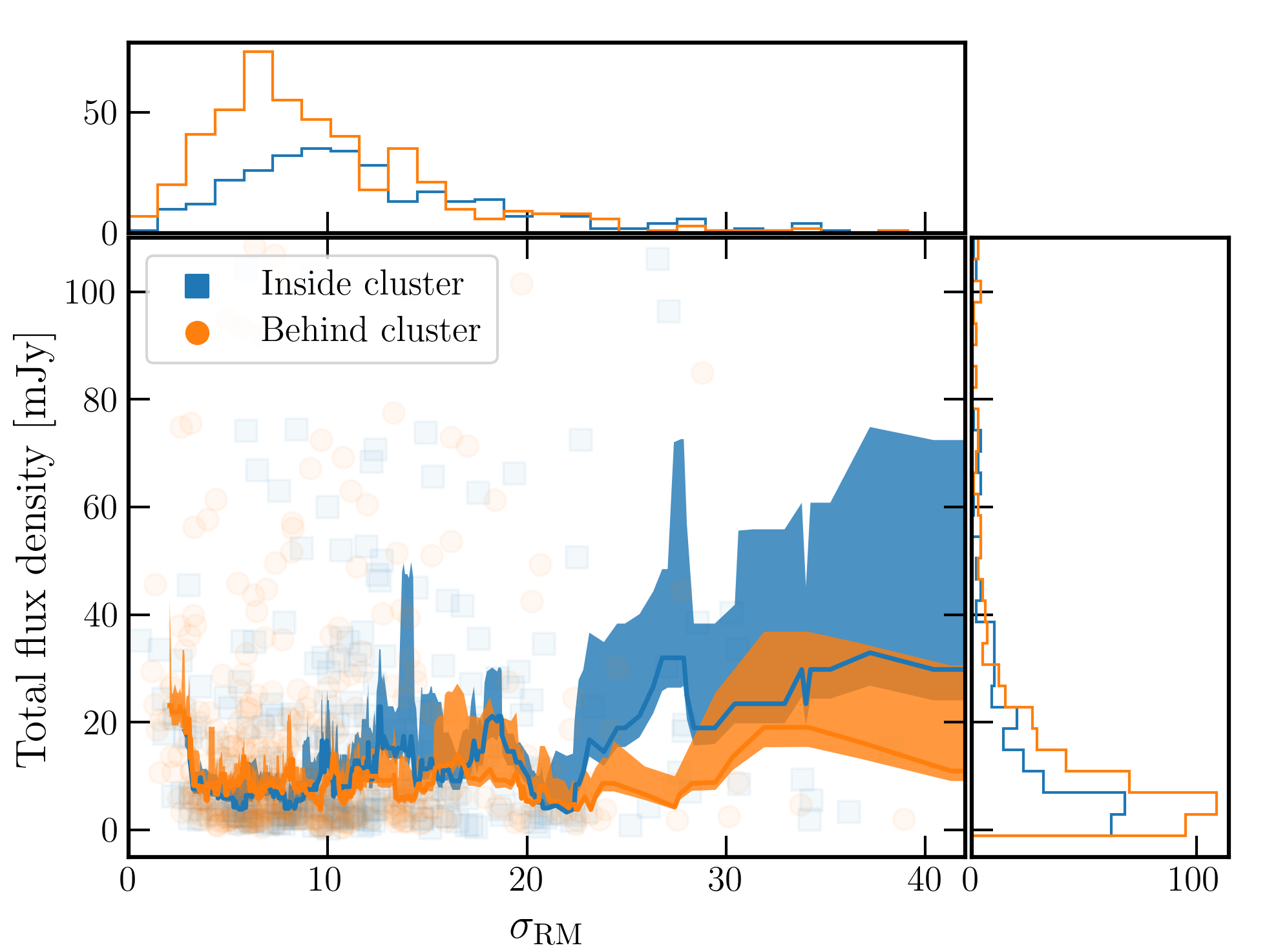}
  \captionsetup{width=0.8\linewidth}
  \captionof{figure}{Same as Figure \ref{fig:A1}, but for total Stokes I flux density against depolarisation parameter $\sigma_\mathrm{RM}$.}
  \label{fig:Iv_sigmaRM}
\end{minipage}%
\begin{minipage}{.5\textwidth}
  \centering
  \includegraphics[width=1.0\linewidth]{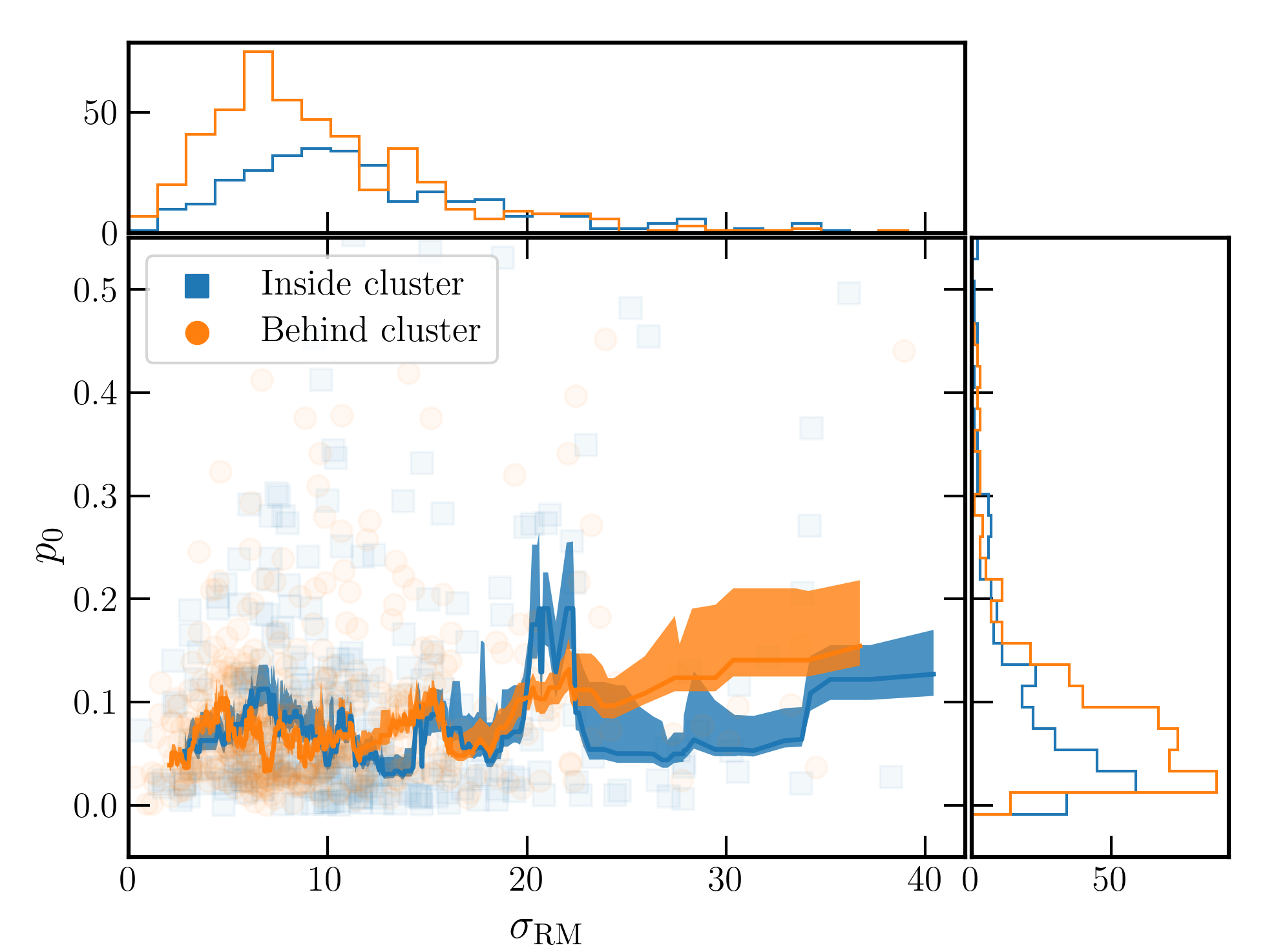}
  \captionsetup{width=0.8\linewidth}
  \captionof{figure}{Same as Figure \ref{fig:A1}, but for intrinsic polarisation fraction against depolarisation parameter $\sigma_\mathrm{RM}$.}
  \label{fig:p0_sigmaRM}
\end{minipage}
\end{figure*}

Now that we have established that the cluster population has a higher median flux density in the centre of the images and that the intrinsic polarisation fraction of cluster members is generally lower in the centre of the images, it is important to know whether these variables correlate with the depolarisation parameter $\sigma_\mathrm{RM}$. Figures \ref{fig:Iv_sigmaRM} and \ref{fig:p0_sigmaRM} show the trend of total flux density and intrinsic polarisation fraction with $\sigma_\mathrm{RM}$. We see for background sources no clear correlation between total flux density and depolarisation. Cluster members do show that the brightest sources show more depolarisation. However, this is to be expected in the case of a magnetised depolarising intracluster medium (ICM) in the cluster centre if the brightest cluster members are also found preferentially in the cluster centre, which is indeed the case as shown by Figure \ref{fig:flux_thetap}. The question remains how much of this effect is a real effect and how much is caused by biases such as only picking up the most depolarised sources near the centre of the cluster. This is addressed in the next section. Figure \ref{fig:p0_sigmaRM} shows the degeneracy between $p_0$ and $\sigma_\mathrm{RM}$, particularly for large values of $\sigma_\mathrm{RM}$. It is interesting that this trend implies that more depolarised sources have larger $p_0$, while Figure \ref{fig:p0_thetap} showed that smaller values of $p_0$ are generally found more towards the cluster centre. This trend would therefore cause a bias in the direction opposite to the trend expected from a depolarising ICM.

\subsection{A2. Monte Carlo experiment}\label{sec:appendix_montecarlo}
\begin{table*}[]
\centering
\caption{Pearson and Spearman correlation coefficients and accompanying p-values for various combinations of parameters shown in Figures \ref{fig:A1}, \ref{fig:Iv_sigmaRM} and \ref{fig:p0_sigmaRM}.}
\label{tab:properties1}
\begin{tabular}{@{}llll@{}}
\toprule
Population                                           & Pearson (r, p-value)         & Spearman (r, p-value)        & Conclusion\tablefootmark{a}                 \\ \midrule
Major axis size - $\theta_p$ cluster members         & (0.15, $7.4\times10^{-3}$)   & (0.22, $8.5\times10^{-5}$)   & weak correlation           \\
Major axis size - $\theta_p$ background sources      & (0.018, $6.9\times10^{-1}$)  & (0.070, $1.2\times10^{-1}$)  & no significant correlation \\
Total Flux - $\theta_p$ cluster members              & (-0.087, $6.6\times10^{-4}$) & (0.0080, $7.5\times10^{-1}$) & weak correlation           \\
Total Flux - $\theta_p$ background sources           & (-0.021, $1.4\times10^{-1}$) & (0.26, $8.3\times10^{-79}$)  & non-monotonic correlation  \\
$p_0$ - $\theta_p$ cluster members                   & (0.15, $7.7\times10^{-3}$)   & (0.33, $1.6\times10^{-9}$)   & weak correlation           \\
$p_0$ - $\theta_p$ background sources                & (-0.041, $3.7\times10^{-1}$) & (0.0020, $9.7\times10^{-1}$) & no significant correlation \\
$p_0$ - $\theta_p$ multi-component sources           & (0.083, $1.28\times10^{-1}$) & (0.116, $3.35\times10^{-2}$) & no significant correlation \\
$p_0$ - $\theta_p$ single-component sources          & (0.001, $9.87\times10^{-1}$) & (0.183, $6.62\times10^{-5}$) & no significant correlation \\
Total Flux - $\sigma_\mathrm{RM}$ cluster members    & (-0.19, $1.0\times10^{-3}$)  & (-0.15, $8.1\times10^{-3}$)  & weak correlation           \\
Total Flux - $\sigma_\mathrm{RM}$ background sources & (0.0060, $9.0\times10^{-1}$) & (0.062, $1.8\times10^{-1}$)  & no significant correlation \\
$p_0$ - $\sigma_\mathrm{RM}$ cluster members         & (0.097, $8.8\times10^{-2}$)  & (0.045, $4.3\times10^{-1}$)  & no significant correlation \\
$p_0$ - $\sigma_\mathrm{RM}$ background sources      & (0.25, $5.2\times10^{-8}$)   & (0.19, $4.0\times10^{-5}$)   & weak correlation           \\ \bottomrule
\end{tabular}
\tablefoot{\tablefoottext{a}{A cutoff $p$-value of 0.05 is used for statistical significance, and the correlation is defined as weak  for values of Pearson $|r| \leq 0.3$.}}
\end{table*}

\begin{figure*}
    \centering
    \includegraphics[width=1.0\textwidth]{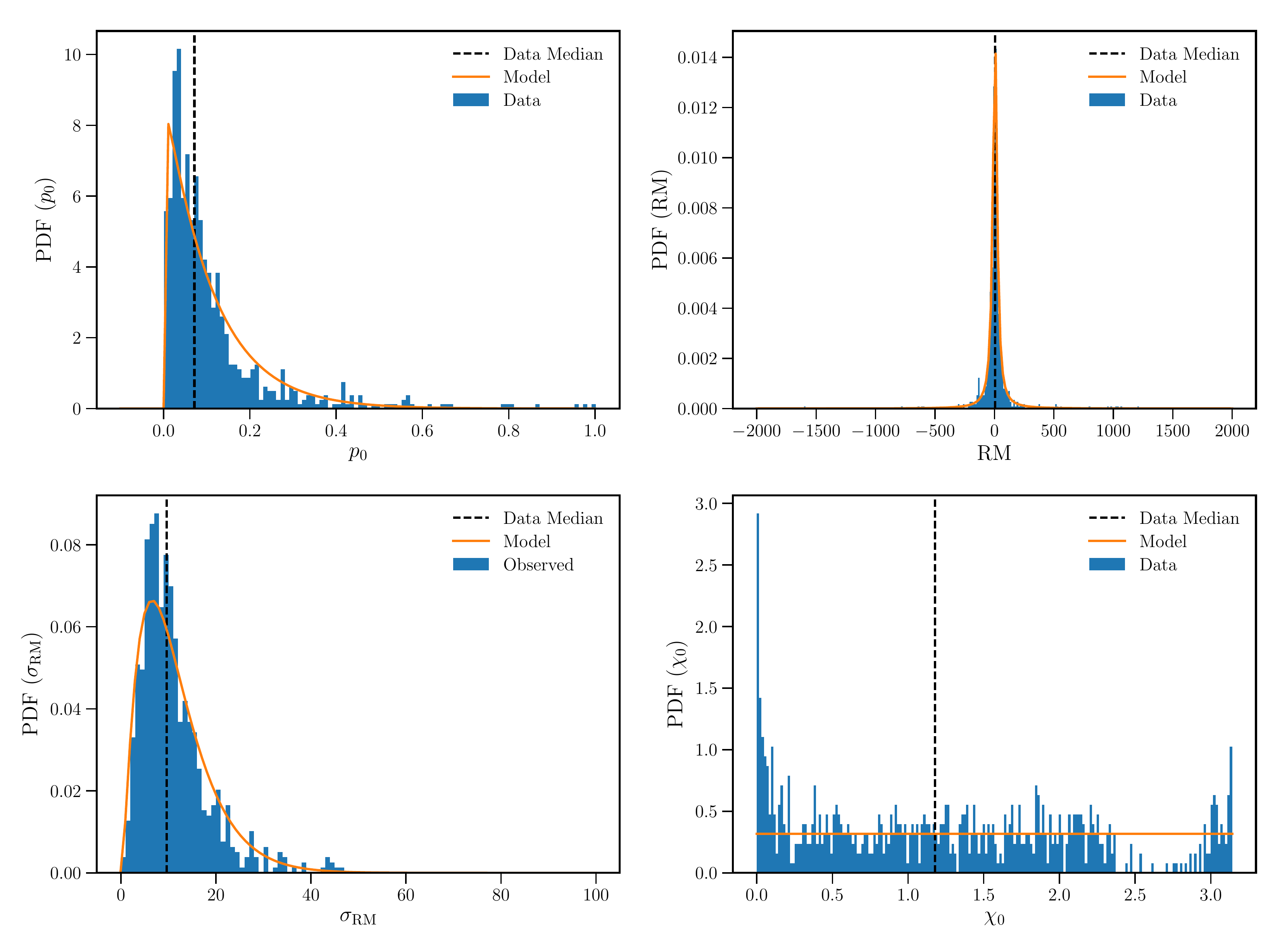}
    \vspace{-0.4cm}
    \caption{Probability distribution functions used to generate polarised radio sources during the Monte Carlo analysis shown in orange. The distribution of the real data is shown in blue. }
    \label{fig:montecarloparams}
    \centering
    \includegraphics[width=1.0\columnwidth]{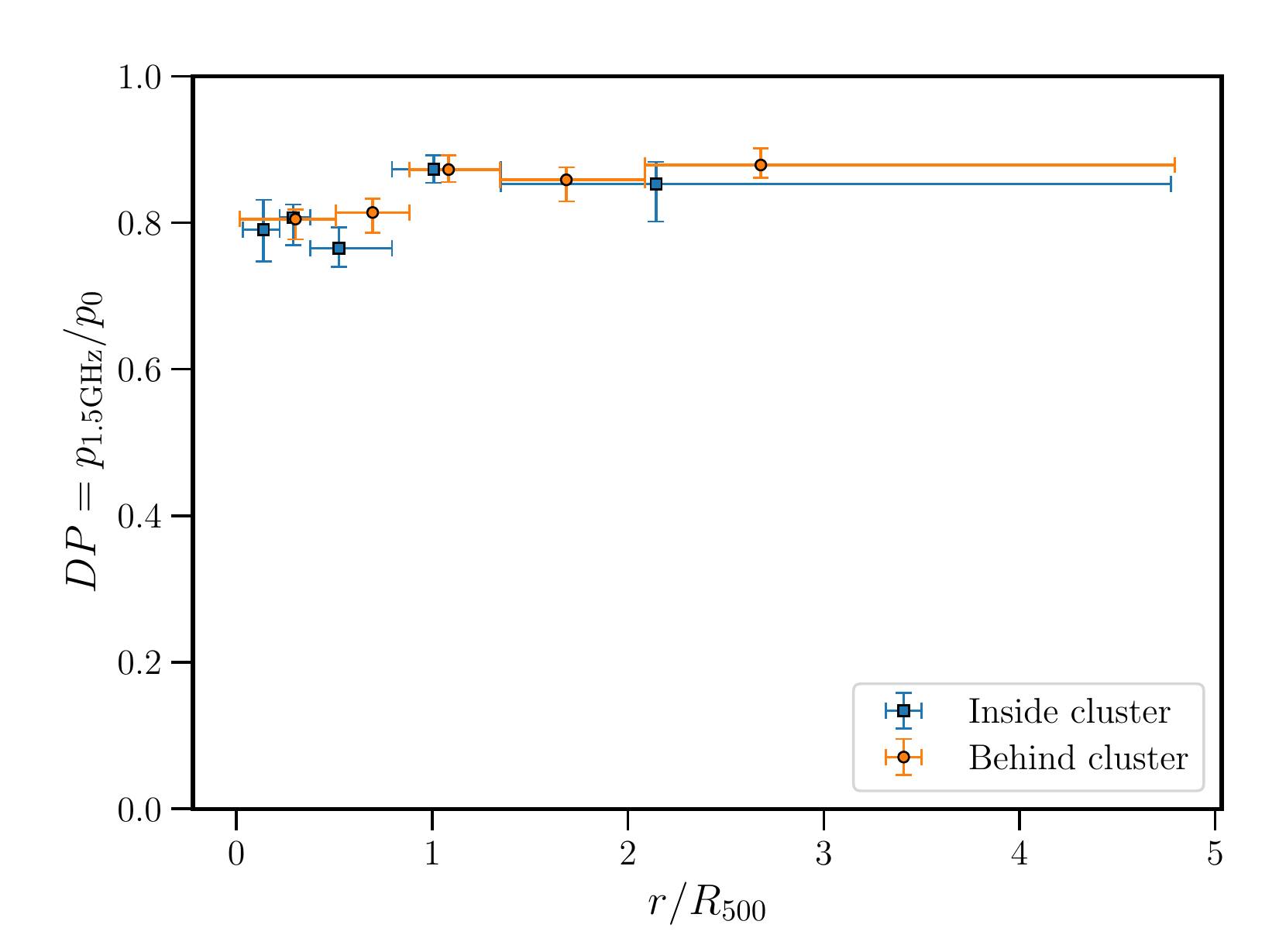}
    \vspace{-0.2cm}
    \caption{Median of the Kaplan-Meier estimate of the depolarisation ratio survival function in different bins of radius for simulated data. The bin width is chosen such that each bin contains an equal number of polarised sources and is denoted by the horizontal lines. The points are plotted at the median radius in each bin}
    \label{fig:DP_rnorm_SIM}
\end{figure*}

The previous section showed that there are no strong trends detected between source properties and polarisation properties or angular radius to the pointing centre, but there are weak trends in the data that possibly bias the results. Most notably, we have seen that the brightest cluster members are preferentially detected in the centre of the images and it is this class of sources that shows the most depolarisation. To quantify the bias introduced by selection effects and the fitting procedure, we took a Monte Carlo approach, simulating polarised radio sources with random properties that are taken from distributions that are representative of the data.  If, through the effect of the choices made during the analysis or because of the radio source properties such as size and flux density a bias is introduced in the depolarisation curve, we should find that bias when employing the same methods on a sample of completely randomly (de)polarised sources. The distributions used to generate random polarised sources for this experiment are shown in Figure \ref{fig:montecarloparams}, where we have fit gamma distributions to the strictly positive values $p_0$ and $\sigma_\mathrm{RM}$ and used a Cauchy distribution for RM to account for the large peak around RM=0. The initial polarisation angle $\chi_0$ was simply drawn from a uniform distribution between 0 and $\pi$.
The steps taken in the Monte Carlo experiment were as follows. First, we calculated the total flux density per channel of all detected total intensity sources. Then, for each total intensity source, we randomly drew a value of $p_0$, RM, $\sigma_\mathrm{RM}$ and $\chi_0$ from the representative probability distribution functions. Third, we computed simulated Stokes Q and U emission using the external depolarisation model given in Eq. \ref{eq:extdepol} with the randomly drawn parameters. Then, we determined which sources are detected in polarisation at a 5$\sigma$ level given the varying background noise due to the primary beam response. Finally, we fitted the detected sources with the MCMC IQU fitting code and calculated upper limits on the undetected sources as explained in Section \ref{sec:fracpolmeasurement}. The resulting best-fit parameters and upper limits were again used to find the median depolarisation in bins of projected radius.

This approach resulted in 1050 simulated sources detected in polarisation.
The resulting median depolarisation trend with projected distance to the cluster centre is shown in Figure \ref{fig:DP_rnorm_SIM} for simulated sources. Figure \ref{fig:DP_rnorm_SIM} shows that there is indeed a very small bias from selection effects or fitting, with a minor trend showing slightly more depolarisation near the cluster centres than at the cluster outskirts. However, this trend is only a small fraction of the real detected trend in Figure \ref{fig:KMestimate_redshift}. Thus, even though there are significant correlations as shown in Table \ref{tab:properties1}, they cause only a minimal bias because they have small correlation coefficients. 
This means that if the observed sources were a population with random polarisation parameters, there would not be a strongly detected depolarisation trend with radius. The striking difference between Figure \ref{fig:DP_rnorm_SIM} and the depolarisation trend detected in the real data shown in Figure \ref{fig:KMestimate_redshift} makes a strong case that the observed depolarisation trend cannot be explained only by selection effects or biases.


\FloatBarrier

\section{Full sample plots}\label{sec:AppendixB}
\begin{figure*}[htb]
    \centering
    \includegraphics[width=1.0\textwidth]{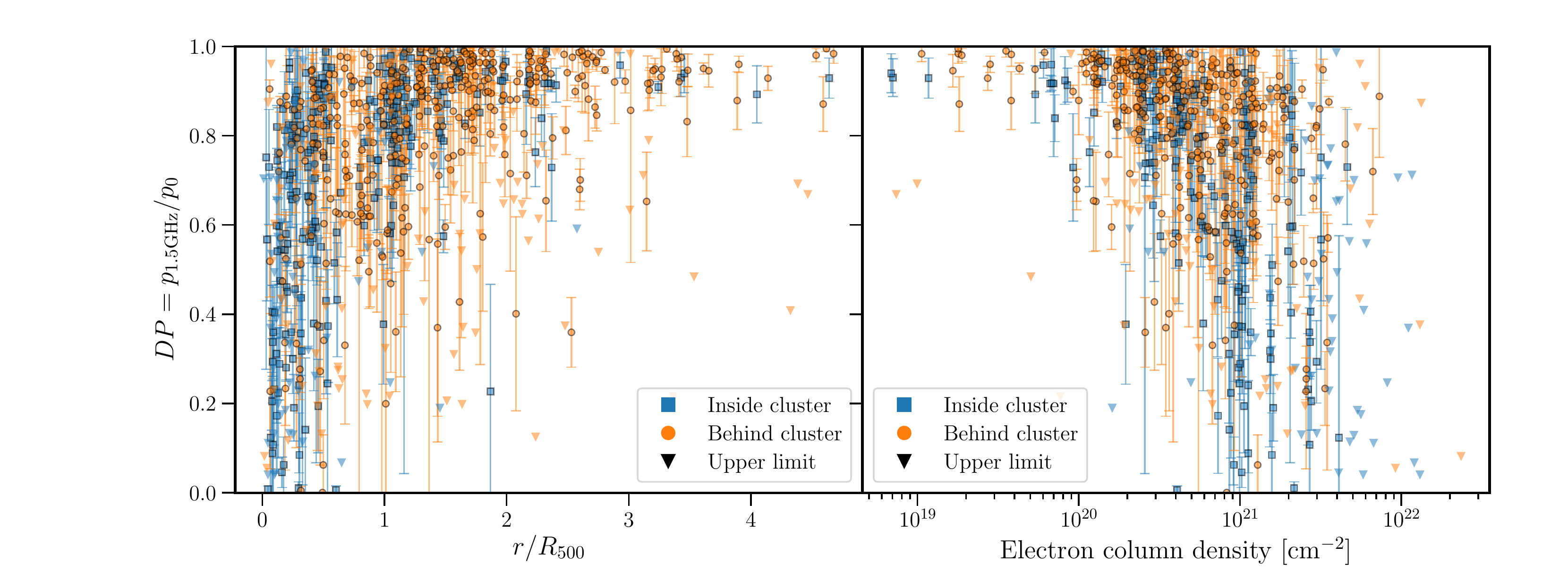}
    \caption{Depolarisation ratio and relevant upper limits for all sources as a function of projected radius and column density. Points are coloured by their position along the line-of-sight with respect to the nearest cluster.}
    \label{fig:DP_vs_rnorm_sourcez_scatter}
\end{figure*}

\begin{figure*}[htb]
    \centering
    \includegraphics[width=1.0\textwidth]{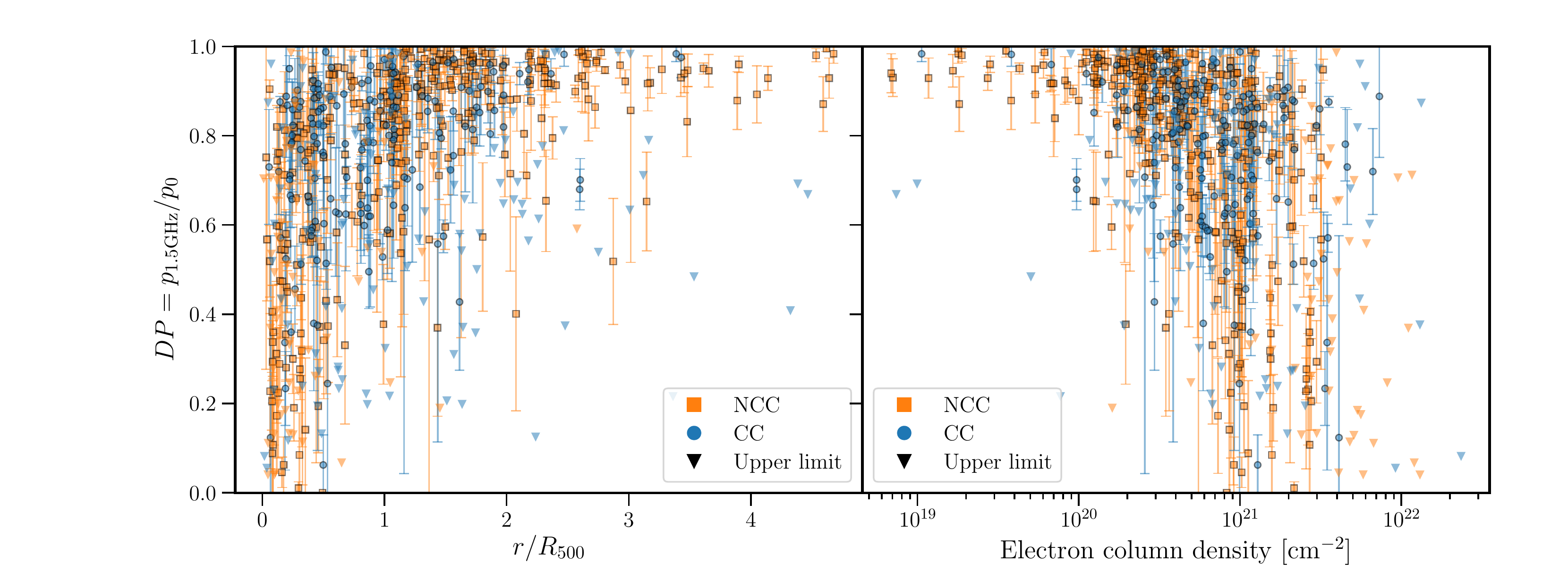}
    \caption{Depolarisation ratio and relevant upper limits for all sources as a function of projected radius and column density. Points are coloured by the dynamic state of the cluster, indicated by non-cool-core (NCC) and cool-core (CC).}
    \label{fig:DP_vs_dynstate_scatter}
\end{figure*}

\begin{figure*}[htb]
    \centering
    \includegraphics[width=1.0\textwidth]{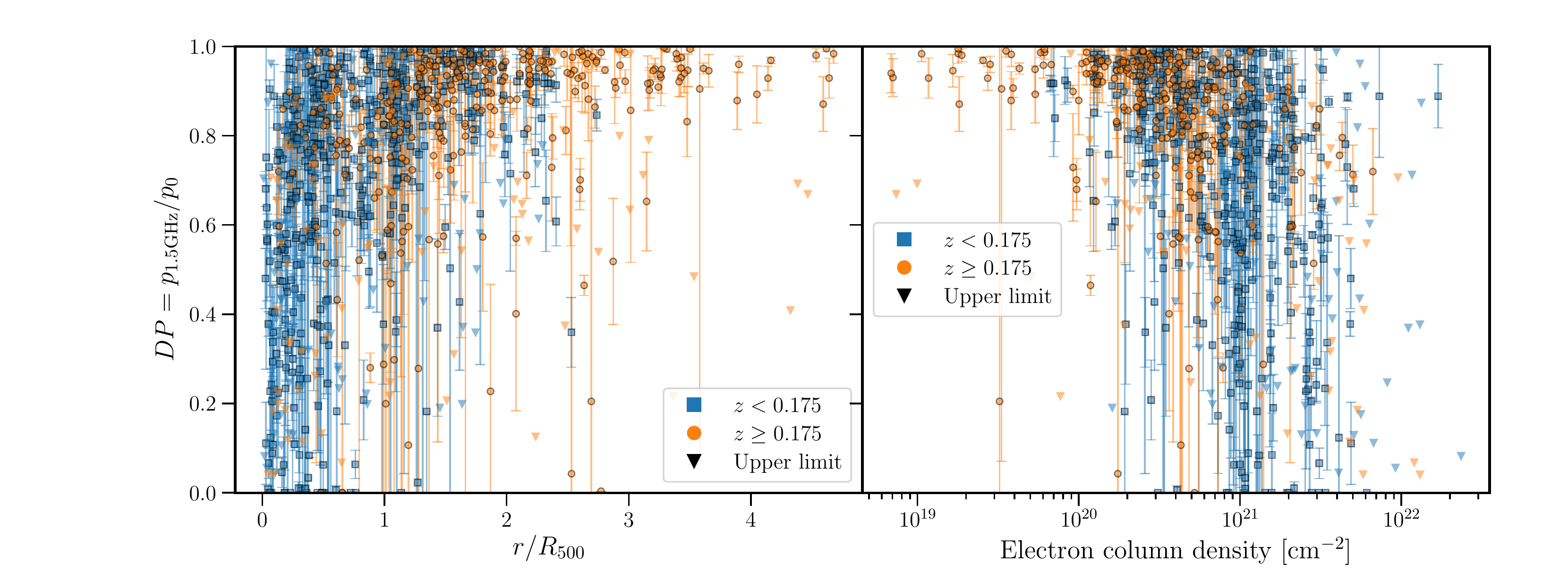}
    \caption{Depolarisation ratio and relevant upper limits for all sources as a function of projected radius and column density. Points are coloured according to the redshift of the cluster.}
    \label{fig:DP_vs_clusterz_scatter}
\end{figure*}

\begin{figure}[htb]
    \centering
    \includegraphics[width=1.0\columnwidth]{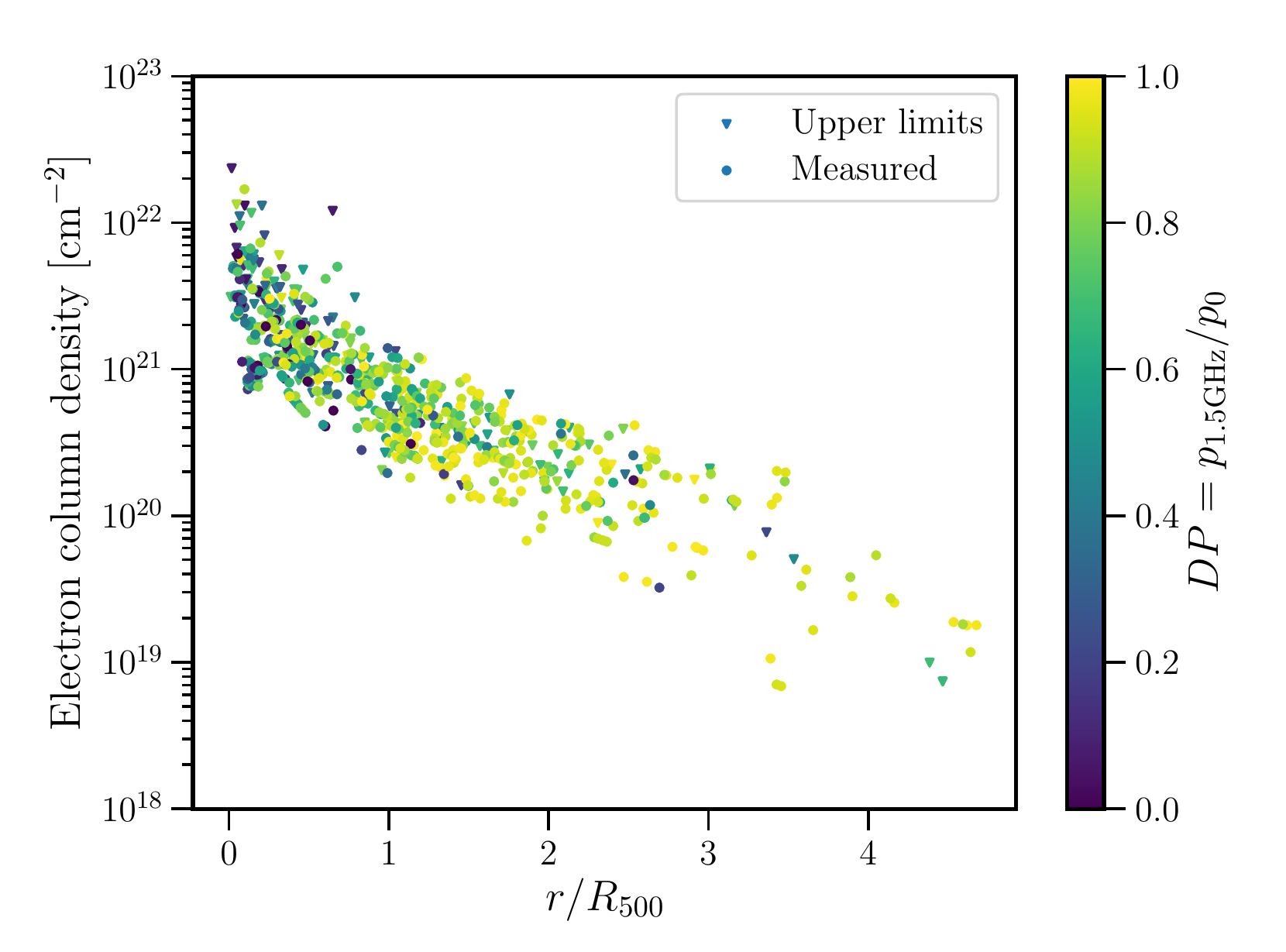}
    \caption{Full sample of measured values and upper limits on the depolarisation of radio sources as a function of both electron column density and projected radius. The points are coloured by the depolarisation value.}
    \label{fig:fullsample}
\end{figure}

\begin{figure}[htb]
    \centering
    \includegraphics[width=1.0\columnwidth]{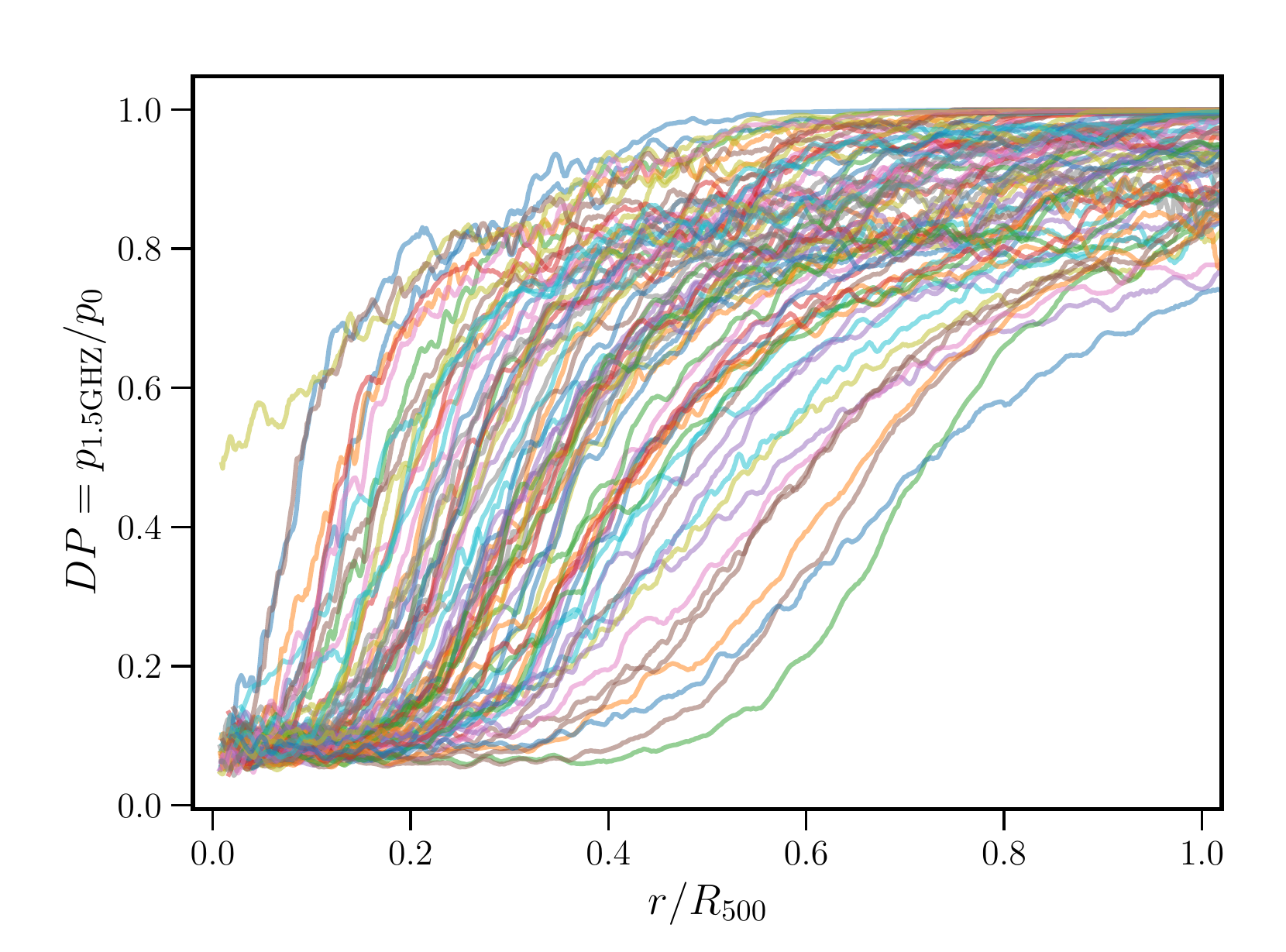}
    \caption{All simulated depolarisation ratio profiles for the full sample of simulated clusters. The assumed parameters for the magnetic field are $B_0=5.0\, \mu$G, $n=3$ and $\eta=0.5$.}
    \label{fig:all_models}
\end{figure}

For completeness, we plot in Figures \ref{fig:DP_vs_rnorm_sourcez_scatter} to \ref{fig:DP_vs_clusterz_scatter} the full sample of data-points that have been summarised with the KM estimator in Figures \ref{fig:KMestimate_redshift} to \ref{fig:redshiftbins}. Figure \ref{fig:fullsample} shows an alternative visualisation of the full sample of data points, which also clearly shows the trend of sources becoming more depolarised as the projected radius and column densities increase. Finally, Figure \ref{fig:all_models} shows the modelled depolarisation profiles for all clusters where X-ray data was available, for a single set of magnetic field parameters.  

\clearpage
\newpage

\section{Polarised source catalogue}

\setcounter{table}{0}
\renewcommand*\thetable{\Alph{section}.\arabic{table}}
\begingroup
\fontsize{6pt}{8pt}\selectfont 
\vfill
\begin{sideways}
  \setlength{\tabcolsep}{3pt}
  \begin{threeparttable}
    \caption{First 30 rows of the catalogue of 819 polarised radio sources that were detected in this work. The full table is available in electronic form at \red{TODO}.}
    \label{tab:polsources}
\begin{tabular}{@{}llllllllllllllllllllll@{}}
\toprule
\begin{tabular}[c]{@{}l@{}}RA\\ {[deg]}\end{tabular} & \begin{tabular}[c]{@{}l@{}}DEC\\ {[deg]}\end{tabular} & \begin{tabular}[c]{@{}l@{}}Maj\\ {[$^{\prime\prime}$]}\end{tabular} & \begin{tabular}[c]{@{}l@{}}Min\\ {[$^{\prime\prime}$]}\end{tabular} & \begin{tabular}[c]{@{}l@{}}PA\\ {[deg]}\end{tabular} & $p_0$ & \begin{tabular}[c]{@{}l@{}}$\chi_0$\\ {[rad]}\end{tabular} & \begin{tabular}[c]{@{}l@{}}RM\\ {[rad m$^{-2}$]}\end{tabular} & \begin{tabular}[c]{@{}l@{}}$\sigma_\mathrm{RM}$\\ {[rad m$^{-2}$]}\end{tabular} & \begin{tabular}[c]{@{}l@{}}$I_0$\\ {[mJy]}\end{tabular} & $\alpha$ & $\chi^2_\mathrm{QU}$ & $z_\mathrm{best}$ & $z_\mathrm{best}$ source & \begin{tabular}[c]{@{}l@{}}$\theta_p$\\ {[arcmin]}\end{tabular} & $r/R_{500}$ & Cluster & \begin{tabular}[c]{@{}l@{}}RA$_\mathrm{opt}$\\ {[deg]}\end{tabular} & \begin{tabular}[c]{@{}l@{}}DEC$_\mathrm{opt}$\\ {[deg]}\end{tabular} & Multi component & Flagged & Note \\ \midrule
-132.2610 & 5.6140 & 8.2 & 5.8 & 90 & $0.102_{-0.005}^{+0.005}$ & $0.911_{-0.036}^{+0.036}$ & $-137_{-1}^{+1}$ & $10_{-0}^{+0}$ & $25.0_{-0.3}^{+0.3}$ & $-0.81_{-0.03}^{+0.03}$ & 171 & $0.584 \pm 0.040$ & 2 & $8.06$ & $0.6$ & G006.47+50.54 & 227.7304 & 5.6153 & True & True &  \\
-132.2780 & 5.6184 & 4.9 & 4.0 & 52 & $0.040_{-0.008}^{+0.010}$ & $3.030_{-0.137}^{+0.080}$ & $-68_{-2}^{+3}$ & $7_{-2}^{+3}$ & $5.6_{-0.1}^{+0.1}$ & $-0.95_{-0.05}^{+0.05}$ & 56 & $0.584 \pm 0.040$ & 2 & $7.80$ & $0.6$ & G006.47+50.54 & 227.7304 & 5.6153 & True & False &  \\
-132.2767 & 5.6172 & 5.4 & 3.8 & 68 & $0.048_{-0.012}^{+0.015}$ & $0.091_{-0.067}^{+0.131}$ & $-113_{-4}^{+3}$ & $13_{-3}^{+4}$ & $7.3_{-0.1}^{+0.1}$ & $-0.98_{-0.04}^{+0.04}$ & 66 & $0.584 \pm 0.040$ & 2 & $7.87$ & $0.6$ & G006.47+50.54 & 227.7304 & 5.6153 & True & False &  \\
-111.9212 & 5.5698 & 6.2 & 5.7 & 59 & $0.058_{-0.009}^{+0.010}$ & $1.943_{-0.130}^{+0.125}$ & $8_{-3}^{+4}$ & $17_{-2}^{+2}$ & $20.2_{-0.2}^{+0.2}$ & $-1.20_{-0.03}^{+0.03}$ & 75 & $0.581 \pm 1.268$ & 2 & $6.10$ & $0.8$ & G021.09+33.25 & 248.0775 & 5.5694 & False & False & Counterpart unsure \\
-111.9898 & 5.4658 & 7.8 & 5.9 & 41 & $0.033_{-0.003}^{+0.003}$ & $2.209_{-0.124}^{+0.092}$ & $4_{-2}^{+2}$ & $3_{-1}^{+2}$ & $25.8_{-0.3}^{+0.3}$ & $-0.86_{-0.03}^{+0.03}$ & 104 & $1.675 \pm 0.611$ & 2 & $12.44$ & $1.7$ & G021.09+33.25 & 248.0106 & 5.4644 & True & True &  \\
-111.7294 & 5.5442 & 12.0 & 6.5 & 76 & $0.561_{-0.091}^{+0.100}$ & $2.254_{-2.220}^{+0.442}$ & $-36_{-8}^{+10}$ & $15_{-1}^{+1}$ & $3.2_{-0.2}^{+0.2}$ & $-2.30_{-0.18}^{+0.17}$ & 151 & - & - & $6.00$ & $0.8$ & G021.09+33.25 & - & - & False & True & No optical counterpart. \\
-111.9211 & 5.4755 & 7.1 & 6.3 & 39 & $0.039_{-0.003}^{+0.003}$ & $0.458_{-0.080}^{+0.078}$ & $28_{-2}^{+2}$ & $14_{-1}^{+1}$ & $38.0_{-0.4}^{+0.4}$ & $-0.43_{-0.03}^{+0.03}$ & 131 & $2.389 \pm 0.208$ & 2 & $8.98$ & $1.2$ & G021.09+33.25 & - & - & False & True & No optical counterpart. \\
-111.9888 & 5.4639 & 7.7 & 6.2 & 49 & $0.033_{-0.002}^{+0.002}$ & $0.015_{-0.011}^{+0.023}$ & $-2_{-1}^{+0}$ & $3_{-1}^{+2}$ & $27.8_{-0.3}^{+0.3}$ & $-0.77_{-0.03}^{+0.03}$ & 152 & $1.675 \pm 0.611$ & 2 & $12.46$ & $1.7$ & G021.09+33.25 & 248.0106 & 5.4644 & True & True &  \\
-48.0417 & -17.7068 & 7.7 & 7.1 & 111 & $0.053_{-0.011}^{+0.014}$ & $0.914_{-0.303}^{+0.302}$ & $-31_{-6}^{+6}$ & $7_{-2}^{+3}$ & $5.5_{-0.1}^{+0.2}$ & $-1.63_{-0.07}^{+0.07}$ & 57 & - & - & $8.50$ & $1.4$ & G028.77-33.56 & - & - & False & True & No optical counterpart. \\
-47.9835 & -17.7248 & 7.2 & 7.2 & 103 & $0.093_{-0.023}^{+0.028}$ & $0.961_{-0.364}^{+0.354}$ & $-24_{-8}^{+9}$ & $14_{-2}^{+3}$ & $4.6_{-0.1}^{+0.1}$ & $-2.04_{-0.09}^{+0.09}$ & 73 & - & - & $5.66$ & $0.9$ & G028.77-33.56 & - & - & False & False & No optical counterpart. \\
-48.0167 & -17.7220 & 7.8 & 6.5 & 89 & $0.023_{-0.007}^{+0.008}$ & $1.854_{-0.437}^{+0.389}$ & $-42_{-9}^{+9}$ & $11_{-3}^{+3}$ & $11.9_{-0.2}^{+0.2}$ & $-1.51_{-0.04}^{+0.04}$ & 59 & - & - & $6.87$ & $1.1$ & G028.77-33.56 & - & - & False & False & No optical counterpart. \\
-119.4219 & 16.0547 & 8.8 & 8.5 & 27 & $0.108_{-0.003}^{+0.004}$ & $1.829_{-0.030}^{+0.031}$ & $-136_{-1}^{+1}$ & $14_{-0}^{+0}$ & $40.2_{-0.6}^{+0.6}$ & $-0.78_{-0.03}^{+0.03}$ & 215 & $0.282 \pm 0.077$ & 2 & $0.95$ & $0.0$ & G029.00+44.56 & 240.5791 & 16.0553 & False & True & Counterpart unsure \\
-119.3140 & 16.1258 & 7.9 & 5.5 & 74 & $0.020_{-0.006}^{+0.535}$ & $1.718_{-0.568}^{+0.472}$ & $-100_{-12}^{+24}$ & $8_{-3}^{+2997}$ & $11.3_{-0.2}^{+0.3}$ & $-0.96_{-0.05}^{+0.05}$ & 68 & - & - & $8.37$ & $0.3$ & G029.00+44.56 & - & - & False & True & No optical counterpart. \\
-119.6253 & 16.0712 & 8.4 & 8.0 & 94 & $0.073_{-0.007}^{+0.010}$ & $0.012_{-0.010}^{+2.539}$ & $4_{-1}^{+10}$ & $4_{-1}^{+2}$ & $14.1_{-0.3}^{+0.3}$ & $-0.77_{-0.06}^{+0.06}$ & 235 & $1.285 \pm 0.435$ & 2 & $10.83$ & $0.4$ & G029.00+44.56 & - & - & False & True & No optical counterpart. \\
-119.4438 & 15.8343 & 7.9 & 6.2 & 40 & $0.024_{-0.005}^{+0.007}$ & $1.497_{-0.236}^{+0.249}$ & $34_{-4}^{+4}$ & $6_{-2}^{+4}$ & $16.7_{-0.4}^{+0.3}$ & $-1.07_{-0.05}^{+0.06}$ & 61 & - & - & $13.16$ & $0.5$ & G029.00+44.56 & - & - & False & False & No optical counterpart. \\
-119.4234 & 16.1195 & 8.5 & 8.3 & 66 & $0.050_{-0.005}^{+0.004}$ & $2.229_{-0.111}^{+0.056}$ & $141_{-2}^{+1}$ & $10_{-1}^{+1}$ & $32.6_{-0.5}^{+0.5}$ & $-0.86_{-0.03}^{+0.03}$ & 163 & $0.385 \pm 0.000$ & 0 & $4.05$ & $0.2$ & G029.00+44.56 & 240.5783 & 16.1194 & False & True &  \\
-152.5871 & 26.5863 & 6.1 & 5.3 & 137 & $0.084_{-0.018}^{+0.023}$ & $0.840_{-0.220}^{+0.217}$ & $-26_{-4}^{+4}$ & $8_{-2}^{+3}$ & $3.8_{-0.1}^{+0.1}$ & $-1.17_{-0.11}^{+0.11}$ & 57 & - & - & $9.07$ & $0.6$ & G033.78+77.16 & - & - & False & False & No optical counterpart. \\
-152.7815 & 26.5929 & 18.0 & 9.2 & 176 & $0.001_{-0.000}^{+0.000}$ & $1.239_{-0.109}^{+0.108}$ & $21_{-2}^{+2}$ & $7_{-2}^{+2}$ & $1238.8_{-14.5}^{+14.5}$ & $-0.99_{-0.03}^{+0.03}$ & 733 & $0.250 \pm 0.114$ & 2 & $1.45$ & $0.1$ & G033.78+77.16 & 207.2186 & 26.5931 & False & True &  \\
-152.6178 & 26.5824 & 6.1 & 6.0 & 176 & $0.098_{-0.034}^{+0.062}$ & $0.941_{-0.519}^{+0.557}$ & $54_{-26}^{+19}$ & $33_{-4}^{+6}$ & $24.2_{-0.3}^{+0.3}$ & $-0.51_{-0.03}^{+0.03}$ & 107 & $0.793 \pm 0.078$ & 2 & $7.43$ & $0.5$ & G033.78+77.16 & 207.3823 & 26.5824 & False & False &  \\
-152.5220 & 26.5313 & 7.0 & 6.1 & 82 & $0.045_{-0.006}^{+0.006}$ & $1.807_{-0.117}^{+0.117}$ & $2_{-3}^{+3}$ & $14_{-1}^{+1}$ & $34.8_{-0.4}^{+0.5}$ & $-0.86_{-0.03}^{+0.03}$ & 80 & $1.187 \pm 0.378$ & 2 & $12.97$ & $0.9$ & G033.78+77.16 & - & - & False & False & No optical counterpart. \\
-152.8105 & 26.5899 & 5.4 & 5.0 & 68 & $0.031_{-0.006}^{+0.008}$ & $2.863_{-0.222}^{+0.177}$ & $1024_{-4}^{+5}$ & $7_{-2}^{+4}$ & $3.7_{-0.1}^{+0.1}$ & $0.02_{-0.06}^{+0.05}$ & 71 & $0.570 \pm 0.000$ & 0 & $2.93$ & $0.2$ & G033.78+77.16 & 207.1893 & 26.5903 & False & False &  \\
-152.5875 & 26.6306 & 7.5 & 6.9 & 62 & $0.046_{-0.003}^{+0.003}$ & $0.003_{-0.002}^{+0.005}$ & $-43_{-0}^{+0}$ & $4_{-1}^{+1}$ & $31.4_{-0.4}^{+0.4}$ & $-0.89_{-0.03}^{+0.03}$ & 160 & - & - & $9.46$ & $0.6$ & G033.78+77.16 & - & - & False & True & No optical counterpart. \\
-2.0992 & -25.9382 & 7.7 & 5.8 & 102 & $0.038_{-0.007}^{+0.009}$ & $3.073_{-2.581}^{+0.054}$ & $-18_{-9}^{+2}$ & $7_{-2}^{+3}$ & $16.5_{-0.2}^{+0.2}$ & $-0.55_{-0.03}^{+0.04}$ & 68 & $0.187 \pm 0.133$ & 3 & $8.47$ & $2.0$ & G034.03-76.59 & 357.9006 & -25.9388 & False & False &  \\
-89.0409 & 10.0803 & 7.5 & 4.7 & 66 & $0.030_{-0.007}^{+0.011}$ & $0.713_{-0.213}^{+0.215}$ & $58_{-4}^{+4}$ & $8_{-2}^{+5}$ & $11.6_{-0.2}^{+0.2}$ & $-0.93_{-0.05}^{+0.04}$ & 82 & $0.559 \pm 0.089$ & 3 & $10.82$ & $1.7$ & G036.72+14.92 & - & - & False & False & No optical counterpart. \\
-88.7551 & 10.1938 & 7.6 & 5.5 & 150 & $0.022_{-0.006}^{+0.008}$ & $0.267_{-0.200}^{+2.351}$ & $-1_{-5}^{+8}$ & $7_{-2}^{+5}$ & $13.1_{-0.2}^{+0.2}$ & $-0.80_{-0.05}^{+0.05}$ & 70 & $0.416 \pm 0.214$ & 3 & $11.41$ & $1.8$ & G036.72+14.92 & 271.2446 & 10.1943 & False & False &  \\
-88.8378 & 9.9030 & 6.4 & 5.3 & 59 & $0.022_{-0.003}^{+0.003}$ & $0.453_{-0.153}^{+0.154}$ & $47_{-3}^{+3}$ & $5_{-2}^{+3}$ & $11.3_{-0.2}^{+0.2}$ & $-0.05_{-0.04}^{+0.04}$ & 77 & $0.186 \pm 0.073$ & 3 & $8.13$ & $1.3$ & G036.72+14.92 & 271.1620 & 9.9031 & False & False &  \\
-88.7312 & 9.9649 & 8.3 & 4.7 & 48 & $0.218_{-0.057}^{+0.050}$ & $2.139_{-2.081}^{+0.308}$ & $19_{-12}^{+5}$ & $4_{-2}^{+3}$ & $1.2_{-0.2}^{+0.2}$ & $-1.02_{-0.35}^{+0.35}$ & 81 & - & - & $8.90$ & $1.4$ & G036.72+14.92 & 271.2703 & 9.9618 & True & False &  \\
-88.7910 & 10.1261 & 13.6 & 7.3 & 57 & $0.108_{-0.020}^{+0.023}$ & $2.184_{-2.099}^{+0.132}$ & $13_{-17}^{+3}$ & $11_{-1}^{+2}$ & $13.5_{-0.3}^{+0.3}$ & $-1.63_{-0.06}^{+0.06}$ & 108 & $0.130 \pm 0.032$ & 3 & $6.88$ & $1.1$ & G036.72+14.92 & 271.2022 & 10.1263 & True & False &  \\
-88.7946 & 10.1282 & 13.6 & 6.4 & 67 & $0.030_{-0.005}^{+0.005}$ & $0.098_{-0.069}^{+0.106}$ & $-16_{-2}^{+2}$ & $9_{-2}^{+2}$ & $29.6_{-0.4}^{+0.4}$ & $-1.30_{-0.03}^{+0.03}$ & 99 & $0.130 \pm 0.032$ & 3 & $6.85$ & $1.1$ & G036.72+14.92 & 271.2022 & 10.1263 & True & False &  \\
-88.8257 & 10.0041 & 14.9 & 8.0 & 172 & $0.018_{-0.000}^{+0.000}$ & $0.988_{-0.006}^{+0.006}$ & $-23_{-0}^{+0}$ & $3_{-0}^{+0}$ & $836.2_{-9.4}^{+9.5}$ & $-0.81_{-0.03}^{+0.03}$ & 10001 & - & - & $2.94$ & $0.5$ & G036.72+14.92 & 271.1741 & 10.0012 & True & True &  \\ \bottomrule
\end{tabular}
\tablefoot{The columns `Maj' and `Min' indicate the major and minor axis of the polarised sources and `PA' the position angle. The results of the fitting as described in Section \ref{sec:fracpolmeasurement} are given in the columns $p_0$ to $\chi^2_\mathrm{QU}$ with uncertainties given by the 16th and 84th percentile of the MCMC chains. The columns starting with $z$ are the results of the redshift estimates as detailed in Section \ref{sec:redshifts}. The angular distance to the pointing centre and projected radius to the cluster centre are given by the $\theta_p$ and $r/R_{500}$ columns respectively. The `Flagged' column indicates True when the source was omitted for the final analysis, for example because it was a bad fit. }
  \end{threeparttable}
\end{sideways}
\vfill
\restoregeometry
\endgroup

\section{Full source catalogue}

\setcounter{table}{0}
\renewcommand*\thetable{\Alph{section}.\arabic{table}}
\begingroup
\fontsize{6pt}{8pt}\selectfont 
\vfill
\begin{sideways}
  \setlength{\tabcolsep}{3pt}
  \begin{threeparttable}
    \caption{First 30 rows of the catalogue of 6\,807 total intensity sources that were detected in this work. The full table is available in electronic form at \red{TODO}.}
    \label{tab:unpolsources}
\begin{tabular}{@{}llllllllllllllllll@{}}
\toprule
\begin{tabular}[c]{@{}l@{}}RA\\ {[deg]}\end{tabular} & \begin{tabular}[c]{@{}l@{}}DEC\\ {[deg]}\end{tabular} & \begin{tabular}[c]{@{}l@{}}Maj\\ {[$^{\prime\prime}$]}\end{tabular} & \begin{tabular}[c]{@{}l@{}}Min\\ {[$^{\prime\prime}$]}\end{tabular} & \begin{tabular}[c]{@{}l@{}}PA\\ {[deg]}\end{tabular} & \begin{tabular}[c]{@{}l@{}}Total flux\\ {[mJy]}\end{tabular} & \begin{tabular}[c]{@{}l@{}}Peak Flux\\ {[mJy beam$^{-1}$]}\end{tabular} & $p_\mathrm{1.5GHz}$ uplim & $z_\mathrm{best}$ & $z_\mathrm{best}$ source & \begin{tabular}[c]{@{}l@{}}$\theta_p$\\ {[arcmin]}\end{tabular} & $r/R_{500}$ & Cluster & \begin{tabular}[c]{@{}l@{}}RA$_\mathrm{opt}$\\ {[deg]}\end{tabular} & \begin{tabular}[c]{@{}l@{}}DEC$_\mathrm{opt}$\\ {[deg]}\end{tabular} & Visual counterpart id & Multi component & Note \\ \midrule
19.7206 & -26.9792 & 13.6 & 10.0 & 142 & $1.059 \pm 0.206$ & $0.587 \pm 0.078$ & 0.48 & - & - & $9.39$ & $2.1$ & G212.97-84.04 & - & - & False & False &  \\
19.6823 & -27.0611 & 9.7 & 6.8 & 99 & $0.328 \pm 0.100$ & $0.376 \pm 0.062$ & 0.87 & - & - & $8.99$ & $2.2$ & G212.97-84.04 & 19.6824 & -27.0613 & False & False &  \\
19.6530 & -26.9383 & 8.4 & 7.9 & 73 & $0.721 \pm 0.090$ & $0.814 \pm 0.056$ & 0.32 & - & - & $6.17$ & $1.3$ & G212.97-84.04 & 19.6528 & -26.9385 & False & False &  \\
19.6515 & -26.9198 & 9.4 & 8.8 & 113 & $1.106 \pm 0.109$ & $1.008 \pm 0.059$ & 0.24 & - & - & $6.57$ & $1.4$ & G212.97-84.04 & - & - & False & False &  \\
19.6284 & -27.0460 & 9.2 & 9.0 & 178 & $3.411 \pm 0.108$ & $3.120 \pm 0.059$ & 0.08 & $0.871 \pm 0.162$ & 3 & $6.19$ & $1.6$ & G212.97-84.04 & 19.6285 & -27.0461 & False & False &  \\
19.6115 & -27.0112 & 9.8 & 8.7 & 106 & $0.401 \pm 0.100$ & $0.357 \pm 0.053$ & 0.54 & - & - & $4.18$ & $1.1$ & G212.97-84.04 & 19.6110 & -27.0110 & False & False &  \\
19.6104 & -27.0795 & 13.2 & 9.0 & 63 & $0.500 \pm 0.151$ & $0.318 \pm 0.063$ & 0.92 & - & - & $7.21$ & $1.8$ & G212.97-84.04 & - & - & False & False &  \\
19.6064 & -26.7558 & 9.9 & 8.6 & 112 & $1.004 \pm 0.194$ & $0.891 \pm 0.103$ & 0.28 & - & - & $13.52$ & $2.9$ & G212.97-84.04 & 19.6062 & -26.7559 & False & False &  \\
19.6056 & -27.0319 & 9.8 & 8.9 & 10 & $0.881 \pm 0.102$ & $0.757 \pm 0.053$ & 0.23 & - & - & $4.73$ & $1.2$ & G212.97-84.04 & 19.6054 & -27.0319 & False & False &  \\
19.5740 & -26.9712 & 24.0 & 10.2 & 69 & $44.649 \pm 0.373$ & $16.358 \pm 0.073$ & 0.01 & - & - & $1.57$ & $0.4$ & G212.97-84.04 & 19.5700 & -26.9724 & True & False &  \\
19.5825 & -26.9747 & 15.9 & 11.9 & 176 & $1.901 \pm 0.261$ & $0.757 \pm 0.077$ & 0.20 & - & - & $2.01$ & $0.5$ & G212.97-84.04 & - & - & True & False & No optical counterpart. \\
19.5664 & -27.1593 & 11.3 & 9.8 & 63 & $5.909 \pm 0.197$ & $4.007 \pm 0.087$ & 0.06 & - & - & $11.14$ & $2.8$ & G212.97-84.04 & 19.5606 & -27.1622 & True & True &  \\
19.5616 & -27.1625 & 18.3 & 11.2 & 63 & $1.635 \pm 0.304$ & $0.600 \pm 0.084$ & 0.19 & - & - & $11.31$ & $2.8$ & G212.97-84.04 & 19.5606 & -27.1622 & True & True &  \\
19.5540 & -27.1671 & 14.0 & 11.1 & 95 & $7.832 \pm 0.464$ & $3.410 \pm 0.084$ & - & - & - & $11.57$ & $2.9$ & G212.97-84.04 & 19.5606 & -27.1622 & True & True &  \\
19.5530 & -26.8093 & 9.4 & 8.2 & 167 & $0.784 \pm 0.112$ & $0.771 \pm 0.064$ & 0.31 & - & - & $9.92$ & $2.1$ & G212.97-84.04 & 19.5531 & -26.8093 & False & False &  \\
19.5561 & -26.8117 & 10.9 & 9.6 & 92 & $0.343 \pm 0.147$ & $0.247 \pm 0.068$ & 0.75 & - & - & $9.79$ & $2.0$ & G212.97-84.04 & - & - & False & False &  \\
19.5467 & -26.9700 & 9.8 & 8.8 & 118 & $7.547 \pm 0.102$ & $6.596 \pm 0.054$ & 0.03 & $0.173 \pm 0.000$ & 0 & $0.29$ & $0.2$ & G212.97-84.04 & 19.5463 & -26.9700 & False & False &  \\
19.5359 & -27.0165 & 8.8 & 8.0 & 90 & $0.519 \pm 0.093$ & $0.554 \pm 0.056$ & 0.34 & - & - & $2.56$ & $0.8$ & G212.97-84.04 & - & - & False & False &  \\
19.5328 & -27.0496 & 8.6 & 8.4 & 27 & $0.814 \pm 0.087$ & $0.848 \pm 0.052$ & 0.22 & - & - & $4.55$ & $1.3$ & G212.97-84.04 & 19.5330 & -27.0496 & False & False &  \\
19.5250 & -26.7445 & 9.7 & 8.6 & 167 & $4.825 \pm 0.232$ & $4.327 \pm 0.124$ & 0.05 & - & - & $13.84$ & $3.0$ & G212.97-84.04 & 19.5249 & -26.7447 & True & True &  \\
19.5218 & -26.7448 & 9.5 & 9.0 & 59 & $3.102 \pm 0.236$ & $2.735 \pm 0.125$ & 0.06 & - & - & $13.84$ & $3.0$ & G212.97-84.04 & 19.5249 & -26.7447 & True & True &  \\
19.5167 & -26.9539 & 10.0 & 8.0 & 5 & $0.372 \pm 0.096$ & $0.351 \pm 0.053$ & 0.63 & - & - & $1.96$ & $0.4$ & G212.97-84.04 & 19.5166 & -26.9531 & False & False &  \\
19.4966 & -27.0013 & 9.4 & 8.9 & 116 & $0.310 \pm 0.105$ & $0.277 \pm 0.056$ & 0.70 & $0.225 \pm 0.000$ & 0 & $3.05$ & $0.9$ & G212.97-84.04 & 19.4964 & -27.0013 & False & False &  \\
19.4759 & -26.8784 & 9.3 & 8.5 & 126 & $1.379 \pm 0.099$ & $1.308 \pm 0.055$ & 0.19 & $0.538 \pm 0.035$ & 3 & $6.85$ & $1.4$ & G212.97-84.04 & 19.4759 & -26.8783 & False & False &  \\
19.4656 & -26.8930 & 9.2 & 8.3 & 161 & $1.790 \pm 0.087$ & $1.770 \pm 0.050$ & 0.14 & - & - & $6.48$ & $1.4$ & G212.97-84.04 & 19.4656 & -26.8931 & False & False &  \\
19.4630 & -26.9808 & 13.5 & 10.1 & 83 & $0.581 \pm 0.155$ & $0.323 \pm 0.059$ & 0.68 & $0.062 \pm 0.035$ & 3 & $4.40$ & $1.1$ & G212.97-84.04 & 19.4630 & -26.9812 & False & False &  \\
19.4531 & -26.7984 & 9.3 & 8.8 & 25 & $0.681 \pm 0.147$ & $0.628 \pm 0.080$ & 0.43 & $0.648 \pm 0.060$ & 3 & $11.66$ & $2.5$ & G212.97-84.04 & 19.4530 & -26.7984 & False & False &  \\
19.4408 & -27.0741 & 9.7 & 7.9 & 89 & $0.577 \pm 0.110$ & $0.566 \pm 0.062$ & 0.44 & - & - & $8.17$ & $2.1$ & G212.97-84.04 & - & - & False & False &  \\
19.4236 & -26.9546 & 9.1 & 8.8 & 167 & $2.697 \pm 0.110$ & $2.515 \pm 0.061$ & 0.09 & - & - & $6.60$ & $1.5$ & G212.97-84.04 & 19.4236 & -26.9547 & False & False &  \\
19.3981 & -27.0851 & 8.9 & 7.8 & 160 & $0.608 \pm 0.115$ & $0.664 \pm 0.070$ & 0.41 & - & - & $10.28$ & $2.5$ & G212.97-84.04 & - & - & False & False &  \\ \bottomrule
\end{tabular}
\tablefoot{The columns `Maj' and 'Min' indicate the major and minor axis of the polarised sources and `PA' the position angle. The upper limit on the polarisation fraction $p_\mathrm{1.5GHz}$ was determined for unpolarised radio sources as described in section \ref{sec:fracpolmeasurement}. The columns starting with $z$ are the results of the redshift estimates as detailed in Section \ref{sec:redshifts}. The angular distance to the pointing centre and projected radius to the cluster centre are given by the $\theta_p$ and $r/R_{500}$ columns respectively. The `Visual counterpart id' column indicates whether the source has undergone visual inspection to mark the optical counterpart or whether it was determined automatically as described in Section \ref{sec:optical}.}
  \end{threeparttable}
\end{sideways}
\vfill
\restoregeometry
\endgroup

\section{Cluster catalogue}
\setcounter{table}{0}
\renewcommand*\thetable{\Alph{section}.\arabic{table}}
\begingroup
\fontsize{11pt}{12pt}\selectfont 
\vfill
\begin{sideways}
  \setlength{\tabcolsep}{3pt}
  \begin{threeparttable}
    \caption{First 30 rows of the catalogue of 124 clusters used in this work. The full table is available in electronic form at \red{TODO}.}
    \label{tab:clusters}

\begin{tabular}{@{}lllllllllll@{}}
\toprule
Cluster & PSZ2 Name & \begin{tabular}[c]{@{}l@{}}cRA\\ {[deg]}\end{tabular} & \begin{tabular}[c]{@{}l@{}}cDEC\\ {[deg]}\end{tabular} & \begin{tabular}[c]{@{}l@{}}pRA\\ {[deg]}\end{tabular} & \begin{tabular}[c]{@{}l@{}}pDEC\\ {[deg]}\end{tabular} & cz & \begin{tabular}[c]{@{}l@{}}$M_\mathrm{500}$\\ {[10$^{14}M_\odot$]}\end{tabular} & \begin{tabular}[c]{@{}l@{}}$R_\mathrm{500}$\\ {[Mpc]}\end{tabular} & Dynamical state ($C_\mathrm{SB}$) & Radio Halo \\ \midrule
G006.47+50.54 & G006.49+50.56 & 227.7319 & 5.7481 & -132.2679 & 5.7481 & 0.08 & 7.0 & 1.3 & CC & False \\
G021.09+33.25 & G021.10+33.24 & 248.1795 & 5.5866 & -111.8204 & 5.5866 & 0.15 & 7.8 & 1.3 & CC & False \\
G028.77-33.56 & G028.77-33.56 & 312.0614 & -17.8090 & -47.9387 & -17.8090 & 0.15 & 4.5 & 1.1 & - & False \\
G029.00+44.56 & G029.06+44.55 & 240.5617 & 16.0536 & -119.4383 & 16.0536 & 0.04 & 3.5 & 1.1 & NCC & False \\
G033.78+77.16 & G033.81+77.18 & 207.2437 & 26.5846 & -152.7562 & 26.5846 & 0.06 & 4.5 & 1.1 & CC & False \\
G034.03-76.59 & G033.97-76.61 & 357.9274 & -26.0953 & -2.0904 & -26.0791 & 0.23 & 7.6 & 1.3 & - & False \\
G036.72+14.92 & G036.73+14.93 & 271.1365 & 10.0362 & -88.8633 & 10.0362 & 0.15 & 5.3 & 1.2 & CC & False \\
G039.85-39.98 & G039.85-39.96 & 321.7764 & -12.1686 & -38.2238 & -12.1686 & 0.18 & 5.9 & 1.2 & NCC & False \\
G040.63+77.13 & G040.58+77.12 & 207.3622 & 28.0946 & -152.6517 & 28.1053 & 0.07 & 2.6 & 1.0 & - & True \\
G042.82+56.61 & G042.81+56.61 & 230.6103 & 27.7033 & -129.3896 & 27.7033 & 0.07 & 4.1 & 1.1 & NCC & False \\
G044.22+48.68 & G044.20+48.66 & 239.5878 & 27.2290 & -120.4121 & 27.2290 & 0.09 & 8.8 & 1.4 & CC & False \\
G046.50-49.43 & G046.47-49.44 & 332.5733 & -12.1582 & -27.4267 & -12.1582 & 0.08 & 4.4 & 1.1 & NCC & False \\
G046.88+56.49 & G046.88+56.48 & 231.0292 & 29.9118 & -128.9708 & 29.9118 & 0.11 & 5.3 & 1.2 & NCC & True \\
G048.05+57.17 & G048.10+57.16 & 230.2998 & 30.5986 & -129.7000 & 30.5986 & 0.08 & 3.6 & 1.1 & NCC & False \\
G049.20+30.86 & G049.22+30.87 & 260.0316 & 26.6138 & -99.9683 & 26.6138 & 0.16 & 5.9 & 1.2 & CC & False \\
G049.33+44.38 & G049.32+44.37 & 245.1297 & 29.9055 & -114.8704 & 29.9055 & 0.10 & 3.7 & 1.1 & NCC & True \\
G049.66-49.50 & G049.69-49.46 & 333.6475 & -10.3919 & -26.3525 & -10.3919 & 0.10 & 3.6 & 1.1 & CC & False \\
G053.44-36.26 & G053.44-36.25 & 323.8120 & -1.0503 & -36.1879 & -1.0503 & 0.32 & - & - & NCC & False \\
G053.52+59.54 & G053.53+59.52 & 227.5341 & 33.4851 & -132.4658 & 33.4851 & 0.11 & 5.9 & 1.2 & NCC & False \\
G055.60+31.86 & G055.59+31.85 & 260.6118 & 32.1431 & -99.3883 & 32.1431 & 0.22 & 7.8 & 1.3 & CC & True \\
G055.97-34.88 & G055.95-34.89 & 323.8269 & 1.4312 & -36.1729 & 1.4312 & 0.12 & 6.7 & 1.2 & NCC & False \\
G056.81+36.31 & G056.77+36.32 & 255.6779 & 34.0808 & -104.3221 & 34.0808 & 0.10 & 4.4 & 1.1 & CC & True \\
G057.33+88.01 & G057.80+88.00 & 194.9093 & 27.9350 & -165.0908 & 27.9350 & 0.02 & 7.2 & 1.4 & NCC & False \\
G057.61+34.94 & G057.61+34.93 & 257.4608 & 34.4586 & -102.5392 & 34.4586 & 0.08 & 3.7 & 1.1 & NCC & False \\
G057.92+27.64 & G057.92+27.64 & 266.0695 & 33.0054 & -93.9304 & 33.0054 & 0.08 & 2.7 & 1.0 & CC & False \\
G058.28+18.59 & G058.29+18.55 & 276.2835 & 30.4342 & -83.7167 & 30.4342 & 0.06 & 3.9 & 1.1 & NCC & False \\
G062.42-46.41 & G062.44-46.43 & 335.9655 & -1.6173 & -24.0346 & -1.6173 & 0.09 & 3.5 & 1.0 & NCC & False \\
G062.92+43.70 & G062.94+43.69 & 247.1383 & 39.5458 & -112.8617 & 39.5458 & 0.03 & 2.9 & 1.0 & CC & False \\
G067.23+67.46 & G067.17+67.46 & 216.4929 & 37.8344 & -143.5071 & 37.8344 & 0.17 & 7.2 & 1.3 & CC & False \\
G068.23+15.20 & G068.22+15.18 & 284.3945 & 38.0050 & -75.6237 & 38.0223 & 0.06 & 2.2 & 0.9 & - & False \\ \bottomrule
\end{tabular}
\tablefoot{The cluster centre coordinates from the X-ray profiles are given by \textit{cRA} and \textit{cDEC} and pointing coordinates are given by \textit{pRA} and \textit{pDEC}. The dynamical state was determined as described in Section \ref{sec:datadynstate} and the presence of a radio halo by a literature search.}
  \end{threeparttable}
\end{sideways}
\vfill
\restoregeometry
\endgroup

\end{appendix}

\end{document}